\def\figsize{9.5cm}

\def\rn{}
\def\nn#1 #2{#2. #1}				
\def\nnn#1 #2 #3{#2. #3. #1}			
\def\nnnn#1 #2 #3 #4{#2. #3. #4 #1}		
\def\nnnnn#1 #2 #3 #4 #5{#2. #3. #4 #5. #1}	
\def\dualand{ and\hbox{ }}				
\def\multiand{, and\hbox{ }}				
\def\rf#1;#2;#3;#4;#5 {{\frenchspacing\par\rn#1, #3 {\bf #4}, #5 (#2). \par}}
\def\rg#1;#2;#3;#4;#5;#6 {{\frenchspacing\par\rn#1, #3 {\bf #4}, #5 (#2). \par}}
\def\rfbook#1;#2;#3;#4;#5 {{\frenchspacing\par\rn#1, {\it #3} (#5, #4, #2).\par}}
\def\rfprep#1;#2;#3 {{\par\frenchspacing\rn#1, #3 (#2).\par}}
\def\rfproc#1;#2;#3;#4;#5;#6 {{\frenchspacing\par\rn#1 #2, in {\it #3}, ed. #4 (#5: #6)\par}}
\def\rfprocp#1;#2;#3;#4;#5;#6;#7 {{\frenchspacing\par\rn#1 #2, in {\it #3}, ed. #4 (#5: #6), p#7\par}}

\def\rg#1;#2;#3;#4;#5;#6 {\par\rn#1 #2, {\it #3}, {\bf #4}, #5 (``#6'') \par}
\def\rf#1;#2;#3;#4;#5 {\par\rn#1, {\it #3}, {\bf #4}, #5 (#2)\par}
\def\rfbook#1;#2;#3;#4;#5 {{\frenchspacing\par\rn#1, {\it #3} (#4: #5, #2)\par}}
\def\rfproc#1;#2;#3;#4;#5;#6 {{\frenchspacing\par\rn#1 #2, in {\it #3}, ed. #4 (#5: #6)\par}}
\def\rfprocp#1;#2;#3;#4;#5;#6;#7 {{\frenchspacing\par\rn#1 #2, in {\it #3}, ed. #4 (#5: #6), p#7\par}}
\def\rfprep#1;#2;#3  {{\par\rn#1, #3 (#2)\par}}
\def\rfprepp#1;#2;#3 {{\par\rn#1 #2, #3\par}}

\def\K{{\rm K}}

\def\Mpc{{\rm Mpc}}
\def\hperMpc{\,h/\Mpc}

\def\eV{{\rm eV}}

\def\expec#1{\langle#1\rangle}

\def\etal{{\frenchspacing\it et al.}}
\def\ie{{\frenchspacing\it i.e.}}
\def\eg{{\frenchspacing\it e.g.}}
\def\etc{{\frenchspacing\it etc.}}

\def\beq#1{\begin{equation}\label{#1}}
\def\eeq{\end{equation}}
\def\beqa#1{\begin{eqnarray}\label{#1}}
\def\eeqa{\end{eqnarray}}
\def\eq#1{equation~(\ref{#1})}
\def\Eq#1{Equation~(\ref{#1})}
\def\eqn#1{~(\ref{#1})}

\def\fig#1{Figure~\ref{#1}}
\def\Fig#1{Figure~\ref{#1}}

\def\PowerTable{1}
\def\ParameterTable{2}
\def\ComparisonTable{3}

\def\Sec#1{Section~\ref{#1}}
\def\Sec#1{Section~\ref{#1}}

\def\nskip{\hskip-2mm}

\def\spose#1{\hbox to 0pt{#1\hss}}
\def\simlt{\mathrel{\spose{\lower 3pt\hbox{$\mathchar"218$}}
     \raise 2.0pt\hbox{$\mathchar"13C$}}}
\def\simgt{\mathrel{\spose{\lower 3pt\hbox{$\mathchar"218$}}
     \raise 2.0pt\hbox{$\mathchar"13E$}}}
\def\simpropto{\mathrel{\spose{\lower 3pt\hbox{$\mathchar"218$}}
     \raise 2.0pt\hbox{$\propto$}}}

\def\ed{\end{document}}

\def\Ob{\Omega_b}
\def\Oc{\Omega_{\rm c}}
\def\Ok{\Omega_k}
\def\Ol{\Omega_\Lambda}
\def\Om{\Omega_m}

\def\On{\Omega_\nu}
\def\Or{\Omega_r}
\def\ob{\omega_b}

\def\ocdm{\omega_{\rm c}}
\def\od{\omega_d}

\def\om{\omega_{\rm m}}
\def\on{\omega_\nu}
\def\fb{f_b}
\def\fn{f_\nu}
\def\ns{{n_s}}
\def\nt{{n_t}}
\def\al{\alpha}
\def\Ot{\Omega_{\rm tot}}
\def\As{A_s}
\def\At{A_t}
\def\Ap{A_{\rm peak}}
\def\Apivot{A_{\rm pivot}}
\def\zeq{{z_{\rm eq}}}
\def\zrec{{z_{\rm rec}}}
\def\zion{{z_{\rm ion}}}
\def\zacc{{z_{\rm acc}}}
\def\teq{t_{\rm eq}}
\def\trec{t_{\rm req}}
\def\tion{t_{\rm ion}}
\def\tacc{t_{\rm acc}}
\def\tnow{t_{\rm now}}
\def\age{\tnow}
\def\Th{\Theta_s}
\def\lA{\l_A}
\def\dA{d_A}
\def\dV{d_V}
\def\Mnu{M_\nu}
\def\Nnu{N_\nu}
\def\Qnl{Q_{\rm nl}}
\def\Op{A_\tau}
\def\kstar{k_*}
\def\Tcmb{T_{\rm cmb}}

\def\rhob{\rho_{\rm b}}
\def\rhoc{\rho_{\rm c}}

\def\rhon{\rho_\nu}

\def\rhol{\rho_\Lambda}

\def\rhom{\rho_{\rm m}}

\def\rhohalo{\rho_{\rm halo}}

\def\xib{\xi_{\rm b}}
\def\xic{\xi_{\rm c}}

\def\xin{\xi_\nu}

\def\ng{n_\gamma}

\def\Avac{A_\Lambda}

\def\Teq{T_{\rm eq}}

\def\rdamp{r_{\rm damp}}
\def\rsound{r_{\rm s}}
\def\req{r_{\rm eq}}
\def\sigmagal{\sigma^*_{\rm gal}}

\def\Ms{M_\odot}

\def\Veff{V_{\rm eff}}

\def\beq#1{\begin{equation}\label{#1}}
\def\eeq{\end{equation}}
\def\beqa#1{\begin{eqnarray}\label{#1}}
\def\eeqa{\end{eqnarray}}
\def\eq#1{equation~(\ref{#1})}
\def\Eq#1{Equation~(\ref{#1})}
\def\eqn#1{~(\ref{#1})}

\def\nbar{{\bar n}}

\def\w{w}

\def\k{{\bf k}}
\def\r{{\bf r}}
\def\s{{\bf s}}

\def\rhat{\widehat{\bf r}}

\def\deltahat{\widehat{\delta}}
\def\xihat{\widehat{\xi}}

\def\p{{\bf p}}

\def\q{{\bf q}}

\def\r{{\bf r}}

\def\F{{\bf F}}

\def\L{{\cal L}}

\def\kmax{k_{\rm max}}

\def\l{\ell}

\def\Pg{P_{\rm g}}
\def\Pghat{\widehat{P}_{\rm g}}
\def\Pgg{P_{\rm gg}}
\def\Pgv{P_{\rm gv}}
\def\Pvv{P_{\rm vv}}
\def\Pgghat{\widehat{P}_{\rm gg}}
\def\Pgvhat{\widehat{P}_{\rm gv}}
\def\Pvvhat{\widehat{P}_{\rm vv}}

\def\Pmono{P^{\rm s}_0}
\def\Pquad{P^{\rm s}_2}
\def\Phexa{P^{\rm s}_4}
\def\Pmonohat{\widehat{P}^{\rm s}_0}
\def\rgv{r_{\rm gv}}

\def\ignore#1{}

\def\simless{\mathbin{\lower 3pt\hbox
        {$\,\rlap{\raise 5pt\hbox{$\char'074$}}\mathchar"7218\,$}}} 
\def\simgreat{\mathbin{\lower 3pt\hbox
        {$\,\rlap{\raise 5pt\hbox{$\char'076$}}\mathchar"7218\,$}}} 

\documentclass[twocolumn,amsmath,nofootinbib]{revtex4} 
\usepackage{url}
\begin{document}
\input{epsf.sty}

\def\mit{1}
\def\tucson{2}
\def\pton{3}
\def\osu{4}
\def\nyu{5}
\def\chicago{6}
\def\fnal{7}
\def\tokyo{8}
\def\colorado{9}
\def\portsmouth{10}
\def\jadwin{11}
\def\lbnl{12}
\def\psu{13}
\def\ictp{14}
\def\hopkins{15}
\def\drexel{16}
\def\case{17}
\def\lanl{18}
\def\washington{19}
\def\saao{20}
\def\uct{21}
\def\apo{22}
\def\barcelona{23}
\def\pitt{24}
\def\tokyoastro{25}
\def\harvard{26}
\def\flagstaff{27}
\def\penn{28}
\def\jpl{29}
\def\caltech{30}
\def\efi{31}
\def\gatan{32}
\def\sussex{33}
\def\seoul{34}
\def\rochester{35}
\def\hawaii{36}

\def\affilmrk#1{$^{#1}$}
\def\affilmk#1#2{$^{#1}$#2;}

\title{Cosmological Constraints from the SDSS Luminous Red Galaxies}

\author{
Max Tegmark\affilmrk{\mit}, 
Daniel J.~Eisenstein\affilmrk{\tucson},
Michael A.~Strauss\affilmrk{\pton},
David H.~Weinberg\affilmrk{\osu},
Michael R.~Blanton\affilmrk{\nyu}, 
Joshua A. Frieman\affilmrk{\chicago,\fnal},
Masataka Fukugita\affilmrk{\tokyo},
James E.~Gunn\affilmrk{\pton},
Andrew J.~S.~Hamilton\affilmrk{\colorado},
Gillian R.~Knapp\affilmrk{\pton},
Robert C.~Nichol\affilmrk{\portsmouth},
Jeremiah P.~Ostriker\affilmrk{\pton},
Nikhil Padmanabhan\affilmrk{\jadwin},
Will J.~Percival\affilmrk{\portsmouth},
David J.~Schlegel\affilmrk{\lbnl}, 
Donald P.~Schneider\affilmrk{\psu}, 
Roman Scoccimarro\affilmrk{\nyu},
Uro\v s Seljak\affilmrk{\ictp,\jadwin}, 
Hee-Jong Seo\affilmrk{\tucson},
Molly Swanson\affilmrk{\mit},
Alexander S.~Szalay\affilmrk{\hopkins},
Michael S.~Vogeley\affilmrk{\drexel},
Jaiyul Yoo\affilmrk{\osu},
Idit Zehavi\affilmrk{\case}, 
Kevork Abazajian\affilmrk{\lanl},
Scott F.~Anderson\affilmrk{\washington}, 
James Annis\affilmrk{\fnal}, 
Neta A.~Bahcall\affilmrk{\pton},
Bruce Bassett\affilmrk{\saao,\uct},
Andreas Berlind\affilmrk{\nyu}, 
Jon Brinkmann\affilmrk{\apo},
Tam\'as Budavari\affilmrk{\hopkins}, 
Francisco Castander\affilmrk{\barcelona},
Andrew Connolly\affilmrk{\pitt},
Istvan Csabai\affilmrk{\hopkins},
Mamoru Doi\affilmrk{\tokyoastro},  
Douglas P.~Finkbeiner\affilmrk{\pton,\harvard},
Bruce Gillespie\affilmrk{\apo},
Karl Glazebrook\affilmrk{\hopkins},
Gregory S.~Hennessy\affilmrk{\flagstaff},
David W.~Hogg\affilmrk{\nyu},
\v Zeljko Ivezi\'c\affilmrk{\washington,\pton},
Bhuvnesh Jain\affilmrk{\penn},
David Johnston\affilmrk{\jpl,\caltech}, 
Stephen Kent\affilmrk{\fnal},
Donald Q.~Lamb\affilmrk{\chicago,\efi},
Brian C.~Lee\affilmrk{\gatan,\lbnl},
Huan Lin\affilmrk{\fnal},
Jon Loveday\affilmrk{\sussex},
Robert H.~Lupton\affilmrk{\pton},
Jeffrey A.~Munn\affilmrk{\flagstaff}, 
Kaike Pan\affilmrk{\apo},
Changbom Park\affilmrk{\seoul}, 
John Peoples\affilmrk{\fnal}, 
Jeffrey R.~Pier\affilmrk{\flagstaff},
Adrian Pope\affilmrk{\hopkins}, 
Michael Richmond\affilmrk{\rochester},
Constance Rockosi\affilmrk{\chicago}, 
Ryan Scranton\affilmrk{\pitt},
Ravi K.~Sheth\affilmrk{\penn}, 
Albert Stebbins\affilmrk{\fnal},
Christopher Stoughton\affilmrk{\fnal}, 
Istv\'an Szapudi\affilmrk{\hawaii}, 
Douglas L. Tucker\affilmrk{\fnal},
Daniel E. Vanden Berk\affilmrk{\pitt},
Brian Yanny\affilmrk{\fnal},
Donald G.~York\affilmrk{\chicago,\efi}
}

\address{
\parshape 1 -3cm 24cm
\affilmk{\mit}{Dept. of Physics, Massachusetts Institute of Technology, Cambridge, MA 02139, USA}
\affilmk{\tucson}{Department of Astronomy, University of Arizona, Tucson, AZ 85721, USA}
\affilmk{\pton}{Princeton University Observatory, Princeton, NJ 08544, USA}
\affilmk{\osu}{Dept.~of Astronomy, Ohio State University, Columbus, OH 43210, USA}
\affilmk{\nyu}{Center for Cosmology and Particle Physics, Dept.~of Physics, New York University, 4 Washington Pl., New York, NY 10003, USA}
\affilmk{\chicago}{Center for Cosmological Physics and Department of Astronomy \& Astrophysics, Univ.~of Chicago, Chicago, IL 60637, USA}
\affilmk{\fnal}{Fermi National Accelerator Laboratory, P.O. Box 500, Batavia, IL 60510, USA}
\affilmk{\tokyo}{Inst.~for Cosmic Ray Research, Univ.~of Tokyo, 5-1-5, Kashiwanoha, Kashiwa, Chiba, 277-8582, Japan}
\affilmk{\colorado}{JILA and Dept.~of Astrophysical and Planetary Sciences, Univ. of Colorado, Boulder, CO 80309, USA}
\affilmk{\portsmouth}{Inst. of Cosmology \& Gravitation, Univ.~of Portsmouth, Portsmouth, P01 2EG, United Kingdom}
\affilmk{\jadwin}{Dept.~of Physics, Princeton University, Princeton, NJ 08544, USA}
\affilmk{\lbnl}{Lawrence Berkeley National Laboratory, Berkeley, CA 94720, USA}
\affilmk{\psu}{Dept.~of Astronomy and Astrophysics, Pennsylvania State University, University Park, PA 16802, USA}
\affilmk{\ictp}{International Center for Theoretical Physics, Strada Costiera 11, 34014 Trieste, Italy}
\affilmk{\hopkins}{Department of Physics and Astronomy, The Johns Hopkins University, 3701 San Martin Drive, Baltimore, MD 21218, USA}
\affilmk{\drexel}{Dept.~of Physics, Drexel University, Philadelphia, PA 19104, USA}
\affilmk{\case}{Dept.~of Astronomy, Case Western Reserve University, 10900 Euclid Avenue, Cleveland, OH 44106-7215}
\affilmk{\lanl}{Theoretical Division, Los Alamos National Laboratory, Los Alamos, NM 87545, USA}
\affilmk{\washington}{Dept.~of Astronomy, Univ.~of Washington, Box 351580, Seattle, WA 98195}
\affilmk{\saao}{South African Astronomical Observatory, Cape Town, South Africa}
\affilmk{\uct}{Applied Mathematics Dept., Univ.~of Cape Town, Cape Town, South Africa}
\affilmk{\apo}{Apache Point Observatory, 2001 Apache Point Rd, Sunspot, NM 88349-0059, USA}
\affilmk{\barcelona}{Institut d'Estudis Espacials de Catalunya/CSIC, Campus UAB, 08034 Barcelona, Spain}
\affilmk{\pitt}{University of Pittsburgh, Department of Physics and Astronomy, 3941 O'Hara Street, Pittsburgh, PA 15260, USA}
\affilmk{\tokyoastro}{Inst.~of Astronomy, Univ.~of Tokyo, Osawa 2-21-1, Mitaka, Tokyo, 181-0015, Japan}
\affilmk{\harvard}{Harvard-Smithsonian Center for Astrophysics, 60 Garden Street, MS46, Cambridge, MA 02138}
\affilmk{\flagstaff}{U.S. Naval Observatory, Flagstaff Station, 10391 W.~Naval Obs.~Rd., Flagstaff, AZ 86001-8521, USA}
\affilmk{\penn}{Department of Physics, University of Pennsylvania, Philadelphia, PA 19104, USA}
\affilmk{\jpl}{Jet Propulsion Laboratory, 4800 Oak Grove Dr., Pasadena CA, 91109, USA}
\affilmk{\caltech}{California Inst.~of Technology, 1200 East California Blvd., Pasadena, CA 91125, USA}
\affilmk{\efi}{Enrico Fermi Institute, University of Chicago, Chicago, IL 60637, USA}
\affilmk{\gatan}{Gatan Inc., Pleasanton, CA 94588}
\affilmk{\sussex}{Sussex Astronomy Centre, University of Sussex, Falmer, Brighton BN1 9QJ, UK}
\affilmk{\seoul}{Department of Astronomy, Seoul National University, 151-742, Korea}
\affilmk{\rochester}{Physics Dept., Rochester Inst. of Technology, 1 Lomb Memorial Dr, Rochester, NY 14623, USA}
\affilmk{\hawaii}{Institute for Astronomy, University of Hawaii, 2680, Woodlawn Drive, Honolulu, HI 96822, USA}
}

\date{Submitted to Phys.~Rev.~D. August 22 2006, revised October 10, accepted October 26}

\begin{abstract}
\clearpage
We measure the large-scale real-space power spectrum $P(k)$ using luminous red 
galaxies (LRGs) in the Sloan Digital Sky Survey (SDSS) and use this measurement to 
sharpen constraints on cosmological parameters from the Wilkinson Microwave Anisotropy Probe (WMAP). 
We employ a matrix-based power spectrum estimation method using Pseudo-Karhunen-Lo\`eve eigenmodes,
producing 
uncorrelated minimum-variance measurements in 20 $k$-bands of 
both the clustering power and its anisotropy due to redshift-space distortions,
with narrow and well-behaved window functions in the range
$0.01\hperMpc < k < 0.2\hperMpc$. 
Results from the LRG and main galaxy samples are consistent, with the former providing higher signal-to-noise.
Our results are robust to omitting angular and 
radial density fluctuations and are consistent between different
parts of the sky. They provide a striking 
confirmation of the predicted large-scale $\Lambda$CDM 
power spectrum.
Combining only SDSS LRG and WMAP data places robust constraints on many cosmological parameters
that complement prior analyses of multiple data sets.
The LRGs provide independent cross-checks on $\Om$ and the baryon fraction in good agreement with WMAP.
Within the context of flat $\Lambda$CDM models, our LRG measurements complement WMAP by sharpening 
the constraints on the matter density, the neutrino density and the tensor amplitude by about a factor of two, giving 
$\Om=0.24\pm 0.02$ ($1\sigma$), $\sum m_\nu\simlt 0.9$ eV (95\%) and $r<0.3$ (95\%).
Baryon oscillations are clearly detected and provide a robust measurement of the comoving distance
to the median survey redshift $z=0.35$ independent of curvature and dark energy properties. 
Within the $\Lambda$CDM framework, our power spectrum measurement 
improves the evidence for spatial flatness,
sharpening the curvature constraint $\Ot=1.05\pm 0.05$ from WMAP alone to $\Ot=1.003\pm 0.010$.
Assuming $\Ot=1$, the equation of state parameter is constrained to $w=-0.94\pm 0.09$,
indicating the potential for more ambitious future LRG measurements to provide precision tests of the nature of dark energy.
All these constraints are essentially independent of scales $k>0.1h$/Mpc and associated nonlinear complications, 
yet agree well with more aggressive published analyses where nonlinear modeling is crucial. 
\end{abstract}


  
\maketitle

\def\v{{\bf v}}

\setcounter{footnote}{0}

\def\percent{\%}

\section{Introduction}
\label{IntroSec}

The dramatic
recent progress by the Wilkinson Microwave Anisotropy Probe (WMAP) and other experiments \cite{Hinshaw06,Page06,Masi05,Sievers05}
measuring the cosmic microwave background (CMB) has made non-CMB experiments even more important in the 
quest to constrain cosmological models and their free parameters.
These non-CMB constraints are crucially needed for breaking CMB degeneracies
\cite{parameters2,EfstathiouBond99}; for instance, 
WMAP alone is consistent with a closed universe with
Hubble parameter $h=0.3$ and no cosmological constant \cite{Spergel06}.
As long as the non-CMB constraints are less reliable and precise than the CMB, they will be the limiting factor and 
weakest link in the precision cosmology endeavor.
Much of the near-term progress in cosmology will therefore be driven by
reductions in statistical and systematic uncertainties of
non-CMB probes of the cosmic expansion history (e.g., SN Ia) and the matter power spectrum
(e.g., Lyman $\alpha$ Forest, galaxy clustering and motions,
gravitational lensing, cluster studies and 21 cm tomography).

The cosmological constraining power of 
three-dimensional maps of the Universe provided by
galaxy redshift surveys has motivated ever more ambitious
observational efforts such as 
the CfA/UZC \cite{Huchra90,Falco99},
LCRS \cite{Shectman96}, PSCz \cite{Saunders00}, 
DEEP \cite{Cooper06},
2dFGRS \cite{Colless03} and SDSS \cite{York00} projects,
resulting in progressively more accurate measurements of the 
galaxy power spectrum $P(k)$ 
\cite{CFApower1,Fisher93,CFApower2,SSRSpower,FKP,QDOTpower,STROMLOpower,LCRSpower,uzc,pscz,pscz2,Percival01,2df,sdsspower,Cole05,Huetsi06a}.
Constraints on cosmological models from these data sets have been most robust when the galaxy clustering could be
measured on large scales where one has confidence in the modeling of nonlinear clustering and biasing
(\eg, \cite{consistent,Spergel03,sdsspars,sdss3dkl,sdsslyaf,sdssbump,Sanchez06,Huetsi06b,Spergel06,Seljak06,Viel06,sdsslrgcl,Blake06}).

Our goal in this paper therefore is to measure $P(k)$ on large scales using the SDSS galaxy redshift survey 
in a way that is maximally useful for cosmological parameter estimation, 
and to explore the resulting constraints on cosmological models.
The emphasis of our cosmological analysis will be on 
elucidating the links between cosmological parameters and 
observable features of the WMAP and SDSS power spectra, and on 
how these two data sets alone provide tight and robust constraints on many parameters
that complement more aggressive but more 
systematics-prone analyses of multiple data sets.

In a parallel paper, Percival {\etal} \cite{Percival06} present a power spectrum
analysis of the Main Galaxy and LRG samples from the SDSS DR5 data set \cite{sdssdr5},
which is a superset of the data used here. There are a number of differences in the analysis methods. 
Percival {\etal} use an FFT-based method to estimate the angle-averaged (monopole)
redshift-space galaxy power spectrum. We use a Pseudo-Karhunen-Lo\`eve method \cite{VogeleySzalay96,karhunen}
(see further discussion and references below)
to estimate the real space (as opposed to redshift space) galaxy power spectrum, using 
finger-of-god compression and linear theory to remove
redshift-space distortion effects. In addition, the many
technical decisions that go into these analyses, regarding
completeness corrections, angular masks, K-corrections and
so forth, were made independently for the two papers, and they present different tests for systematic uncertainties.
Despite these many differences of detail, our conclusions agree
to the extent that they overlap (as discussed in \Sec{PercivalCompSec} and Appendix~\ref{MethodComparisonSec}), a
reassuring indication of the robustness of the results.

\subsection{Relation between different samples}

The amount of information in a galaxy redshift survey about the galaxy power spectrum $\Pg(k)$ and cosmological parameters 
depends not on the number of galaxies {\it per se}, but on the {\it effective volume} of the survey, defined by \cite{galfisher} as
\beq{VeffDefEq}
\Veff(k)\equiv\int\left[{\nbar(\r)\Pg(k)\over 1+\nbar(\r)\Pg(k)}\right]^2 d^3r,
\eeq
where $\nbar(\r)$ is the expected 
number density of galaxies in the survey in the absence of clustering, and the FKP approximation of 
\cite{FKP} has been used.
The power spectrum error bars scale approximately as $\Delta\Pg(k)\propto\Veff(k)^{-1/2}$, which for a 
fixed power $\Pg$ is minimized if a fixed total number of galaxies are spaced 
with density $\nbar\sim\Pg^{-1}$ \cite{Kaiser86}.
The SDSS Luminous Red Galaxy (LRG) sample was designed \cite{EisensteinLRGselection,StraussTargetSelection} to contain such 
``Goldilocks'' galaxies with a just-right number density for probing the power around the baryon wiggle 
scale $k\sim (0.05-0.1)h/$Mpc.
For comparison, the SDSS main galaxy sample \cite{StraussTargetSelection} is much denser and is 
dominated by sample variance on these scales, whereas the
SDSS quasar sample \cite{RichardsQuasarSelection} is much sparser and is dominated by Poisson shot noise.
As shown in \cite{sdssbump}, the effective volume of the LRG sample is about six times larger than 
that of the SDSS main galaxies even though the number of LRGs is an order of magnitude lower, 
and the LRG volume is over ten times larger than
that of the 2dFGRS. 
These scalings are confirmed by our results below, which show that $(\Delta\Pg/\Pg)^2$ 
on large scales is about six times smaller for the SDSS LRGs than for the main sample galaxies. 
This gain results both from sampling a 
larger volume, and from the fact that the LRG
are more strongly clustered (biased) than are ordinary galaxies; $\Pg$
for LRGs is about 3 times larger than for the main galaxy sample.

We will therefore focus our analysis on the SDSS LRG sample. 
Although we also measure the SDSS main sample power spectrum,
it adds very little in terms of statistical constraining 
power; increasing the effective volume by $15\%$ cuts the error bar $\Delta P$ by only about $(1+0.15)^{1/2}-1\sim 7\%$.
This tiny improvement is easily outweighed by the gain in simplicity from analyzing LRGs alone, 
where (as we will see) complications such as redshift-dependence of clustering properties are substantially smaller.

A complementary approach 
implemented
by \cite{sdsslrgcl,Blake06} is to measure the angular clustering of 
SDSS LRGs with photometric redshifts, compensating for the loss of radial information with 
an order of magnitude more galaxies extending out to higher redshift. 
We will see that this gives comparable or slightly smaller error bars on very large scales $k\simlt 0.02$, but slightly 
larger error bars on the smaller scales that dominate our cosmological constraints; this is because the number
of modes down to a given scale $k$ grows as $k^3$ for our three-dimensional spectroscopic analysis, whereas they grow
only as $k^2$ for a 2-dimensional angular analysis.

\subsection{Relation between different methods}

In the recent literature, two-point galaxy clustering has been quantified using a variety of estimators of both power spectra and 
correlation functions.
The most recent power spectrum measurements for both the 2dFGRS \cite{Percival01,Cole05} 
and the SDSS \cite{Huetsi06a,Huetsi06b,Percival06} have 
all interpolated the galaxy density field onto a cubic grid and measured $P(k)$ using a Fast Fourier Transform (FFT).

Appendix~\ref{MethodComparisonSec} shows that as long as discretization errors from the FFT gridding are negligible, 
this procedure is mathematically equivalent to measuring the correlation function with a weighted version of the
standard ``DD-2DR+RR'' method \cite{LandySzalay93,Hamilton93}, multiplying by ``RR'' and then Fourier transforming. 
Thus the only advantage of the FFT approach is numerical speedup, 
and comparing the results with recent correlation function analyses such as 
\cite{Hawkins03,Eisenstein04,Zehavi05,sdssbump} will provide useful consistency checks.

Another approach, pioneered by \cite{VogeleySzalay96}, has been to 
construct ``lossless'' estimators of the power spectrum with the smallest
error bars that are possible based on information theory 
\cite{Karhunen47,VogeleySzalay96,karhunen,uzc,pscz,2df,sdsspower,sdss3dkl,Hamilton05}.
We will travel this complementary route in the present paper, following the matrix-based 
Pseudo Karhunen-Lo\`eve (PKL) eigenmode
method described in \cite{sdsspower}, as
it has the following advantages:
\begin{enumerate}
\item It produces power spectrum measurements with uncorrelated error bars.
\item It produces narrow and well-behaved window functions.
\item It is lossless in the information theory sense.
\item It treats redshift distortions without the small-angle approximation.
\item It readily incorporates the so-called integral constraint \cite{PeacockNicholson91,Fisher93}, which can 
otherwise artificially suppress large-scale power.
\item It allows testing for systematics that produce excess power in angular or radial modes.
\end{enumerate}
These properties make the results of the PKL-method very easy to interpret and use.
The main disadvantage is that the PKL-method is numerically painful to implement and execute; our PKL analysis described below required 
about 
a terabyte of disk space for matrix storage and about a year of CPU time, which contributed to the long gestation period of this paper.

The rest of this paper is organized as follows.
We describe our galaxy samples and our modeling of them in \Sec{DataSec} and measure their power spectra in \Sec{PowerSec}.
We explore what this does and does not reveal about cosmological parameters in \Sec{CosmoSec}.
We summarize our conclusions and place them in context in \Sec{ConclSec}.
Further details about analysis techniques are given in Appendix A.

\section{Galaxy data}
\label{DataSec}

\begin{figure} 
\centerline{\epsfxsize=\figsize\epsffile{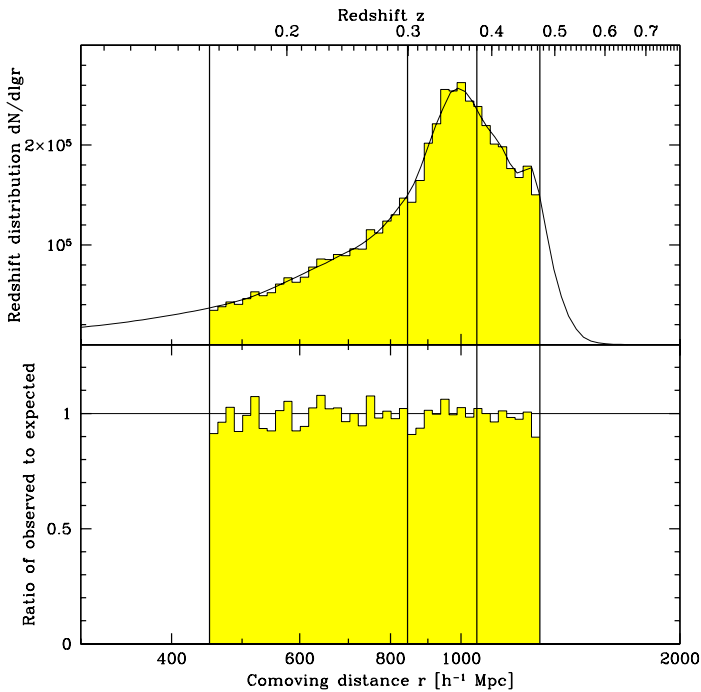}}
\caption[1]{\label{zhistFig}\footnotesize%
The redshift distribution of the luminous red galaxies used
is shown as a histogram and compared with 
the expected distribution in the absence of clustering, 
$\ln 10\times \int \nbar(r)r^3 d\Omega$ (solid curve)
in comoving coordinates assuming a flat $\Omega_\Lambda=0.75$ cosmology.
The bottom panel shows the ratio of observed and expected distributions.
The four vertical lines delimit the NEAR, MID and FAR samples.
}
\end{figure}

\begin{figure*} 
\centerline{\epsfxsize=18cm\epsffile{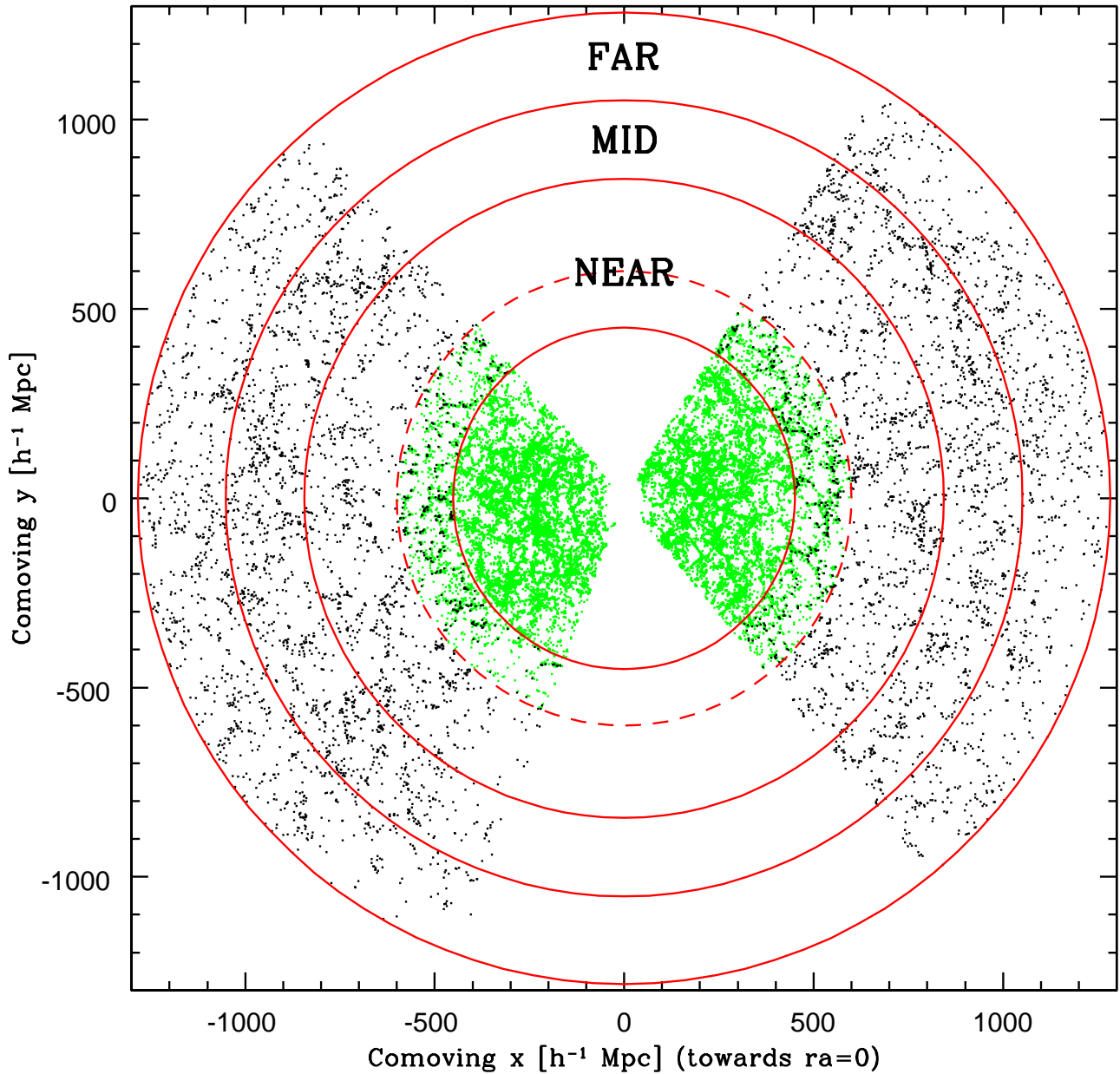}}
\caption[1]{\label{sliceFig}\footnotesize%
The distribution of the 6{,}476 
LRGs (black) and 32{,}417 main galaxies (green/grey) that are within $1.25^\circ$ of the Equatorial plane.
The solid circles indicate the boundaries of our NEAR, MID and FAR subsamples. The ``safe13'' main galaxy sample analyzed here and in \cite{sdsspower} is more local, 
extending out only to $600h^{-1}$ Mpc (dashed circle).
}
\end{figure*}

\begin{figure*} 
\vskip-4.7cm
\centerline{\epsfxsize=18cm\epsffile{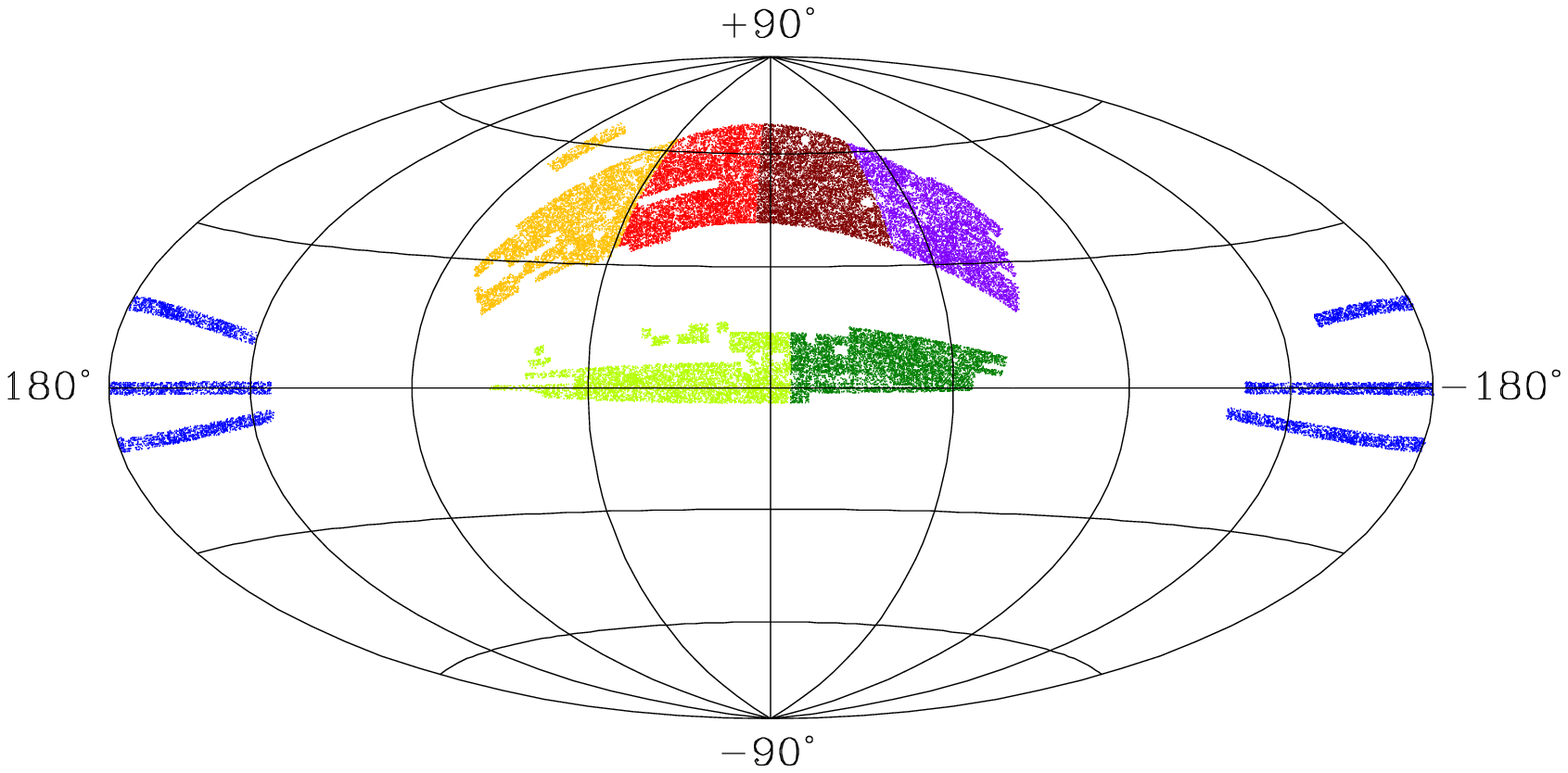}}
\vskip-6cm
\caption[1]{\label{AitoffFig}\footnotesize%
The angular distribution of our LRGs is shown in Hammer-Aitoff projection in celestial coordinates, 
with the seven colors/greys indicating the 
seven angular subsamples that we analyze.
}
\end{figure*}

The SDSS \cite{York00,Stoughton02} uses a mosaic CCD camera \cite{Gunn98} on a dedicated telescope \cite{Gunn06}
to image the sky in five photometric
bandpasses denoted $u$, $g$, $r$, $i$ and $z$ \cite{Fukugita96}.
After astrometric calibration \cite{Pier03}, 
photometric data reduction \cite{Lupton03,Lupton06} and photometric calibration \cite{Hogg01,Smith02,Ivezic04,Tucker06}, 
galaxies are selected for spectroscopic observations \cite{StraussTargetSelection}. To a good approximation, 
the main galaxy sample consists of 
all galaxies with $r$-band apparent Petrosian magnitude $r<17.77$ after correction for 
reddening as per \cite{SFD}; there are about
90 such galaxies per square degree, with a median redshift of 0.1 and a tail out to
$z\sim 0.25$. 
Galaxy spectra are also measured for the LRG sample \cite{EisensteinLRGselection}, 
targeting an additional $\sim 12$ galaxies per square degree, enforcing $r<19.5$ and color-magnitude cuts
described in \cite{EisensteinLRGselection,sdssbump} that select mainly luminous elliptical/early type
galaxies at redshifts up to $\sim 0.5$.
These targets are assigned to spectroscopic plates of diameter $2.98^\circ$ 
into which 640 optical fibers are plugged by an adaptive 
tiling algorithm \cite{Blanton03}, feeding a pair of 
CCD spectrographs \cite{Uomoto04}, after which the
spectroscopic data reduction and redshift determination are performed by 
automated pipelines. 
The rms galaxy redshift errors are of order 
30 km/s for main galaxies and 50 km/s for LRGs \cite{EisensteinLRGselection}, 
hence negligible for the purposes of the present paper.

Our analysis is based on $58,360$ LRGs and $285,804$ main galaxies (the ``safe13'' cut) from the $390,288$ galaxies in the 
4th SDSS data release (``DR4") \cite{sdssdr4}, processed via the SDSS data repository at New York University
\cite{sdssvagc}.   
The details of how these samples were processed and modeled are given
in Appendix A of \cite{sdsspower} and in \cite{sdssbump}. 
The bottom line is that each sample is completely specified by three entities:
\begin{enumerate}
\item The galaxy positions (RA, Dec and comoving redshift
space distance $r$ for each galaxy),
\item The radial selection function $\nbar(r)$, which gives the expected 
number density of galaxies as a function of distance,
\item The angular selection function $\nbar(\rhat)$, which gives the 
completeness as a function of direction in the sky, specified in a set of spherical polygons \cite{mangle}.
\end{enumerate}
Our samples are constructed so that 
their three-dimensional selection function is separable, \ie, simply the product 
$\nbar(\r)=\nbar(\rhat)\nbar(r)$ of an angular and a radial part; 
here $r\equiv |\r|$ and $\rhat \equiv \r/r$ are the comoving radial distance
and the unit vector corresponding to the position $\r$.
The effective sky area covered is
$\Omega\equiv\int\nbar(\rhat)d\Omega\approx 4259$ square degrees, and the 
typical completeness $\nbar(\rhat)$ exceeds 90\%. 
The radial selection function $\nbar(r)$ for the LRGs is the one constructed and described in detail in \cite{Zehavi05,sdssbump},
based on integrating an empirical model of the luminosity function and color distribution of the LRGs
against the luminosity-color selection boundaries of the sample. 
\Fig{zhistFig} shows that it agrees well with the observed galaxy distribution. 
The conversion from redshift $z$ to comoving distance was made for a flat $\Lambda$CDM cosmological model with $\Om=0.25$.
If a different cosmological model is used for this conversion, then our measured dimensionless power spectrum $k^3 P(k)$ 
is dilated very slightly (by $\simlt 1\%$ for models consistent with our measurements) along the $k$-axis; 
we include this dilation effect in our cosmological parameter
analysis as described in Appendix~\ref{LikelihoodSec}.
 
For systematics testing and numerical purposes, we also analyze a variety of sub-volumes in the LRG sample. 
We split the sample into three radial slices, labeled 
NEAR ($0.155<z<0.300$),
MID  ($0.300<z<0.380$) and
FAR  ($0.380<z<0.474$),
containing roughly equal numbers of galaxies, as illustrated in \fig{sliceFig}.
Their galaxy-weighted mean redshifts are $0.235$, $0.342$ and $0.421$, respectively. 
We also split the sample into the seven angular regions illustrated in \fig{AitoffFig}, each again containing 
roughly the same number of galaxies.   

It is worth emphasizing that the LRGs constitute a remarkably clean and uniform galaxy sample,
containing the same type of galaxy (luminous early-types) at all redshifts.
Not only is it nearly complete ($\nbar(\rhat)\sim 1$ as mentioned above), but 
it is close to volume-limited for $z\simlt 0.38$ \cite{EisensteinLRGselection,sdssbump}, 
\ie, for our NEAR and MID slices.
	
\section{Power spectrum measurements}
\label{PowerSec}

We measure the power spectrum of our various samples using the PKL method described in \cite{sdsspower}.
We follow the procedure of \cite{sdsspower} exactly, with some 
additional numerical improvements described in Appendix A, so we merely summarize the process very briefly here.
The first step is to adjust the galaxy redshifts slightly to compress so-called fingers-of-god (FOGs),
virialized galaxy clusters that appear elongated along the line-of-sight in redshift space;
we do this 
with several different thresholds and return to how this affects the results in \Sec{LRGrobustnessSec}.
The LRGs are not just brightest cluster galaxies; about 20\% of them appear to reside in a dark matter
halo with one or more other LRG's.
The second step is to expand the three-dimensional galaxy density field in $N$ three-dimensional functions termed PKL-eigenmodes,
whose variance and covariance retain essentially all the information about the $k<0.2h/$Mpc power spectrum from the galaxy catalog.
We use $N=42{,}000$ modes for the LRG sample and $4000$ modes for the main sample, reflecting their very different effective volumes.
The third step is estimating the power spectrum from quadratic combinations of these PKL mode coefficients by a matrix-based process 
analogous to the standard procedure for measuring CMB power spectra from pixelized CMB maps. 
The second and third steps are mathematically straightforward but, as mentioned, numerically demanding for large $N$.

\begin{figure*} 
\centerline{\epsfxsize=18cm\epsffile{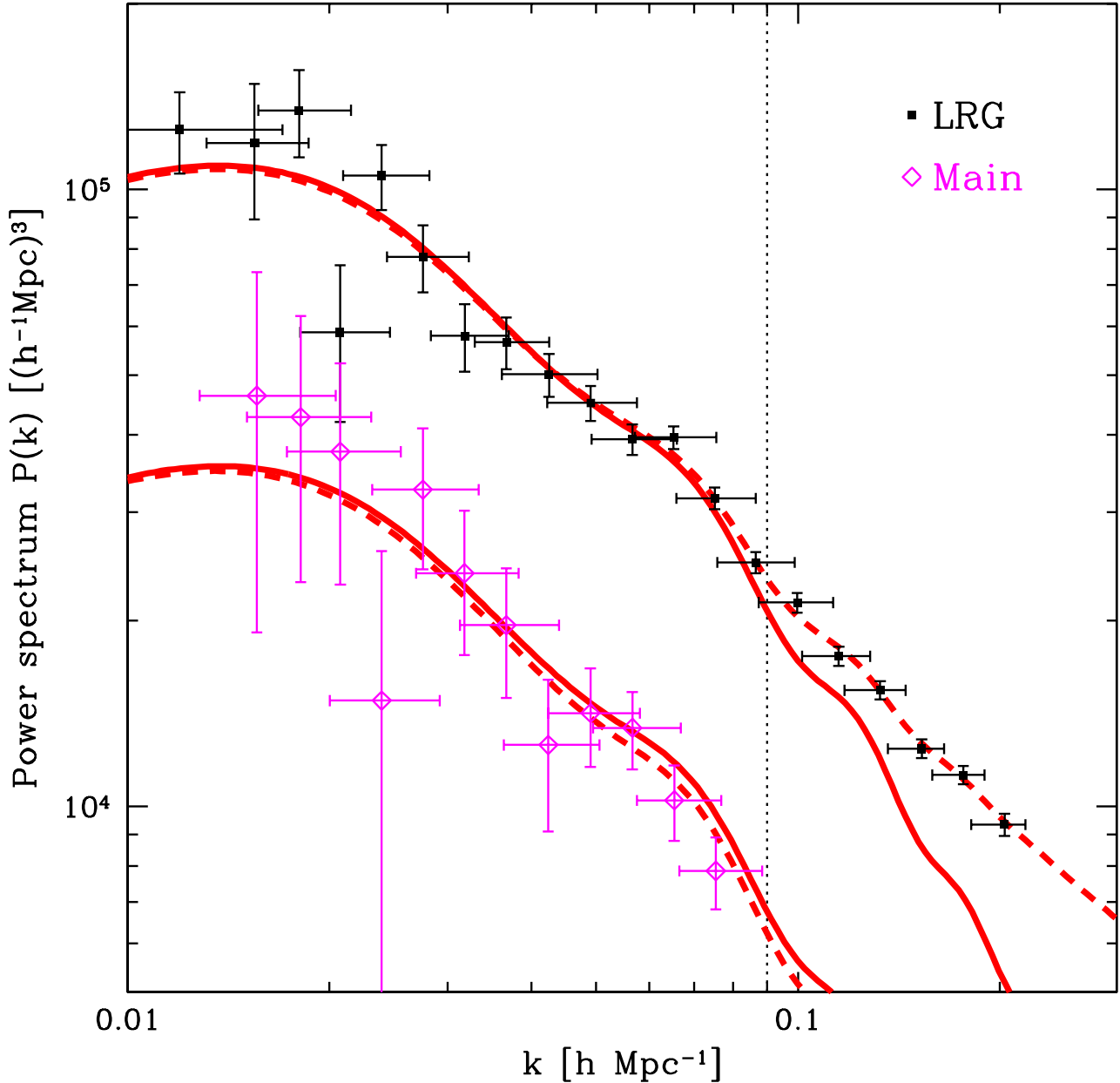}}
\caption[1]{\label{all2powerFig}\footnotesize%
Measured power spectra for the full LRG and main galaxy samples.
Errors are uncorrelated and full window functions are shown in \fig{Wfig}.
The solid curves correspond to 
the linear theory $\Lambda$CDM fits to WMAP3 alone from Table 5 of \cite{Spergel06},
normalized to galaxy bias 
$b=1.9$ (top) and $b=1.1$ (bottom) 
relative to the $z=0$ matter power. 
The dashed curves include the nonlinear correction of \cite{Cole05} for $A=1.4$, 
with 
$\Qnl=30$ 
for the LRGs and $\Qnl=4.6$ for the main galaxies; see \eq{QnlEq}.
The onset of nonlinear corrections is clearly
visible for $k\simgt 0.09 h/$Mpc (vertical line).
}
\end{figure*}

\begin{figure} 
\vskip-0.8cm
\centerline{\epsfxsize=\figsize\epsffile{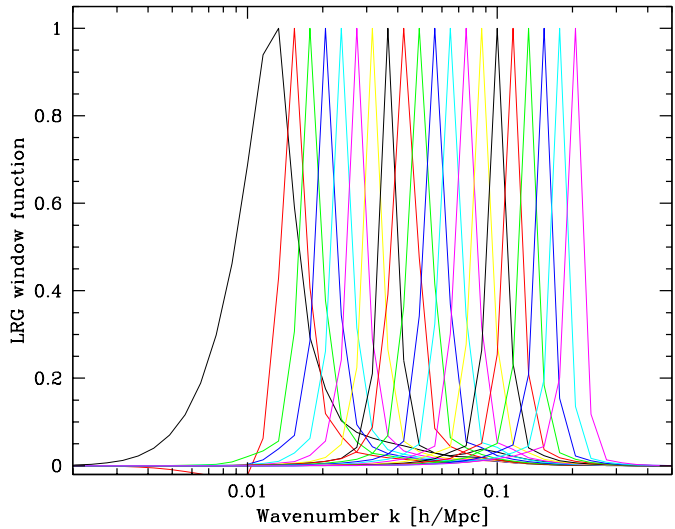}}
\vskip-1cm
\caption[1]{\label{Wfig}\footnotesize%
The window functions corresponding to the LRG band powers in \fig{all2powerFig} are
plotted, normalized to have unit peak height.
Each window function typically peaks at 
the scale $k$ that the corresponding band power 
estimator was designed to probe.
}
\end{figure}

\begin{table}
\bigskip
\noindent
{\footnotesize {\bf Table~\PowerTable} -- The real-space galaxy power spectrum $\Pg(k)$ in 
units of $(h^{-1}\Mpc)^3$ measured from the LRG sample.
The errors on $\Pg$ are $1\sigma$, uncorrelated between bands.
The $k$-column gives 
the median of the window function and its $20^{th}$ and $80^{th}$ percentiles;
the exact window functions from 
\url{http://space.mit.edu/home/tegmark/sdss.html} (see \fig{Wfig}) should
be used for any quantitative analysis. 
Nonlinear modeling is definitely required if the six measurements on the smallest scales
(below the line) are used for model fitting. These error bars do not include an overall calibration uncertainty of 
$3\%$ ($1\sigma$) related to redshift space distortions (see Appendix~\ref{BetaSec}).
\bigskip
\begin{center}   
{\footnotesize
\begin{tabular}{|cc|}
\hline
$k\>[h/$Mpc]		&Power $\Pg$ \\
\hline
$0.012^{+0.005}_{-0.004}$  &$124884\pm 18775$\\
$0.015^{+0.003}_{-0.002}$  &$118814\pm 29400$\\
$0.018^{+0.004}_{-0.002}$  &$134291\pm 21638$\\
$0.021^{+0.004}_{-0.003}$  &$ 58644\pm 16647$\\
$0.024^{+0.004}_{-0.003}$  &$105253\pm 12736$\\
$0.028^{+0.005}_{-0.003}$  &$ 77699\pm  9666$\\
$0.032^{+0.005}_{-0.003}$  &$ 57870\pm  7264$\\
$0.037^{+0.006}_{-0.004}$  &$ 56516\pm  5466$\\
$0.043^{+0.008}_{-0.006}$  &$ 50125\pm  3991$\\
$0.049^{+0.008}_{-0.007}$  &$ 45076\pm  2956$\\
$0.057^{+0.009}_{-0.007}$  &$ 39339\pm  2214$\\
$0.065^{+0.010}_{-0.008}$  &$ 39609\pm  1679$\\
$0.075^{+0.011}_{-0.009}$  &$ 31566\pm  1284$\\
$0.087^{+0.012}_{-0.011}$  &$ 24837\pm   991$\\
\hline
$0.100^{+0.013}_{-0.012}$  &$ 21390\pm   778$\\
$0.115^{+0.013}_{-0.014}$  &$ 17507\pm   629$\\
$0.133^{+0.012}_{-0.015}$  &$ 15421\pm   516$\\
$0.153^{+0.012}_{-0.017}$  &$ 12399\pm   430$\\
$0.177^{+0.013}_{-0.018}$  &$ 11237\pm   382$\\
$0.203^{+0.015}_{-0.022}$  &$  9345\pm   384$\\
\hline  
\end{tabular}
}
\end{center}     
} 
\end{table}

\subsection{Basic results}

The measured real-space power spectra are shown in \fig{all2powerFig} for the LRG and MAIN samples and 
are listed in Table~\PowerTable.
When interpreting them, two points should be borne in mind:
\begin{enumerate}
\item The data points (a.k.a.~band power measurements) probe a weighted average of the true power spectrum $P(k)$ 
defined by the window functions shown in \fig{Wfig}.  
Each point is plotted at the median $k$-value of its window with a horizontal bar ranging from the
$20^{th}$ to the $80^{th}$ percentile.
\item The errors on the points, indicated by the vertical bars, are uncorrelated, even though the horizontal bars overlap.
Other power spectrum estimation methods (see Appendix~\ref{MethodComparisonSec}) effectively produce a smoothed version of 
what we are plotting, with error bars that are smaller but highly correlated.
\end{enumerate}
Our Fourier convention is such that the dimensionless power
$\Delta^2$ of \cite{PeacockDodds94} is given by
$\Delta^2(k) = 4\pi(k/2\pi)^3 P(k)$.

Before using these measurements to constrain cosmological models, one faces important issues regarding their interpretation, related to 
evolution, nonlinearities and systematics.

\subsection{Clustering evolution}

\begin{figure} 
\centerline{\epsfxsize=\figsize\epsffile{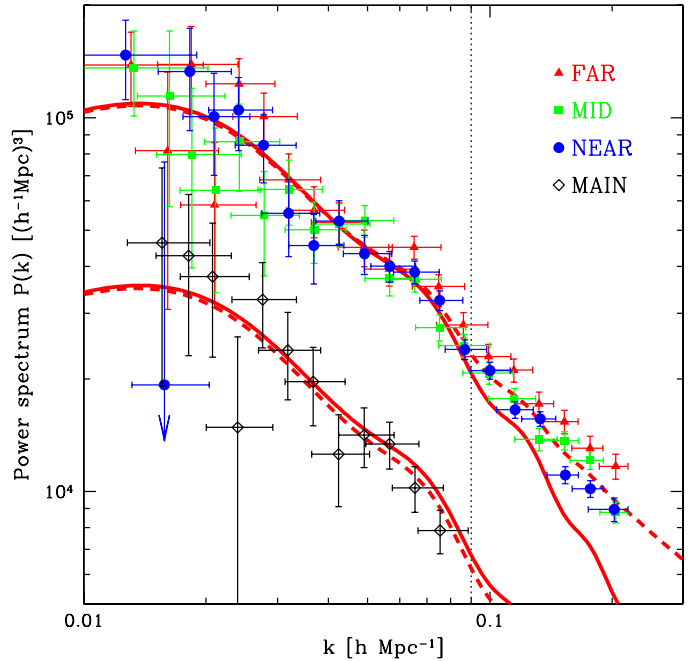}}
\caption[1]{\label{all4powerFig}\footnotesize%
Same as \fig{all2powerFig}, but showing the NEAR (circles), MID (squares) and FAR (triangles) LRG subsamples.
On linear scales, they are all well fit by the WMAP3 model 
with the same clustering amplitude, and there is no sign of clustering evolution.
}
\end{figure}

The standard theoretical expectation is for matter clustering to grow over time and for bias 
(the relative clustering of galaxies and matter)
to decrease over time \cite{Fry96,bias,MoWhite96} for a given class of galaxies.
Bias is also luminosity-dependent, which would be expected to affect the FAR sample but not
the MID and NEAR samples (which are effectively volume limited with a $z$-independent mix of galaxy luminosities \cite{EisensteinLRGselection}).
Since the galaxy clustering amplitude is the product of these two factors, matter clustering and bias, it could 
therefore in principle either increase or decrease across the redshift range $0.155<z<0.474$ of the LRG sample.
We quantify this empirically by measuring the power spectra of the  NEAR, MID and FAR LRG subsamples.
The results are plotted in \fig{all4powerFig} and show no evidence for evolution of
the large-scale galaxy ($k\simlt 0.1h/\Mpc$) power spectrum in either shape or amplitude. 
To better quantify this, we fit the WMAP-only best-fit $\Lambda$CDM model from 
Table 5 of \cite{Spergel06} (solid line in \fig{all4powerFig}) to our power spectra, 
by scaling its predicted $z=0$ matter power spectrum by $b^2$ for a constant bias factor $b$, using only the 14 data points that are  
essentially 
in the linear regime, leftward of the dotted vertical line $k=0.09h$/Mpc.
For the NEAR, MID and FAR subsamples, this 
gives best fit bias factors
$b\approx 1.95$, $1.91$ and $2.02$, respectively.
The fits are all good, giving 
$\chi^2\approx 10.3$, $11.2$ and $15.9$ for the three cases, in agreement with the
expectation $\chi^2=13\pm\sqrt{2\times 13}\approx 13\pm 5$ and consistent with the linear-theory prediction that the large-scale LRG
power spectrum should not change its shape over time, merely (perhaps) its amplitude.

The overall amplitude of the LRG power spectrum is constant within
the errors over this redshift range, in good agreement with the
results of \cite{Zehavi05,sdsslrgcl} at the corresponding mean redshifts. Relative to
the NEAR sample, the clustering amplitude is 
$2.4\%\pm 3\%$ lower in MID and $3.5\%\pm 3\%$ higher in FAR.
In other words, in what appears to be a numerical coincidence, the growth over time in the matter power spectrum 
is approximately canceled by a drop in the bias factor to within our measurement uncertainty.
For a flat $\Om=0.25$ $\Lambda$CDM model, the matter clustering grows by about 10\% from the FAR to NEAR sample mean redshifts,
so this suggests that the bias drops by a similar factor. For a galaxy population evolving 
passively, under the influence of gravity alone \cite{Fry96,bias}, $b$ would be expected to drop by about 5\% over this redshift range; a slight additional 
drop could be caused by luminosity-dependent bias coupling to the slight change in the luminosity function for the FAR sample,
which is not volume limited.

This 
cancellation of LRG clustering evolution is a fortuitous coincidence that simplifies our analysis: 
we can pool all our LRGs and measure a single power spectrum for
this single sample.     
It is not a particularly surprising result: many authors have found
that the galaxy clustering strength is essentially independent of
redshift, even to redshifts $z>3$ \cite{Giavalisco98}, 
and even the effect that is partly canceled (the expected $10\%$ growth in matter clustering) is small, 
because of the limited redshift range probed.

\begin{figure} 
\centerline{\epsfxsize=\figsize\epsffile{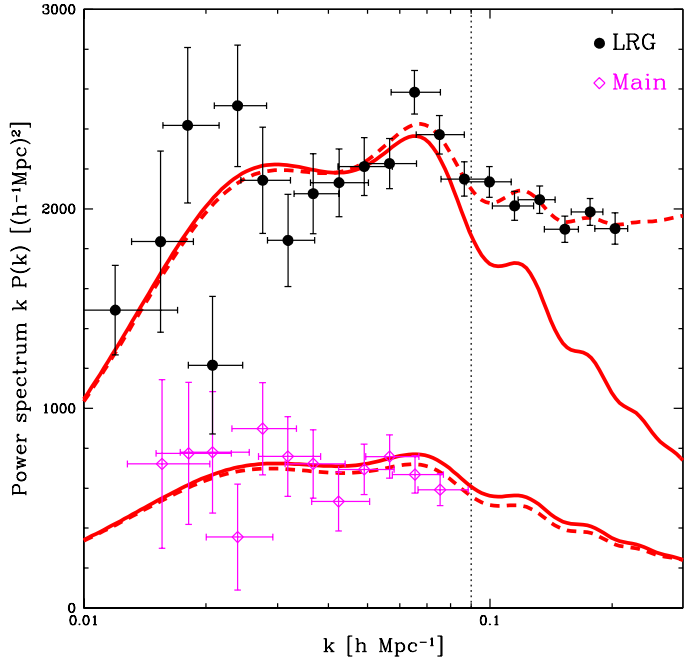}}
\caption[1]{\label{linear_powerFig}\footnotesize%
Same as \fig{all2powerFig}, but multiplied by $k$ and plotted with a linear vertical axis to more clearly 
illustrate departures from a simple power law.
}
\end{figure}

\subsection{Redshift space distortions}

\begin{figure} 
\centerline{\epsfxsize=\figsize\epsffile{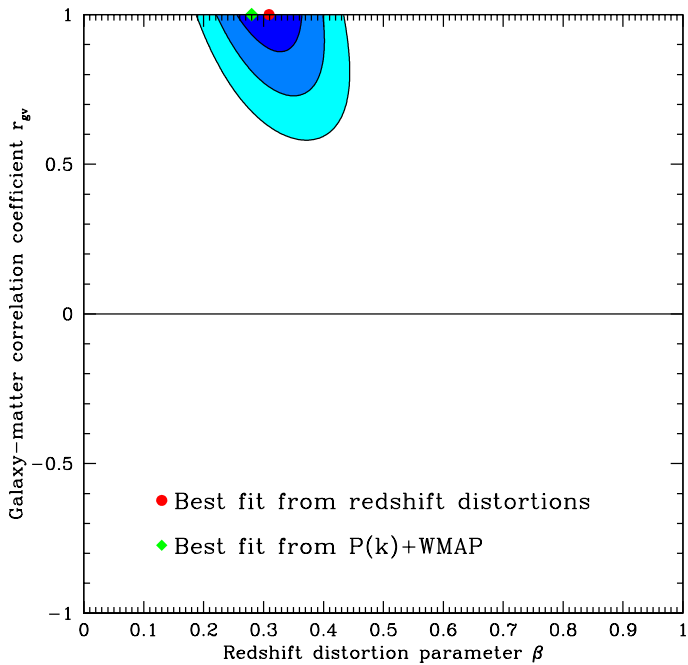}}
\caption[1]{\label{betarFig}\footnotesize%
Constraints on the redshift space distortion parameters $\beta$ and $\rgv$.
The contours show the 1, 2 and 3$\sigma$ constraints from the observed LRG clustering anisotropy, with the 
circular dot indicating the best fit values. The diamond shows the completely independent 
$\beta$-estimate inferred from our analysis of the WMAP3 and LRG power spectra
(it puts no constraints on $\rgv$, but has been plotted at $\rgv=1$).
}
\end{figure}

\label{zspaceSec}

As described in detail in \cite{sdsspower}, an intermediate step in our PKL-method is measuring three separate
power spectra, $\Pgg(k)$, $\Pgv(k)$ and $\Pvv(k)$, which
encode clustering anisotropies due to redshift space distortions. 
Here ``velocity" refers to the negative of the peculiar velocity divergence.
Specifically, 
$\Pgg(k)$ and $\Pvv(k)$ are the power spectra of the galaxy density and velocity fields, respectively,
whereas 
$\Pgv(k)$ is the cross-power between galaxies and velocity,
all defined in real space rather than redshift space.

In linear perturbation theory, these three power spectra are related by \cite{Kaiser87}
\beqa{KaiserLimitEq1}
\Pgv(k) &=& \beta\rgv\Pgg(k),\\
\Pvv(k) &=& \beta^2\Pgg(k),\label{KaiserLimitEq2}
\eeqa
where $\beta\equiv f/b$, $b$ is the bias factor,
$\rgv$ is the dimensionless correlation coefficient between the
galaxy and matter density fields \cite{DekelLahav99,Pen98,bias}, and 
$f\approx\Om^{0.6}$ is the dimensionless linear growth rate for linear density fluctuations. (When computing
$f$ below, we use the more accurate approximation of \cite{growl}.)

The LRG power spectrum $P(k)$ tabulated and plotted above is a minimum-variance estimator of $\Pgg(k)$ that
linearly combines the $\Pgg(k)$, $\Pgv(k)$ and $\Pvv(k)$ estimators as described in \cite{sdsspower} and 
Appendix~\ref{BetaSec}, effectively marginalizing over the redshift space distortion parameters $\beta$ and $\rgv$.
As shown in Appendix~\ref{BetaSec}, this linear combination is roughly proportional to 
the angle-averaged (monopole) redshift-space galaxy power spectrum, so for the purposes of the nonlinear modeling
in the next section, the reader may think of our measured $P(k)$ as essentially a rescaled version of the redshift 
space power spectrum. 
However, unlike the redshift space power spectrum measured with the
FKP and FFT methods (Appendix~\ref{MethodComparisonSec}), our measured $P(k)$ is unbiased on large scales.
This is because linear redshift distortions are treated exactly, 
without resorting to the small-angle approximation, and account is taken of the fact that 
the anisotropic survey geometry can skew the relative abundance of galaxy pairs around a single point 
as a function of angle to the line of sight.

The information about anisotropic clustering that is discarded in our estimation of $P(k)$ allows us to measure $\beta$ and 
perform a powerful consistency test. \Fig{betarFig} shows the joint constraints on $\beta$ and $\rgv$ from fitting 
equations\eqn{KaiserLimitEq1} and\eqn{KaiserLimitEq2} to the $0.01h/\Mpc\le k\le 0.09h/\Mpc$ LRG data, using the best fit 
WMAP3 model from \fig{all2powerFig} for $\Pgg(k)$ and marginalizing over its amplitude.
The data are seen to favor $\rgv\approx 1$ in good agreement with prior work \cite{bias2,Wild05}.
Assuming $\rgv=1$ (that galaxy density linearly traces matter density on these large scales) 
gives the measurement $\beta=0.309\pm 0.035$ ($1\sigma$).
This measurement is rather robust to changing the FOG compression threshold by a notch (\Sec{LRGrobustnessSec}) or 
slightly altering the maximum $k$-band included, both of which affect the central value by of order $0.01$.
As a cross-check, we can compute $\beta=f(\Om,\Ol)/b$ at the median survey redshift based on our multi-parameter analysis presented in 
\Sec{CosmoSec}, which for our vanilla class of models gives
$\beta = 0.280\pm 0.014$ (marked with a diamond in \fig{betarFig})\footnote{Here 
$\beta=f(\Om,\Ol)/b$ is computed with $\Om$, $\Ol$ and $b$ evaluated at the median redshift $z=0.35$,
when 
$b=2.25\pm 0.08$,
taking into account linear growth of matter clustering between then and now.
}.
That these two $\beta$-measurements agree within $1\sigma$ is highly non-trivial, since
the second $\beta$-measurement makes no use whatsoever of redshift space distortions, 
but rather extracts $b$ from the ratio of LRG power to CMB power, and determines $\Om$ from CMB and LRG power spectrum shapes. 

\subsection{Nonlinear modeling}
\label{NonlinearModelingSec}

\begin{figure} 
\centerline{\epsfxsize=\figsize\epsffile{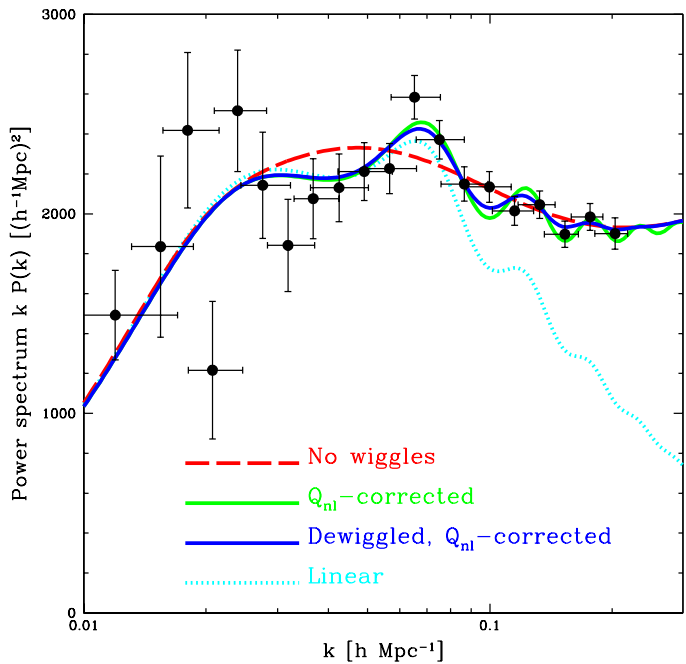}}
\caption[1]{\label{dewiggledPowerFig}\footnotesize%
Power spectrum modeling.
The best-fit WMAP3 model from Table 5 of \cite{Spergel06} is shown with a linear bias $b=1.89$ (dotted curve),
after applying the nonlinear bias correction with $Q=31$ (the more wiggly solid curve),
and after also applying the wiggle suppression of \cite{EisensteinSeoWhite06} (the less wiggly solid curve),
which has no effect on very large scales and asymptotes to the ``no wiggle'' spectrum of \cite{EH99} (dashed curve) on very small scales.
The data points are the LRG measurements from \fig{linear_powerFig}.
}
\end{figure}

Above we saw that our $k<0.09h/$Mpc measurements of the LRG power spectrum were well fit by the linear theory matter power spectrum predicted by WMAP3.
In contrast, Figures~\ref{all2powerFig}, \ref{all4powerFig} and~\ref{linear_powerFig} 
show clear departures from the linear theory prediction on smaller scales.
There are several reasons for this that have been  extensively studied in the literature:
\begin{enumerate}
\item Nonlinear evolution alters the broad shape of the matter power spectrum on small scales.
\item Nonlinear evolution washes out baryon wiggles on small scales.
\item The power spectrum of the dark matter halos in which the galaxies reside
differs from that of the underlying matter power spectrum in both amplitude and shape, causing bias.
\item Multiple galaxies can share the same dark matter halo, enhancing small-scale bias.
\end{enumerate}
We fit these complications using a model involving the three ``nuisance parameters'' ($b,\Qnl,\kstar$) as illustrated in \fig{dewiggledPowerFig}.
Following \cite{Cole05,EisensteinSeoWhite06}, we model our measured galaxy power spectrum as
\beq{QnlEq}
\Pg(k)=P_{\rm dewiggled}(k) b^2{1+\Qnl k^2\over 1+1.4k},
\eeq
where the first factor on the right hand side accounts for the non-linear suppression of baryon wiggles and the last factor accounts
for a combination of the non-linear change of the global matter power spectrum shape and scale-dependent bias of the galaxies relative
to the dark matter.
For $P_{\rm dewiggled}(k)$ we adopt the prescription \cite{EisensteinSeoWhite06}
\beq{DewigglingEq}
P_{\rm dewiggled}(k) = W(k) P(k) + [1-W(k)]P_{\rm nowiggle}(k), 
\eeq
where $W(k)\equiv e^{-(k/\kstar)^2/2}$ and
$P_{\rm nowiggle}(k)$ denotes the ``no wiggle'' power spectrum defined in \cite{EH99} and illustrated in \fig{dewiggledPowerFig}.
In other words, $P_{\rm dewiggled}(k)$ is simply a weighted average of the linear power spectrum and the wiggle-free version thereof.
Since the $k$-dependent weight $W(k)$ transitions from $1$ for $k\ll\kstar$ to $0$ for $k\gg\kstar$, \eq{DewigglingEq} retains wiggles on large scales and gradually fades them out
beginning around $k=\kstar$.
Inspired by \cite{EisensteinSeoWhite06}, we define the wiggle suppression scale $\kstar\equiv 1/\sigma$, 
where $\sigma\equiv\sigma_\perp^{2/3}\sigma_\parallel^{1/3} (\As/0.6841)^{1/2}$ and 
$\sigma_\perp$ and $\sigma_\parallel$ are given by equations (12) and (13) in \cite{EisensteinSeoWhite06} 
based on fits to cosmological N-body simulations. The expression in parenthesis is an 
amplitude scaling factor that equals unity for the best fit WMAP3 normalization $\As=0.6841$ of \cite{Spergel06}.
Essentially, $\sigma$ is the characteristic peculiar-velocity-induced 
displacement of galaxies that causes the wiggle suppression; \cite{EisensteinSeoWhite06}
define it for a fixed power spectrum normalization, and it scales linearly with 
fluctuation amplitude, \ie, $\propto\As^{1/2}$. 
For the cosmological parameter range allowed by WMAP3, we find that $\kstar\sim 0.1h/\Mpc$,
with a rather rather weak dependence on cosmological parameters (mainly $\Om$ and $\As$).

\begin{figure} 
\centerline{\epsfxsize=\figsize\epsffile{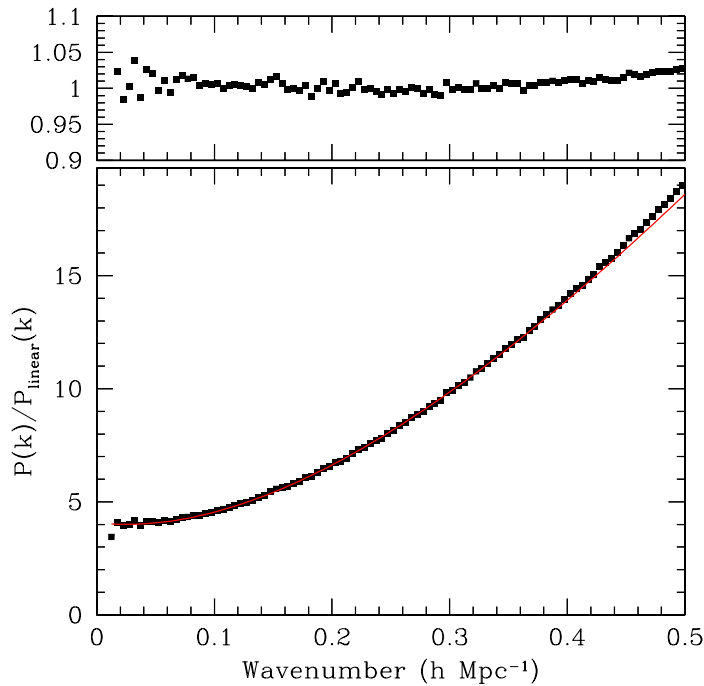}}
\caption[1]{\label{QbiasFig}\footnotesize%
The points in the bottom panel show the ratio of the real-space power spectrum from 51 averaged $n$-body 
simulations (see text) to the linear power spectrum dewiggled with $\kstar=0.1h/$\Mpc.
Here LRGs were operationally defined as halos with mass exceeding $8\times 10^{12}\Ms$, corresponding 
to at least ten simulation particles.
The solid curve shows the prediction from \eq{QnlEq} with $b=2.02$, $\Qnl=27$, seen to be an excellent fit for $k\simlt 0.4h/$\Mpc.
The top panel shows the ratio of the simulation result to this fit.
Although the simulation specifications and the LRG identification prescription can clearly be improved, they constitute the 
first and only that we tried, 
and were in no way adjusted to try to fit our $\Qnl=30\pm 4$ measurement from Table~{\ParameterTable}. 
This agreement suggests that our use of equations\eqn{QnlEq} and\eqn{DewigglingEq} to model nonlinearities is reasonable and that our measured $\Qnl$-value is plausible.
}
\end{figure}

The simulations and analytic modeling described by \cite{Cole05} suggest that the $\Qnl$-prescription given by \eq{QnlEq} 
accurately captures the scale-dependent bias of galaxy populations on the scales that we are interested in, though they
examined samples less strongly biased than the LRGs considered here.
To verify the applicability of this prescription for LRGs in combination with our dewiggling model, we reanalyze the 51 $n$-body simulations 
described in \cite{SeoEisteinstein05}, each of which uses a $512h^{-1}$ Mpc box with $256^3$ particles and WMAP1 parameters.
\Fig{QbiasFig} compares these simulation results with our nonlinear modeling prediction defined by equations~\eqn{QnlEq} and\eqn{DewigglingEq} for 
$b=2.02$, $\Qnl=27.0$, showing excellent agreement (at the 1\% 
level) for $k\simlt 0.4h/$Mpc.
Choosing a $\kstar$ very different from $0.1h/$Mpc causes 5\%
wiggles to appear in the residuals because of a over- or under-suppression of the baryon oscillations.
These simulations are likely to be underresolved and the LRG halo prescription 
used (one LRG for each halo above a threshold mass of $8\times 10^{12}\Ms$)
is clearly overly simplistic, so the true value of $\Qnl$ that best describes LRGs could be somewhat different.
Nonetheless, this test provides encouraging evidence that 
\eq{QnlEq} is accurate in combination with \eq{DewigglingEq} and 
that our $Q=30\pm 4$ measurement from Table~{\ParameterTable} is plausible. 
Further corroboration is provided by the results in \cite{sdsslrgcl} using the 
Millennium Simulation \cite{Springel06}.
Here LRG type galaxies were simulated and selected in an arguably more realistic way, yet giving results nicely consistent 
with \fig{QbiasFig}, with a best-fit value $\Qnl\approx 24$. 
(We will see in \Sec{DataRobustnessSec} that FOG-compression can readily account for these slight 
differences in $\Qnl$-value.)
A caveat to both of these simulation tests is that they were performed in real space, and our procedure for
measuring $P_g(k)$ reconstructs the real space power spectrum exactly only in the linear regime \cite{sdsspower}.
Thus, these results should be viewed as encouraging but preliminary,
and more work is needed to establish the validity of the nonlinear modeling beyond $k\simgt 0.1h/$Mpc; for up-to-date discussions 
and a variety of ideas for paths forward, see, \eg, \cite{Huff06,Yoo06,McDonald06,Smith06}.

In addition to this simulation-based theoretical evidence that our nonlinear modeling method is accurate, 
we have encouraging empirical evidence: \fig{dewiggledPowerFig} shows an excellent fit to our measurements. 
Fitting the best-fit WMAP3 model from \cite{Spergel03} to our first 20 data points (which extend out to $k=0.2h/$Mpc) by varying $(b,\Qnl)$ gives 
$\chi^2=19.2$ for $20-2=18$ degrees of freedom, where the expected $1\sigma$ range is $\chi^2=18\pm(2\times 18)^{1/2}=18\pm 6$,  
so the fit is excellent. Moreover, Figures~\ref{linear_powerFig} and~\ref{dewiggledPowerFig} show that
that main outliers are on large and highly linear scales, not on the smaller scales where our nonlinear modeling has an effect.

The signature of baryons is clearly seen in the measured power spectrum.
If we repeat this fit with baryons replaced by dark matter, $\chi^2$ increases by $8.8$, corresponding to 
a baryon detection at $3.0\sigma$ (99.7\% significance). 
Much of this signature lies in the acoustic oscillations: if we instead repeat the fit with
$\kstar=0$, corresponding to fully removing the wiggles, $\chi^2$ increases by an amount corresponding 
to a detection of wiggles at $2.3\sigma$ (98\% significance). The data are not yet sensitive enough to distinguish between the 
wiggled and dewiggled spectra; dewiggling reduces $\chi^2$ by merely $0.04$.

In summary, the fact that LRGs tend to live in high-mass dark matter halos is a double-edged sword: 
it helps by giving high bias $b\sim 2$ and luminous galaxies observable at great distance,
but it also gives a stronger nonlinear correction (higher $\Qnl$) that becomes important on larger 
scales than for typical galaxies.
Although \fig{QbiasFig} suggests that our nonlinear modeling is highly accurate out to $k=0.4h/$Mpc, we retain only measurements 
with $k\simlt 0.2h/$Mpc for our cosmological parameter analysis to be conservative, and plan further work to test the validity of various nonlinear modeling approaches. 
In \Sec{LRGrobustnessSec}, we will see that our data with $0.09h/$Mpc$<k\simlt 0.2h$/Mpc, where nonlinear 
effects are clearly visible, allow us to constrain the nuisance parameter $\Qnl$ without significantly improving our constraints on cosmological parameters.
In other words, the cosmological constraints that we will report below are quite insensitive to our nonlinear modeling and come mainly from the linear power 
spectrum at $k<0.09h/$Mpc.
More sophisticated treatments of galaxy bias in which $\Qnl$ is effectively computed from theoretical models constrained by small
scale clustering may eventually obviate the need to marginalize over this nuisance parameter, increasing the leverage
of our measurements for constraining the linear power spectrum shape \cite{Yoo06}.

\begin{figure} 
\centerline{\epsfxsize=\figsize\epsffile{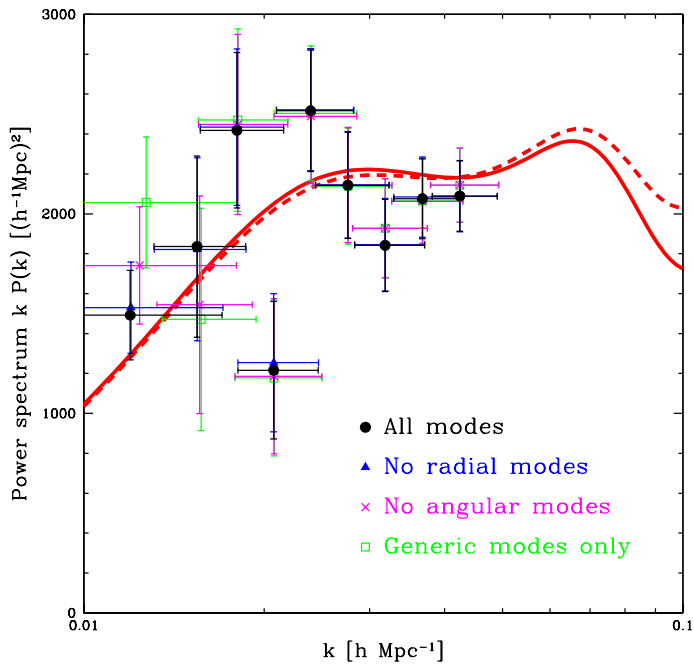}}
\caption[1]{\label{modezapFig}\footnotesize%
Same as \fig{linear_powerFig}, but showing the effect of discarding special modes on the large-scale power.
The circles with associated error bars correspond to our measured power 
spectrum using all 4000 full-sample PKL modes. The other points show the effect of removing 
the 332 purely angular modes (crosses), the 18 purely radial modes (triangles),
and all special modes combined (squares), including seven associated with the motion of the 
local group as described in \cite{sdsspower}.
Any systematic errors adding power to these special modes would cause the black circles to lie
systematically {\it above} the other points. 
These special modes are seen to have less impact at larger $k$ because they are outnumbered:
the number of radial, angular, and generic modes below a given $k$-value scales as
$k$, $k^2$ and $k^3$, respectively.
}
\end{figure}

\subsection{Robustness to systematic errors}
\label{SystematicsSec}

Let us now consider potential systematic errors in the LRG data that could affect our results.
Examples of such effects 
include radial modulations 
(due to mis-estimates of the radial selection function)
and angular modulations (due to effects such as  
uncorrected dust extinction, variable observing conditions, 
photometric calibration errors and fiber collisions) of the density field.
As long as such effects are uncorrelated with the cosmic density field, they will tend to 
add rather than subtract power.

\subsubsection{Analysis of subsets of galaxies}

To test for effects that would be expected to vary across the sky (depending 
on, say, reddening, seasonally variable photometric calibration errors, or 
observing conditions such as seeing and sky brightness), 
we repeat our entire analysis for the seven different angular subsets of the sky
shown in \fig{AitoffFig} in search of inconsistencies. 
To search for potential zero-point offsets and other systematic effects associated with the southern Galactic stripes, they
are defined as one of these seven angular subsets (see \fig{AitoffFig}).
To test for effects that depend on redshift, we use the measurements for our three 
redshift slices, plotted in \fig{all4powerFig}.

To test the null hypothesis that all these subsamples are consistent with having the same power spectrum, 
we fit them all to our WMAP+LRG best-fit vanilla model described in \Sec{CosmoSec}, including our nonlinear correction 
(this $P(k)$ curve is quite similar to the best-fit WMAP3 model plotted above in, \eg, \fig{all2powerFig}).
We include the 20 band-powers with $k\simlt 0.2$ in our fit, so if the null hypothesis is correct, we expect
a mean $\chi^2$ of 20 with a standard deviation of $\sqrt{2\times 20}\approx 6.3$. 
Our seven angular subsamples give a mean $\expec{\chi^2}\approx 22.6$ and 
a scatter $\expec{(\chi^2-20)^2}^{1/2}\approx 6.9$.
Our three radial subsamples give $\expec{\chi^2}\approx 18.6$ and $\expec{(\chi^2-20)^2}^{1/2}\approx 2.4$.
All of the ten $\chi^2$-values are statistically consistent with the null hypothesis at the 95\% 
level.
We also repeated the cosmological parameter analysis reported below with the southern stripes omitted, finding no significant change in 
the measured parameter values.
In other words, all our angular and radial subsamples are consistent with having the same power spectrum, so 
these tests reveal no evidence for systematic errors causing radial or angular power spectrum variations.

\subsubsection{Analysis of subsets of modes}

Because of their angular or radial nature, all 
potential systematic errors discussed above
create excess power mainly in the radial and angular modes.
As mentioned above, one of the advantages of the PKL method is that it allows these modes to be excluded from the analysis,
in analogy to the way potentially contaminated pixels in a CMB map can be excluded from a CMB power spectrum analysis.
To quantify any such excess, we therefore repeat our full-sample analysis with 
radial and/or angular modes deleted.
The results of this test are shown in \fig{modezapFig} and are very
encouraging; the differences are tiny.  
Any systematic errors adding power to these special modes would cause the black circles
to lie systematically {\it above} the other points, but no such trend is seen, 
so there is no indication of excess radial or
angular power in the data. 

The slight shifts seen in the 
power on the largest scales 
are expected, since a non-negligible fraction of the
information has been discarded on those scales.
\Fig{modezapFig} shows that removing the special modes 
results in a noticeable error bar increase 
on the largest scales and essentially no change on smaller scales.
This can be readily understood geometrically.
If we count the number of modes that probe mainly scales $k<k_*$,
then the number of purely radial, purely angular and arbitrary modes
will grow as $k_*$, $k_*^2$ and $k_*^3$, respectively.
This means that ``special'' modes (radial and angular) will make up a larger fraction
of the total pool on large scales (at small $k$), and that the purely 
radial ones will be outnumbered by the purely angular ones.
Conversely, the first 12 modes are all special ones: the monopole, the seven modes related to local-group motion, 
one radial mode and three angular modes. 
This means that almost all information on the very largest scales is lost 
when discarding special modes. \Fig{modezapFig} illustrates this with the leftmost point labeled 
``generic'' both having large error bars and being shifted to the right, where more information 
remains --- yet it is consistent, lying about $1.3\sigma$ above an imaginary 
line between the two leftmost black points.
We also repeated the cosmological parameter analysis reported below with the special modes omitted, finding no significant change in 
the measured parameter values. 

\subsection{Other tests}
\label{PercivalCompSec}

We have found no evidence for systematic errors afflicting our power spectrum, 
suggesting that such effects, if present, 
are substantially smaller than our statistical errors.
For additional bounds on potential systematic errors in the SDSS LRG sample, see \cite{Percival06}.

A direct comparison of our $P(k)$-measurement and that of \cite{Percival06} is complicated because these are 
not measurements
of the same function. \cite{Percival06} measures the angle-averaged redshift-space galaxy power spectrum,
whereas our PKL-method attempts to recover the real space galaxy power spectrum, 
using finger-of-god (FOG) compression and linear theory to remove redshift-space distortion effects \cite{sdsspower}.
The galaxy selection is also different, with \cite{Percival06} mixing main sample galaxies in with the LRGs.
Both of these differences are expected to affect the nonlinear corrections.
In addition, the quantity $P(k)$ plotted in \cite{Percival06} has correlated points 
with broader window functions than our uncorrelated points, and the angular coverage of the sample used in \cite{Percival06} is about 20\% 
larger.
To make 
a direct but approximate
comparison with \cite{Percival06}, we perform our own 
FKP analysis, both with and without FOG-compression, and as described in Appendix~\ref{MethodComparisonSec}, 
we obtain good agreement with \cite{Percival06} on linear scales for the 
case of no defogging.

We further investigate the robustness of our results to systematic errors in 
\Sec{DataRobustnessSec} below, this time focusing on their potential impact on cosmological parameters.

\section{Cosmological parameters}
\label{CosmoSec}

Let us now explore the cosmological implications of our measurements by combining them with those from WMAP.
As there has recently been extensive work on constraining cosmological parameters  
by combining multiple
cosmological data sets involving CMB, galaxy clustering, Lyman $\alpha$ Forest, gravitational lensing, supernovae Ia and other probes 
(see, in particular, \cite{Spergel06,Seljak06}), we will focus more narrowly on what can be learned 
from WMAP and the LRGs alone. This is interesting for two reasons:
\begin{enumerate}
\item Less is more, in the sense that our results hinge on fewer assumptions about data quality and modeling.
The WMAP and LRG power spectra suffice to break all major degeneracies within a broad class of models, yet they are also 
two remarkably clean measurements, probing gravitational clustering only on very large scales where 
complicated nonlinear physics is unlikely to cause problems. 
\item Since the LRG power spectrum is likely to be included (together with WMAP and other data sets) 
in future parameter analyses by other groups, it is 
important to elucidate what information it contains about cosmological parameters.
We will therefore place particular emphasis on clarifying the links between cosmological parameters and 
observable features of both the  LRG and WMAP power spectra, 
notably the LRG matter-radiation equality scale, the LRG acoustic scale, the CMB acoustic scale,
unpolarized CMB peak height ratios and large-scale CMB polarization.
\end{enumerate}
We then compare our constraints with those from other cosmological probes in \Sec{OtherDataSec}.
We also compare our results with the analysis of \cite{sdssbump} below, which had the narrower focus of measuring the
LRG acoustic scale; the correlation function analysis in that paper complements our present analysis, since 
the acoustic oscillations in $P(k)$ correspond to a readily measured single localized feature in real space 
\cite{BashinskyBertschinger02,sdssbump}.

We work within the context of the arguably simplest inflationary scenario that fits our data. 
This is a hot Big Bang cosmology with primordial fluctuations that are adiabatic (\ie,
we do not include isocurvature modes) and Gaussian, with negligible
generation of fluctuations by cosmic strings, textures or domain
walls. We assume the standard model of particle physics with three active
neutrino species, very slightly heated during the era of electron/positron
annihilation~\cite{GnedinGnedin98}.
Within this framework, we parameterize
our cosmological model in terms of 12 parameters that are nowadays rather standard, 
augmented with the two nuisance parameters $b$ and $\Qnl$ from \eq{QnlEq}: 
\beq{pEq}
\p\equiv(\Ot,\Ol,\ob,\ocdm,\on,w,\As,r,\ns,\nt,\al,\tau,b,\Qnl).
\eeq
Table~{\ParameterTable} defines these 14 parameters and another 45 that can be derived from them; 
in essence, $(\Ot,\Ol,\ob,\ocdm,\on,w)$ define the cosmic matter budget, 
$(\As,\ns,\alpha,r,\nt)$ specify the seed fluctuations and $(\tau,b,\Qnl)$ are nuisance parameters.
We will frequently use the term ``vanilla'' to refer to the minimal model space parametrized by 
$(\Ol,\ob,\ocdm,\As,\ns,\tau,b,\Qnl)$, setting $\on=\al=r=\nt=0$, $\Ot=1$ and $w=-1$;
this is the smallest subset of our parameters that provides a good fit to our data.
Since current $\nt$-constraints are too weak to be interesting, we make the slow-roll assumption $\nt=-r/8$ throughout this
paper rather than treat $\nt$ as a free parameter.

All our parameter constraints were computed using the now standard Monte Carlo Markov Chain (MCMC) approach 
\cite{Metropolis,Hastings,Gilks96,Christensen01,CosmoMC,Slosar03,Verde03} as implemented in \cite{sdsspars} 
\footnote{%
\label{LikelihoodFootnote}
To mitigate numerically deleterious degeneracies, the independent MCMC variables are chosen to be
the parameters 
$(\Th,\Ol,\ob,\od,\fn,w,\Ap,\ns,\al,r,\nt,\Op ,b,\Qnl)$ 
from Table~{\ParameterTable}, where $\od\equiv\ocdm+\on$, \ie, 
$(\Ot,\ocdm,\on,\As,\tau)$ are replaced by $(\Th,\od,\fn,\Ap,e^{-2\tau})$ as in \cite{Knox03,sdsspars}.
When imposing a flatness prior $\Ot=1$, we retained $\Th$ as a free parameter and dropped $\Ol$.
The WMAP3 log-likelihoods are computed with the software provided by the WMAP team or 
taken from WMAP team chains on the LAMBDA archive (including all unpolarized and polarized information) 
and fit by a multivariate 4th order polynomial \cite{cmbfit} for 
more rapid MCMC-runs involving galaxies. The SDSS likelihood uses the LRG sample alone and is 
computed with the software available at \url{http://space.mit.edu/home/tegmark/sdss/} and described 
in Appendix~\ref{LikelihoodSec}, employing
only the measurements with $k\le 0.2h/\Mpc$ unless otherwise specified.
Our WMAP3+SDSS chains have $3\times 10^6$ steps each and are thinned by a factor of 10.
To be conservative, we do not use our SDSS measurement of the redshift space distortion parameter 
$\beta$, nor do we use any other information (``priors'') whatsoever unless explicitly stated.
}.

\begin{table*}
\noindent 
{\tiny
Table~\ParameterTable: 
Cosmological parameters measured from WMAP and SDSS LRG data with the Occam's razor approach described in the text:
the constraint on each quantity is marginalized over all other parameters in the vanilla set $(\ob,\ocdm,\Ol,\As,\ns,\tau,b,\Qnl)$.
Error bars are $1\sigma.$
\begin{center}
\begin{tabular}{|l|l|l|l|}
\hline
Parameter		&Value	 &Meaning			&Definition\\
\hline
\multicolumn{3}{|l|}{Matter budget parameters:}&\\
\cline{1-4}
$        \Omega_{\rm{tot}}$      &$ 1.003^{+ 0.010}_{- 0.009}$                                          &Total density/critical density 	&$\Ot=\Om+\Ol=1-\Ok$\\
$           \Omega_\Lambda$      &$ 0.761^{+ 0.017}_{- 0.018}$                                          &Dark energy density parameter  	&$\Ol\approx h^{-2}\rho_\Lambda(1.88\times 10^{-26}$kg$/$m$^3)$\\
$                 \omega_b$      &$ 0.0222^{+ 0.0007}_{- 0.0007}$                                       &Baryon density		 	&$\ob=\Ob h^2 \approx \rho_b/(1.88\times 10^{-26}$kg$/$m$^3)$\\
$                 \omega_c$      &$ 0.1050^{+ 0.0041}_{- 0.0040}$                                       &Cold dark matter density 	 	&$\ocdm=\Oc h^2 \approx \rho_c/(1.88\times 10^{-26}$kg$/$m$^3)$\\
$               \omega_\nu$      &$< 0.010 \>(95\percent)$                                              &Massive neutrino density		&$\on=\On h^2 \approx \rho_\nu/(1.88\times 10^{-26}$kg$/$m$^3)$\\
$                        w$      &$-0.941^{+ 0.087}_{- 0.101}$                                          &Dark energy equation of state		&$p_\Lambda/\rho_\Lambda$ (approximated as constant)\\
\hline
\multicolumn{3}{|l|}{Seed fluctuation parameters:}&\\
\cline{1-4}
$                      A_s$      &$ 0.690^{+ 0.045}_{- 0.044}$                                          &Scalar fluctuation amplitude		&Primordial scalar power at $k=0.05$/Mpc\\
$                        r$      &$< 0.30 \>(95\percent)$                                               &Tensor-to-scalar ratio		&Tensor-to-scalar power ratio at $k=0.05$/Mpc\\
$                      n_s$      &$ 0.953^{+ 0.016}_{- 0.016}$                                          &Scalar spectral index			&Primordial spectral index at $k=0.05$/Mpc\\
$                    n_t+1$      &$ 0.9861^{+ 0.0096}_{- 0.0142}$                                       &Tensor spectral index			&$\nt=-r/8$ assumed\\
$                   \alpha$      &$-0.040^{+ 0.027}_{- 0.027}$                                          &Running of spectral index		&$\alpha=d\ns/d\ln k$ (approximated as constant)\\
\hline
\multicolumn{3}{|l|}{Nuisance parameters:}&\\
\cline{1-4}
$                     \tau$      &$ 0.087^{+ 0.028}_{- 0.030}$                                          &Reionization optical depth		&\\
$                        b$      &$ 1.896^{+ 0.074}_{- 0.069}$                                          &Galaxy bias factor			&$b=[P_{\rm galaxy}(k)/P(k)]^{1/2}$ on large scales, where $P(k)$ refers to today.\\
$              Q_{\rm{nl}}$      &$30.3^{+ 4.4}_{- 4.1}$                                                &Nonlinear correction parameter \cite{Cole05}&$\Pg(k)=P_{\rm dewiggled}(k)b^2(1+Q_{\rm nl}k^2)/(1+1.7k)$\\
\hline\hline
\multicolumn{3}{|l|}{Other popular parameters (determined by those above):}&\\
\cline{1-4}
$                        h$      &$ 0.730^{+ 0.019}_{- 0.019}$                                          &Hubble parameter			&$h = \sqrt{(\ob+\ocdm+\on)/(\Ot-\Ol)}$\\
$                 \Omega_m$      &$ 0.239^{+ 0.018}_{- 0.017}$                                          &Matter density/critical density	&$\Om=\Ot-\Ol$\\
$                 \Omega_b$      &$ 0.0416^{+ 0.0019}_{- 0.0018}$                                       &Baryon density/critical density 	&$\Ob=\ob/h^2$\\
$                 \Omega_c$      &$ 0.197^{+ 0.016}_{- 0.015}$                                          &CDM density/critical density 		&$\Oc=\ocdm/h^2$\\
$               \Omega_\nu$      &$< 0.024 \>(95\percent)$                                              &Neutrino density/critical density 	&$\On=\on/h^2$\\
$                 \Omega_k$      &$-0.0030^{+ 0.0095}_{- 0.0102}$                                       &Spatial curvature			&$\Ok=1-\Ot$\\
$                 \omega_m$      &$ 0.1272^{+ 0.0044}_{- 0.0043}$                                       &Matter density			&$\om=\ob+\ocdm+\on = \Om h^2$\\
$                    f_\nu$      &$< 0.090 \>(95\percent)$                                              &Dark matter neutrino fraction		&$\fn=\rho_\nu/\rho_d$\\
$                      A_t$      &$< 0.21 \>(95\percent)$                                               &Tensor fluctuation amplitude		&$\At=r\As$\\
$                    M_\nu$      &$< 0.94 \>(95\percent)$ eV                                            &Sum of neutrino masses 		&$\Mnu\approx(94.4\>{\rm eV})\times\on$~~~\cite{KolbTurnerBook}\\
$                 A_{.002}$      &$ 0.801^{+ 0.042}_{- 0.043}$                                          &WMAP3 normalization parameter		&$\As$ scaled to $k=0.002$/Mpc: $A_{.002} = 25^{1-\ns}\As$ if $\al=0$\\
$                 r_{.002}$      &$< 0.33 \>(95\percent)$                                               &Tensor-to-scalar ratio (WMAP3)	&Tensor-to-scalar power ratio at $k=0.002$/Mpc\\
$                 \sigma_8$      &$ 0.756^{+ 0.035}_{- 0.035}$                                          &Density fluctuation amplitude		&$\sigma_8=\{4\pi\int_0^\infty [{3\over x^3}(\sin x-x\cos x)]^2 P(k) {k^2 dk\over(2\pi)^3}\}^{1/2}$, $x\equiv k\times 8h^{-1}$Mpc\\
$   \sigma_8\Omega_m^{0.6}$      &$ 0.320^{+ 0.024}_{- 0.023}$                                          &Velocity fluctuation amplitude&\\
\hline
\multicolumn{3}{|l|}{Cosmic history parameters:}&\\
\cline{1-4}
$              z_{\rm{eq}}$      &$3057^{+105}_{-102}$                                                  &Matter-radiation Equality redshift	&$z_{\rm eq}\approx 24074\om - 1$\\
$             z_{\rm{rec}}$      &$1090.25^{+ 0.93}_{- 0.91}$                                           &Recombination redshift		&$z_{\rm rec}(\om,\ob)$ given by eq.~(18) of \cite{Hu04}\\
$             z_{\rm{ion}}$      &$11.1^{+ 2.2}_{- 2.7}$                                                &Reionization redshift (abrupt)	&$\zion\approx 92 (0.03h\tau/\ob)^{2/3}\Om^{1/3}$ (assuming abrupt reionization; \cite{reion})\\
$             z_{\rm{acc}}$      &$ 0.855^{+ 0.059}_{- 0.059}$                                          &Acceleration redshift			&$\zacc=[(-3w-1)\Ol/\Om]^{-1/3w}-1$ if $w<-1/3$\\
$              t_{\rm{eq}}$      &$ 0.0634^{+ 0.0045}_{- 0.0041}$ Myr                                   &Matter-radiation Equality time 	&$\teq \approx$($9.778$ Gyr)$\times h^{-1}\int_{\zeq}^\infty [H_0/H(z)(1+z)]dz$~~~\cite{KolbTurnerBook}\\
$             t_{\rm{rec}}$      &$ 0.3856^{+ 0.0040}_{- 0.0040}$ Myr                                   &Recombination time 			&$\trec\approx$($9.778$ Gyr)$\times h^{-1}\int_{\zrec}^\infty [H_0/H(z)(1+z)]dz$~~~\cite{KolbTurnerBook}\\
$             t_{\rm{ion}}$      &$ 0.43^{+ 0.20}_{- 0.10}$ Gyr                                         &Reionization time			&$\tion\approx$($9.778$ Gyr)$\times h^{-1}\int_{\zion}^\infty [H_0/H(z)(1+z)]dz$~~~\cite{KolbTurnerBook}\\
$             t_{\rm{acc}}$      &$ 6.74^{+ 0.25}_{- 0.24}$ Gyr                                         &Acceleration time 			&$\tacc\approx$($9.778$ Gyr)$\times h^{-1}\int_{\zacc}^\infty [H_0/H(z)(1+z)]dz$~~~\cite{KolbTurnerBook}\\
$             t_{\rm{now}}$      &$13.76^{+ 0.15}_{- 0.15}$ Gyr                                         &Age of Universe now			&$\tnow\approx$($9.778$ Gyr)$\times h^{-1}\int_0^\infty [H_0/H(z)(1+z)]dz$~~~\cite{KolbTurnerBook}\\
\hline
\multicolumn{3}{|l|}{Fundamental parameters (independent of observing epoch):}&\\
\cline{1-4}
$                        Q$      &$ 1.945^{+ 0.051}_{- 0.053}$ $\times10^{-5}$                          &Primordial fluctuation amplitude	&$Q=\delta_h\approx A_{.002}^{1/2}\times 59.2384\mu$K$/T_{\rm CMB}$\\
$                   \kappa$      &$ 1.3^{+ 3.7}_{- 4.3}$ $\times10^{-61}$                               &Dimensionless spatial curvature \cite{Q}&$\kappa=(\hbar c/k_B T_{\rm CMB} a)^2 k$\\ 
$             \rho_\Lambda$      &$ 1.48^{+ 0.11}_{- 0.11}$ $\times10^{-123}\rho_{\rm{Pl}}$             &Dark energy density			&$\rho_\Lambda\approx h^2\Ol\times(1.88\times 10^{-26}$kg$/$m$^3)$\\
$         \rho_{\rm{halo}}$      &$ 6.6^{+ 1.2}_{- 1.0}$ $\times10^{-123}\rho_{\rm{Pl}}$                &Halo formation density		&$\rhohalo=18\pi^2 Q^3\xi^4$\\
$                      \xi$      &$ 3.26^{+ 0.11}_{- 0.11}$ eV                                          &Matter mass per photon 		&$\xi =\rhom/\ng$\\
$                    \xi_b$      &$ 0.569^{+ 0.018}_{- 0.018}$ eV                                       &Baryon mass per photon 		&$\xib=\rhob/\ng$\\
$                    \xi_c$      &$ 2.69^{+ 0.11}_{- 0.10}$ eV                                          &CDM mass per photon 			&$\xic=\rhoc/\ng$\\
$                  \xi_\nu$      &$< 0.26 \>(95\percent)$ eV                                            &Neutrino mass per photon 		&$\xin=\rhon/\ng$\\
$                     \eta$      &$ 6.06^{+ 0.20}_{- 0.19}$ $\times10^{-10}$                            &Baryon/photon ratio			&$\eta=n_b/n_g=\xib/m_p$\\
$                A_\Lambda$      &$2077^{+135}_{-125}$                                                  &Expansion during matter domination	&$(1+\zeq)(\Om/\Ol)^{1/3}$ \cite{anthroneutrino}\\  
$      \sigma^*_{\rm{gal}}$      &$ 0.561^{+ 0.024}_{- 0.023}$ $\times10^{-3}$                          &Seed amplitude on galaxy scale	&Like $\sigma_8$ but on galactic ($M=10^{12} M_\odot$) scale early on\\
\hline
\multicolumn{3}{|l|}{CMB phenomenology parameters:}&\\
\cline{1-4}
$            A_{\rm{peak}}$      &$ 0.579^{+ 0.013}_{- 0.013}$                                          &Amplitude on CMB peak scales		&$\Ap=\As e^{-2\tau}$\\
$           A_{\rm{pivot}}$      &$ 0.595^{+ 0.012}_{- 0.011}$                                          &Amplitude at pivot point		&$A_{\rm{peak}}$ scaled to $k=0.028$/Mpc: $A_{\rm pivot}= 0.56^{\ns-1}A_{\rm{peak}}$ if $\al=0$\\
$                      H_1$      &$ 4.88^{+ 0.37}_{- 0.34}$                                             &1st CMB peak ratio			&$H_1(\Ot,\Ol,\ob,\om,w,\ns,\tau)$ given by \cite{observables}\\
$                      H_2$      &$ 0.4543^{+ 0.0051}_{- 0.0051}$                                       &2nd to 1st CMB peak ratio		&$H_2 = (0.925\om^{0.18} 2.4^{\ns-1})/[1+(\ob/0.0164)^{12\om^{0.52}})]^{0.2}$~~~\cite{observables}\\
$                      H_3$      &$ 0.4226^{+ 0.0088}_{- 0.0086}$                                       &3rd to 1st CMB peak ratio		&$H_3 = 2.17 [1+(\ob/0.044)^2]^{-1} \om^{0.59} 3.6^{\ns-1}/[1 + 1.63(1-\ob/0.071)\om]$\\
$        d_A(z_{\rm{rec}})$      &$14.30^{+ 0.17}_{- 0.17}$ Gpc                                         &Comoving angular diameter distance to CMB&$d_A(\zrec)={c\over H_0}\sinh\left[\Ok^{1/2}\int_0^\zrec [H_0/H(z)]dz\right]/\Ok^{1/2}$~~~\cite{KolbTurnerBook}\\
$        r_s(z_{\rm{rec}})$      &$ 0.1486^{+ 0.0014}_{- 0.0014}$ Gpc                                   &Comoving sound horizon scale		&$\rsound(\om,\ob)$ given by eq.~(22) of \cite{Hu04}\\
$            r_{\rm{damp}}$      &$ 0.0672^{+ 0.0009}_{- 0.0008}$ Gpc                                   &Comoving acoustic damping scale	&$\rdamp(\om,\ob)$ given by eq.~(26) of \cite{Hu04}\\
$                 \Theta_s$      &$ 0.5918^{+ 0.0020}_{- 0.0020}$                                       &CMB acoustic angular scale fit (degrees)&$\Th(\Ot,\Ol,w,\ob,\om)$ given by \protect\cite{observables}\\
$                   \ell_A$      &$302.2^{+ 1.0}_{- 1.0}$                                               &CMB acoustic angular scale		&$\l_A=\pi d_A(\zrec)/\rsound(\zrec)$\\ 
\hline
\end{tabular}
\end{center}     
} 
\end{table*}

\begin{table*}
\noindent 
{\footnotesize
Table~\ComparisonTable: 
How key cosmological parameter constraints depend on data used and on assumptions about other parameters.
The columns compare different theoretical priors indicated by numbers in {\it italics}. 
$w_c$ denotes dark energy that can cluster as in \cite{Spergel06}. Rows labeled ``+SDSS'' combine WMAP3 and SDSS LRG data.
\begin{center}
\begin{tabular}{|l|r|ccccccc|}
\hline
			   &Data  &Vanilla 			  &Vanilla+$\Ot$ 		 &Vanilla+$r$ 		        &Vanilla+$\alpha$ 	       &Vanilla+$\on$ 	      	     &Vanilla+$w$		&Vanilla+$w_c$\\
\hline
$\Omega_{\rm{tot}}$        & WMAP &${\it1}$                      &$1.054^{+0.064}_{-0.046}$     &${\it1}$                      &${\it1}$                      &${\it1}$                      &${\it1}$                      &${\it1}$                      \\
                           &+SDSS &${\it1}$                      &$1.003^{+0.010}_{-0.009}$     &${\it1}$                      &${\it1}$                      &${\it1}$                      &${\it1}$                      &${\it1}$                      \\
$\Omega_\Lambda$           & WMAP &$0.761^{+0.032}_{-0.037}$     &$0.60^{+0.14}_{-0.17}$        &$0.805^{+0.038}_{-0.042}$     &$0.708^{+0.051}_{-0.060}$     &$0.651^{+0.082}_{-0.086}$     &$0.704^{+0.071}_{-0.100}$     &$0.879^{+0.064}_{-0.168}$     \\
                           &+SDSS &$0.761^{+0.017}_{-0.018}$     &$0.757^{+0.020}_{-0.021}$     &$0.771^{+0.018}_{-0.019}$     &$0.750^{+0.020}_{-0.022}$     &$0.731^{+0.024}_{-0.030}$     &$0.757^{+0.019}_{-0.020}$     &$0.762^{+0.020}_{-0.021}$     \\
$\Omega_m$                 & WMAP &$0.239^{+0.037}_{-0.032}$     &$0.46^{+0.23}_{-0.19}$        &$0.195^{+0.042}_{-0.038}$     &$0.292^{+0.060}_{-0.051}$     &$0.349^{+0.086}_{-0.082}$     &$0.30^{+0.10}_{-0.07}$        &$0.12^{+0.17}_{-0.06}$        \\
                           &+SDSS &$0.239^{+0.018}_{-0.017}$     &$0.246^{+0.028}_{-0.025}$     &$0.229^{+0.019}_{-0.018}$     &$0.250^{+0.022}_{-0.020}$     &$0.269^{+0.030}_{-0.024}$     &$0.243^{+0.020}_{-0.019}$     &$0.238^{+0.021}_{-0.020}$     \\
$\omega_m$                 & WMAP &$0.1272^{+0.0082}_{-0.0080}$  &$0.1277^{+0.0082}_{-0.0079}$  &$0.1194^{+0.0096}_{-0.0092}$  &$0.135^{+0.010}_{-0.009}$     &$0.139^{+0.011}_{-0.011}$     &$0.1274^{+0.0083}_{-0.0082}$  &$0.1269^{+0.0082}_{-0.0080}$  \\
                           &+SDSS &$0.1272^{+0.0044}_{-0.0043}$  &$0.1260^{+0.0066}_{-0.0064}$  &$0.1268^{+0.0043}_{-0.0042}$  &$0.1271^{+0.0045}_{-0.0044}$  &$0.1301^{+0.0048}_{-0.0044}$  &$0.1248^{+0.0063}_{-0.0059}$  &$0.1264^{+0.0075}_{-0.0079}$  \\
$\omega_b$                 & WMAP &$0.0222^{+0.0007}_{-0.0007}$  &$0.0218^{+0.0008}_{-0.0008}$  &$0.0233^{+0.0011}_{-0.0010}$  &$0.0210^{+0.0010}_{-0.0010}$  &$0.0215^{+0.0009}_{-0.0009}$  &$0.0221^{+0.0007}_{-0.0007}$  &$0.0222^{+0.0008}_{-0.0007}$  \\
                           &+SDSS &$0.0222^{+0.0007}_{-0.0007}$  &$0.0222^{+0.0007}_{-0.0007}$  &$0.0229^{+0.0009}_{-0.0008}$  &$0.0213^{+0.0010}_{-0.0010}$  &$0.0221^{+0.0008}_{-0.0008}$  &$0.0223^{+0.0007}_{-0.0007}$  &$0.0224^{+0.0008}_{-0.0007}$  \\
$\omega_\nu$               & WMAP &${\it0}$                      &${\it0}$                      &${\it0}$                      &${\it0}$                      &$<0.024\>(95\percent)$        &${\it0}$                      &${\it0}$                      \\
                           &+SDSS &${\it0}$                      &${\it0}$                      &${\it0}$                      &${\it0}$                      &$<0.010\>(95\percent)$        &${\it0}$                      &${\it0}$                      \\
$M_\nu$                    & WMAP &${\it0}$                      &${\it0}$                      &${\it0}$                      &${\it0}$                      &$<2.2\>(95\percent)$          &${\it0}$                      &${\it0}$                      \\
                           &+SDSS &${\it0}$                      &${\it0}$                      &${\it0}$                      &${\it0}$                      &$<0.94\>(95\percent)$         &${\it0}$                      &${\it0}$                      \\
$w$                        & WMAP &${\it-1}$                     &${\it-1}$                     &${\it-1}$                     &${\it-1}$                     &${\it-1}$                     &$-0.82^{+0.23}_{-0.19}$       &$-1.69^{+0.88}_{-0.85}$       \\
                           &+SDSS &${\it-1}$                     &${\it-1}$                     &${\it-1}$                     &${\it-1}$                     &${\it-1}$                     &$-0.941^{+0.087}_{-0.101}$    &$-1.00^{+0.17}_{-0.19}$       \\
$\sigma_8$                 & WMAP &$0.758^{+0.050}_{-0.051}$     &$0.732^{+0.051}_{-0.046}$     &$0.706^{+0.064}_{-0.072}$     &$0.776^{+0.056}_{-0.053}$     &$0.597^{+0.085}_{-0.075}$     &$0.736^{+0.054}_{-0.052}$     &$0.747^{+0.066}_{-0.066}$     \\
                           &+SDSS &$0.756^{+0.035}_{-0.035}$     &$0.747^{+0.046}_{-0.044}$     &$0.751^{+0.036}_{-0.036}$     &$0.739^{+0.036}_{-0.035}$     &$0.673^{+0.056}_{-0.061}$     &$0.733^{+0.048}_{-0.043}$     &$0.745^{+0.057}_{-0.056}$     \\
$r_{.002}$                 & WMAP &${\it0}$                      &${\it0}$                      &$<0.65\>(95\percent)$         &${\it0}$                      &${\it0}$                      &${\it0}$                      &${\it0}$                      \\
                           &+SDSS &${\it0}$                      &${\it0}$                      &$<0.33\>(95\percent)$         &${\it0}$                      &${\it0}$                      &${\it0}$                      &${\it0}$                      \\
$n_s$                      & WMAP &$0.954^{+0.017}_{-0.016}$     &$0.943^{+0.017}_{-0.016}$     &$0.982^{+0.032}_{-0.026}$     &$0.871^{+0.047}_{-0.046}$     &$0.928^{+0.022}_{-0.024}$     &$0.945^{+0.017}_{-0.016}$     &$0.947^{+0.019}_{-0.017}$     \\
                           &+SDSS &$0.953^{+0.016}_{-0.016}$     &$0.952^{+0.017}_{-0.016}$     &$0.967^{+0.022}_{-0.020}$     &$0.895^{+0.041}_{-0.042}$     &$0.945^{+0.017}_{-0.017}$     &$0.950^{+0.016}_{-0.016}$     &$0.953^{+0.018}_{-0.017}$     \\
$\alpha$                   & WMAP &${\it0}$                      &${\it0}$                      &${\it0}$                      &$-0.056^{+0.031}_{-0.031}$    &${\it0}$                      &${\it0}$                      &${\it0}$                      \\
                           &+SDSS &${\it0}$                      &${\it0}$                      &${\it0}$                      &$-0.040^{+0.027}_{-0.027}$    &${\it0}$                      &${\it0}$                      &${\it0}$                      \\
$h$                        & WMAP &$0.730^{+0.033}_{-0.031}$     &$0.53^{+0.15}_{-0.10}$        &$0.782^{+0.058}_{-0.047}$     &$0.679^{+0.044}_{-0.040}$     &$0.630^{+0.065}_{-0.044}$     &$0.657^{+0.085}_{-0.086}$     &$1.03^{+0.46}_{-0.37}$        \\
                           &+SDSS &$0.730^{+0.019}_{-0.019}$     &$0.716^{+0.047}_{-0.043}$     &$0.744^{+0.022}_{-0.021}$     &$0.713^{+0.022}_{-0.022}$     &$0.695^{+0.025}_{-0.028}$     &$0.716^{+0.031}_{-0.029}$     &$0.727^{+0.037}_{-0.034}$     \\
$t_{\rm{now}}$             & WMAP &$13.75^{+0.16}_{-0.16}$       &$16.0^{+1.5}_{-1.8}$          &$13.53^{+0.21}_{-0.25}$       &$13.98^{+0.20}_{-0.20}$       &$14.31^{+0.24}_{-0.33}$       &$13.96^{+0.34}_{-0.28}$       &$13.44^{+0.49}_{-0.27}$       \\
                           &+SDSS &$13.76^{+0.15}_{-0.15}$       &$13.93^{+0.59}_{-0.58}$       &$13.65^{+0.17}_{-0.18}$       &$13.90^{+0.19}_{-0.19}$       &$13.98^{+0.22}_{-0.20}$       &$13.80^{+0.18}_{-0.17}$       &$13.77^{+0.26}_{-0.24}$       \\
$\tau$                     & WMAP &$0.090^{+0.029}_{-0.029}$     &$0.083^{+0.029}_{-0.029}$     &$0.091^{+0.031}_{-0.032}$     &$0.101^{+0.031}_{-0.031}$     &$0.082^{+0.029}_{-0.030}$     &$0.087^{+0.030}_{-0.031}$     &$0.087^{+0.030}_{-0.030}$     \\
                           &+SDSS &$0.087^{+0.028}_{-0.030}$     &$0.088^{+0.029}_{-0.031}$     &$0.085^{+0.029}_{-0.031}$     &$0.101^{+0.032}_{-0.032}$     &$0.087^{+0.028}_{-0.029}$     &$0.090^{+0.030}_{-0.031}$     &$0.089^{+0.030}_{-0.032}$     \\
$b$                        & WMAP &                              &                              &                              &                              &                              &                              &                              \\
                           &+SDSS &$1.896^{+0.074}_{-0.069}$     &$1.911^{+0.092}_{-0.086}$     &$1.919^{+0.078}_{-0.072}$     &$1.853^{+0.081}_{-0.077}$     &$2.03^{+0.11}_{-0.10}$        &$1.897^{+0.076}_{-0.072}$     &$1.92^{+0.10}_{-0.08}$        \\
$Q_{\rm{nl}}$              & WMAP &                              &                              &                              &                              &                              &                              &                              \\
                           &+SDSS &$30.3^{+4.4}_{-4.1}$          &$30.0^{+4.6}_{-4.2}$          &$30.9^{+4.5}_{-4.1}$          &$34.7^{+6.1}_{-5.4}$          &$34.9^{+6.9}_{-5.3}$          &$31.0^{+4.7}_{-4.3}$          &$31.0^{+5.0}_{-4.4}$          \\
\hline
$\Delta\chi^2$             & WMAP  &$   0.0$                    &$  -2.0$                    &$   0.0$                    &$  -3.6$                    &$  -1.0$                    &$  -1.0$                    &$   0.0$                     \\
                           &+SDSS  &$   0.0$                    &$   0.0$                    &$  -0.5$                    &$  -2.4$                    &$  -0.5$                    &$  -0.9$                    &$  -0.3$                     \\
\hline
\end{tabular}
\end{center}     
} 
\end{table*}

\begin{figure*} 
\centerline{\epsfxsize=18cm\epsffile{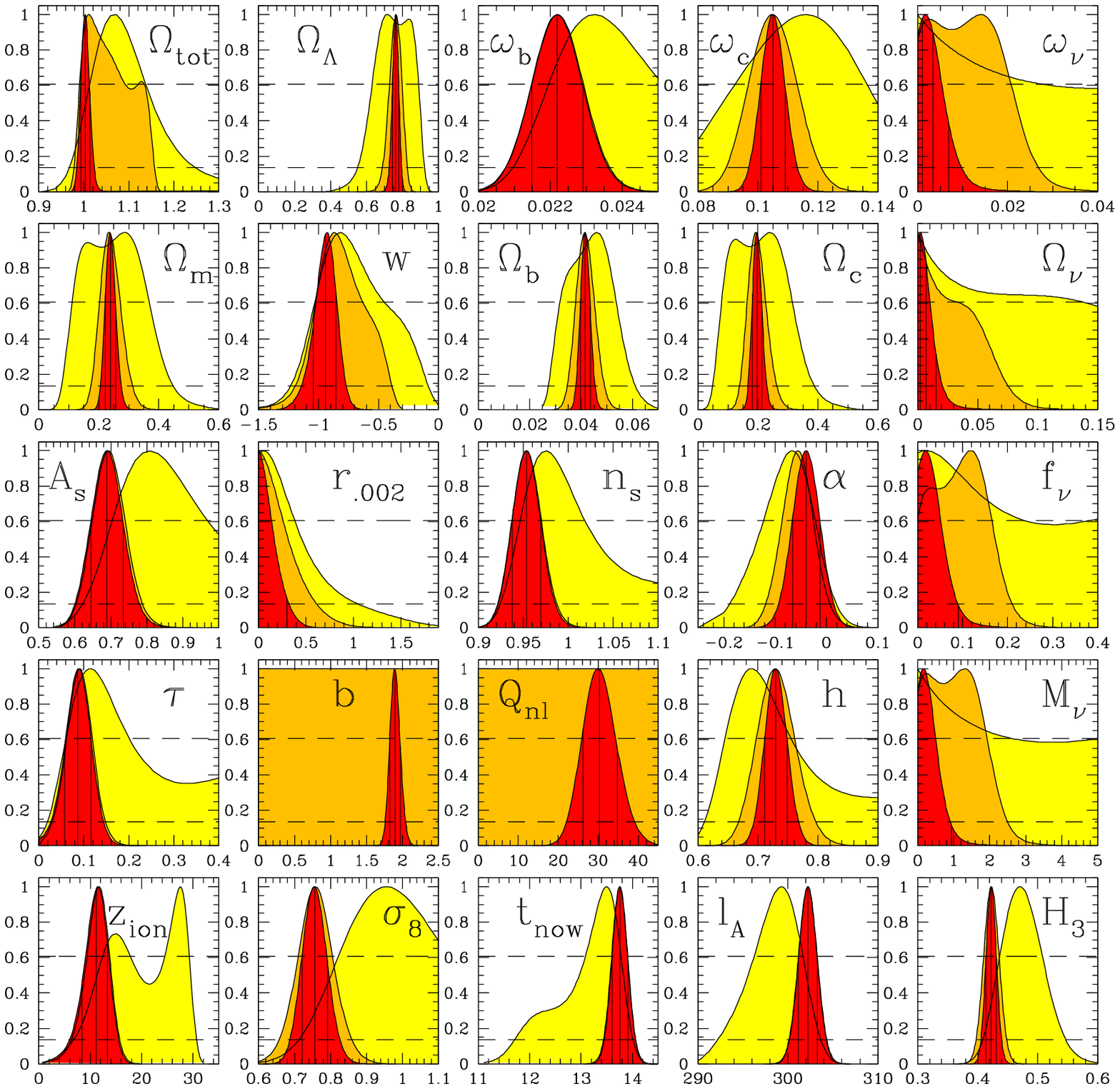}}
\caption[1]{\label{1d_fig}\footnotesize%
Constraints on key individual cosmological quantities using 
WMAP1 (yellow/light grey distributions),
WMAP3 (narrower orange/grey distributions) and 
including SDSS LRG information 
(red/dark grey distributions). If the orange/grey is completely hidden behind the red/dark grey, the
LRGs thus add no information.
Each distribution shown has been marginalized over all other quantities
in the ``vanilla'' class of models parametrized by $(\Ol,\ob,\ocdm,\As,\ns,\tau,b,\Qnl)$.
The parameter measurements and error bars quoted in the tables correspond
to the median and the central 68\% of the distributions, indicated by 
three vertical lines for the WMAP3+SDSS case above.
When the distribution peaks near zero (like for $r$), we instead quote an upper limit
at the 95th percentile (single short vertical line).
The horizontal dashed lines indicate $e^{-x^2/2}$ for
$x=1$ and $2$, respectively, so if the distribution were Gaussian, 
its intersections with these lines would correspond to 
$1\sigma$ and $2\sigma$ limits, respectively.
}
\end{figure*}

\subsection{Basic results}

Our constraints on individual cosmological parameters are given 
in Tables~{\ParameterTable} and {\ComparisonTable} and illustrated in \fig{1d_fig},
both for WMAP alone and when including our SDSS LRG information.
Table~{\ParameterTable} and \fig{1d_fig} take the Occam's razor approach of marginalizing only over 
``vanilla'' parameters $(\Ol,\ob,\ocdm,\As,\ns,\tau,b,\Qnl)$, whereas Table~{\ComparisonTable} shows 
how key results depend on assumptions about the non-vanilla parameters
$(\Ot,\on,w,r,\al)$ introduced one at a time. 
In other words, Table~{\ParameterTable} and \fig{1d_fig} use the vanilla assumptions by default; for example, models with $\on\ne 0$ are
used only for the constraints on $\on$ and other neutrino parameters ($\On$, $\xin$, $\fn$ and $\Mnu$).

The parameter measurements and error bars quoted in the tables correspond
to the median and the central 68\% of the probability distributions, indicated by 
three vertical lines in \fig{1d_fig}.
When a distribution peaks near zero, we instead quote an upper limit at the 95th percentile.
Note that the tabulated median values are near but not identical to those of the maximum likelihood model.
Our best fit vanilla model has $\Ol=0.763$, $\ob=0.0223$, $\ocdm=0.105$, $\As=0.685$, $\ns=0.954$, $\tau=0.0842$, $b=1.90$, $\Qnl=31.0$.
As customary, the $2\sigma$
contours in the numerous two-parameter figures below are drawn where the likelihood
has dropped to $0.0455$ of its maximum value, which corresponds to  
$\Delta\chi^2\approx 6.18$ and $95.45\%\approx 95\%$
enclosed probability for a two-dimensional Gaussian distribution.

We will spend most of the remainder of this paper digesting this information one 
step at a time, focusing on what WMAP and SDSS do and don't tell us about 
the underlying physics, and on how robust the constraints are to assumptions about physics and data sets.
The one-dimensional constraints in the tables and \fig{1d_fig} 
fail to reveal important information hidden in parameter correlations and degeneracies, so 
we will study
the joint constraints on key 2-parameter 
pairs. We will begin with the vanilla 6-parameter space of models, 
then introduce additional parameters (starting in \Sec{VanillaSec}) to quantify both how accurately we can
measure them and to what extent they weaken the constraints on the other parameters.

First, however, some of the parameters in Table~{\ParameterTable} deserve comment.
The additional parameters below the double line in Table~{\ParameterTable} are all determined by those above
the double line by simple functional relationships, and fall into several groups.

Together with the usual suspects under the heading ``other popular parameters'', 
we have included alternative fluctuation amplitude  parameters:
to facilitate comparison with other work, we quote the seed fluctuation amplitudes
not only at the
scale $k=0.05$/Mpc employed by CMBfast \cite{cmbfast}, CAMB \cite{CAMB} and CosmoMC \cite{CosmoMC} (denoted $\As$ and $r$), but also at the
scale $k=0.002$/Mpc employed by the WMAP team in \cite{Spergel06} (denoted $A_{.002}$ and $r_{.002}$).

The ``cosmic history parameters'' specify when our universe 
became matter-dominated, recombined, reionized, started accelerating ($\ddot{a}>0$), 
and produced us.

Those labeled ``fundamental parameters'' are intrinsic properties of our universe that are independent of
our observing epoch $\tnow$. (In contrast, most other parameters would have different numerical values if
we were to measure them, say, 10 Gyr from now. 
For example, $\tnow$ would be about $24$ Gyr, $\zeq$ and $\Ol$ would be larger, and
$h$, $\Om$ and $\om$ would all be smaller. Such parameters are therefore not properties of our universe, but
merely alternative time variables.)

The $Q$-parameter (not to be confused with $\Qnl$!) is the primordial density fluctuation amplitude $\sim 10^{-5}$.
The curvature parameter $\kappa$ is the 
curvature that the Universe would have had at the Planck time
if there was no inflationary epoch, and its small numerical value $\sim 10^{-61}$ 
constitutes the flatness problem that inflation solves.
$(\xi,\xib,\xic,\xin)$ are the fundamental parameters corresponding to the cosmologically popular quartet
$(\Om,\Ob,\Oc,\On)$, giving the densities per CMB photon.
The current densities are $\rho_i=\rho_h\omega_i$, where $i=m,b,c,\nu$ and $\rho_h$ denotes the constant
reference density
$3(H/h)^2/8\pi G=3{(100\,\hbox{km}\,\hbox{s}^{-1}\Mpc^{-1})}^2/8\pi G\approx 1.87882\times 10^{-26}\hbox{kg}/\hbox{m}^3$,
so the conversion between the conventional and fundamental density parameters is 
$\xi_i\equiv\rho_i/\ng\approx 25.646\,\eV \times (\Tcmb/2.726\K)\omega_i$ in units where $c=1$.
The parameter $\xi_m$ is of the same order as the temperature at 
matter-radiation equality temperature, 
$k\Teq\approx 0.22\xi$ \cite{axions}\footnote{The 
matter-radiation equality temperature is given by 
\beq{TeqEq}
k\Teq = {30\zeta(3)\over\pi^4}{\left[1+{7\over 8}\Nnu\left({4\over 11}\right)^{4/3}\right]^{-1}}\xi\approx 0.2195\xi,
\eeq
where $\zeta(3)\approx 1.202$, and the effective number of neutrino species in the standard model 
is $\Nnu\approx 3.022$ \cite{GnedinGnedin98} when taking into account the effect of electron-positron annihilation 
on the relic neutrino energy density.
}. 

The tiny value $\sim 10^{-123}$ of the vacuum density $\rhol$ in Planck units where $c=G=\hbar=1$ constitutes the well-known
cosmological constant problem, and the tiny yet similar value of the parameter combination $Q^3\xi^4$ explains the origin of
attempts to explain this value anthropically 
\cite{BarrowTipler,LindeLambda,Weinberg87,Efstathiou95,Vilenkin95,Martel98,GarrigaVilenkin03,inflation}:
$Q^3\xi^4$ is roughly the density of the universe at the time 
when the first nonlinear 
dark matter halos would form if $\rhol=0$ \cite{axions}, so if $\rhol\gg Q^3\xi^4$, 
dark energy freezes fluctuation growth before then and no nonlinear structures ever form.

The parameters $(\Avac,\sigmagal)$ are useful for anthropic buffs, since 
they directly determine the density fluctuation history on galaxy scales 
through equation (5) in \cite{anthroneutrino} (where $\sigmagal$ is denoted $\sigma_M(0)$).
Roughly, fluctuations grow from the initial level $\sigmagal$ by a factor $\Avac$.
Marginalizing over the neutrino fraction gives $\Avac=2279^{+240}_{-182}$, $\sigmagal=0.538^{+ 0.024}_{- 0.022}\times 10^{-3}$.

The group labeled ``CMB phenomenology parameters'' contains parameters that correspond rather closely to the quantities most accurately measured
by the CMB, such as heights and locations of power spectrum peaks. Many are seen to be measured at the percent level or better.
These parameters are useful for both numerical and intuition-building 
purposes \cite{observables,Knox03,cmbfit,Kosowsky02,cmbwarp,pico}.
Whereas CMB constraints suffer from severe degeneracies involving physical parameters further up in the table 
(involving, \eg, $\Ot$ and $\Ol$ as discussed below),
these phenomenological parameters are all constrained with small and fairly uncorrelated measurement errors.
By transforming the 
multidimensional WMAP3 log-likelihood function into the space spanned by 
$(H_2,\om,\fn,\Ol,w,\Th,\Apivot,H_3,\al,r,\nt,\Op ,b,\Qnl)$, it becomes better approximated by our quartic polynomial fit described in
Footnote \ref{LikelihoodFootnote} and \cite{cmbfit}: 
for example, 
the rms error is a negligible $\Delta\ln\L\approx 0.03$ for the vanilla case.
Roughly speaking, this transformation replaces the curvature parameter $\Ot$ by the characteristic peak scale $\Th$, 
the baryon fraction by the ratio $H_2$ of the first two peak heights,
the spectral index $\ns$ by the ratio $H_3$ of the third to first peak heights, and the overall peak amplitude $\Ap$ by the amplitude $\Apivot$
at the pivot scale where it is uncorrelated with $\ns$.
Aside from this numerical utility, these parameters also help demystify the ``black box'' aspect of CMB parameter constraints, elucidating 
their origin in terms of features in the data and in the physics \cite{observables}.

\subsection{Vanilla parameters}
\label{VanillaSec}

\Fig{1d_fig} compares the constraints on key parameters from the 1-year WMAP data (``WMAP1''), the 3-year WMAP data (``WMAP3'') and 
WMAP3 combined with our SDSS LRG measurements (``WMAP+LRG''). 
We include the WMAP1 case because it constitutes a well-tested baseline 
and illustrates both the dramatic progress in the field and what the key new WMAP3 information is, particularly from $E$-polarization.

\subsubsection{What WMAP3 adds}
\label{WMAPaddsSec}

The first thing to note is the dramatic improvement from WMAP1 to WMAP3 emphasized in \cite{Spergel06}. (Plotted WMAP1 constraints are from \cite{sdsspars}.) 
As shown in \cite{Lewis06},
this stems almost entirely from the new measurement of the low-$\l$ E power spectrum, 
which detects the reionization signature at about $3\sigma$ and determines the corresponding optical depth $\tau=0.09\pm 0.03$.
This measurement breaks the severe vanilla degeneracy in the WMAP1 data \cite{Spergel03,sdsspars} (see \fig{2d_Omh_6par_fig}) and causes the dramatic tightening of the constraints on 
$(\ob,\ocdm,\Ol,\As,\ns)$ seen in the figures;
essentially, with $\tau$ well constrained, the ratio of large scale power to the acoustic peaks determines $\ns$, and the relative heights of the acoustic
peaks then determine $\ob$ and $\ocdm$ without residual uncertainty due to $\ns$.
Indeed, \cite{Lewis06} has shown that discarding all the WMAP3 polarization data (both TE and EE) and replacing it with 
a Gaussian prior $\tau=0.09\pm 0.03$ recovers parameter constraints essentially identical to those from the full WMAP3 data set. 
In \Sec{CMBrobustnessSec}, we will return to the issue of what happens if this $\tau$-measurement is 
compromised by polarized foreground contamination.

The second important change from WMAP1 to WMAP3 is that the central values of some parameters have shifted noticeably \cite{Spergel06}.
Improved modeling of noise correlations and polarized foregrounds have lowered the low-$\l$ TE power and thus eliminated the 
WMAP1 evidence for $\tau\sim 0.17$. Since the fluctuation amplitude scales as $e^{\tau}$ times the CMB peak amplitude, this $\tau$ drop of 0.08
would push $\sigma_8$ down by about $8\%$. 
In addition, better measurements around the 3rd peak and a change in analysis procedure (marginalizing over the SZ-contribution) have lowered $\om$ by about 13\%, 
causing fluctuation growth to start later ($\zeq$ decreases) and end earlier ($\zacc$ increases), 
reducing $\sigma_8$ by another 8\%.
These effects combine to lower $\sigma_8$ by about 21\% when also taking into account the slight lowering of $\ns$.

\subsubsection{What SDSS LRGs add}

\begin{figure}[T]
\centerline{\epsfxsize=\figsize\epsffile{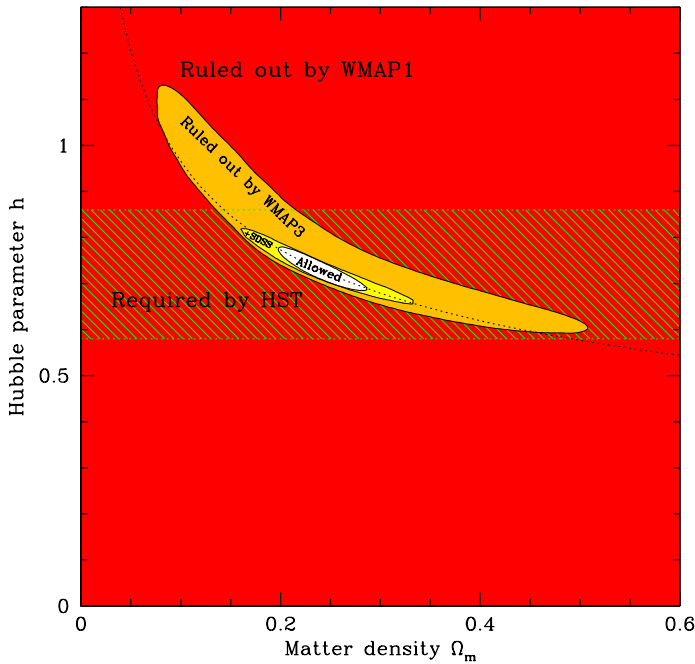}}
\caption[1]{\label{2d_Omh_6par_fig}\footnotesize%
95\% constraints in the $(\Om,h)$ plane. 
For 6-parameter ``vanilla'' models, 
the shaded red/grey region is ruled out by WMAP1 and the
shaded orange/grey region by WMAP3; the main source of the dramatic improvement 
is the measurement of E-polarization breaking the degeneracy involving 
$\tau$. Adding SDSS LRG information further constrains the parameters to the white region marked ``Allowed''.
The horizontal hatched band is required by the HST key project
\protect\cite{Freedman01}. 
The dotted line shows the fit
$h = 0.72(\Om/0.25)^{-0.32}$, explaining the origin of the
percent-level constraint $h(\Om/0.25)^{0.32} = 0.719\pm 0.008$ $(1\sigma)$.
}
\end{figure}

A key reason that non-CMB datasets such as the 2dFGRS and the SDSS 
improved WMAP1 constraints so dramatically was that they helped break the vanilla banana degeneracy seen in \fig{2d_Omh_6par_fig}, so the fact that WMAP3 now mitigates this internally with its
E-polarization measurement of $\tau$ clearly reduces the value added by other datasets.
However, Table~{\ComparisonTable} shows that our LRG measurements nonetheless give substantial improvements, cutting error bars on $\Om$, $\om$ and $h$ by about a factor of two for vanilla models and by up to almost an order of magnitude
when curvature, tensors, neutrinos or $w$ are allowed.

\begin{figure}[T]
\centerline{\epsfxsize=\figsize\epsffile{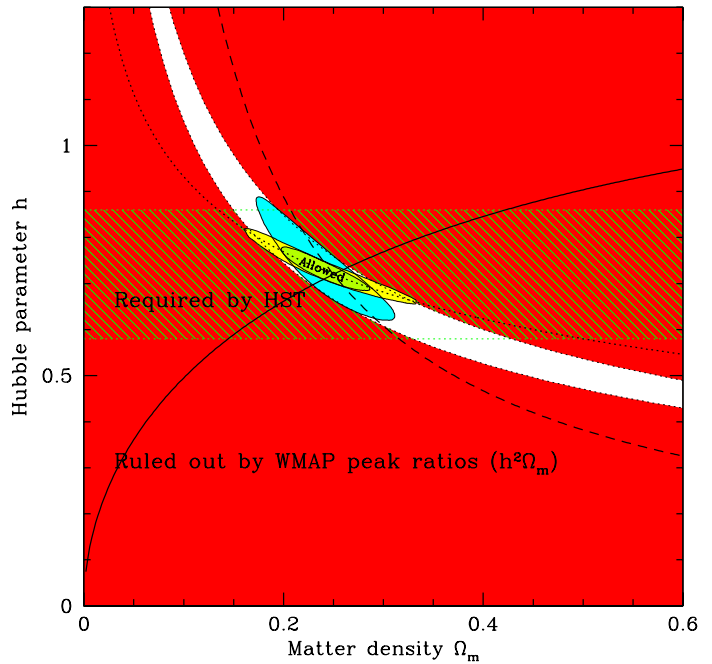}}
\caption[1]{\label{Omh_assumptionsFig}\footnotesize%
Illustration of the physics underlying the previous figure.
Using only WMAP CMB peak height ratios constrains $(\om,\ob,\ns)$ independently of $\As$, $\tau$,
curvature and late-time dark energy 
properties. This excludes 
all but the white band 
$\om\equiv h^2\Om=0.127\pm 0.017$ ($2\sigma$). 
If we assume $\Ot=1$ and vanilla dark energy, we can supplement this 
with independent ``standard ruler'' information from either
WMAP CMB (thin yellow/light grey ellipse) giving 
$\Om=0.239\pm 0.034$ ($1\sigma$), or SDSS galaxies (thicker blue/grey ellipse) giving 
$\Om=0.239\pm 0.027$ ($1\sigma$).
These two rulers are not only beautifully consistent, but also complementary, with the joint constraints
(small ellipse marked ``allowed'') being tighter than those from using either separately, giving 
$\Om=0.238\pm 0.017$ ($1\sigma$). The plotted 2-dimensional constraints are all $2\sigma$.
The three black curves correspond to constant ``horizontal'' observables: 
constant angular scales 
for the acoustic peaks in the CMB power (dotted, $h\simpropto\Om^{-0.3}$), 
for the acoustic peaks in the galaxy power (solid, $h\simpropto\Om^{0.37}$) and 
 for the turnover in the galaxy power spectrum (dashed, $h\simpropto\Om^{-0.93}$).
This illustrates why the galaxy acoustic scale is even more helpful than that of the CMB 
for measuring $\Om$: although it is currently less accurately measured, its degeneracy direction
is more perpendicular to the CMB peak ratio measurement of $h^2\Om$.
}
\end{figure}

The physics underlying these improvements is illustrated in \fig{Omh_assumptionsFig}.
The cosmological information in the CMB splits naturally into two parts, one ``vertical'' and one ``horizontal'', corresponding to the vertical and
horizontal positions of the power spectrum peaks.

\clearpage

\clearpage

By vertical information, we mean the relative heights of the acoustic peaks, which depend only on the physical matter densities 
$(\om,\ob,\on)$ and the scalar primordial power spectrum shape $(\ns,\al)$. 
They are independent of curvature and dark energy, since $\Ol(z)\approx\Ok(z)\approx 0$ at $z\simgt 10^3$. They are independent of $h$, 
since the physics at those early times 
depended only on the expansion rate as a function of temperature back then, 
which is simply $\xi^{1/2} T^{3/2}$ times a known numerical constant, where $\xi$ is given 
by $\om$ and the current CMB temperature (see Table 3 in \cite{axions}).
They are also conveniently independent of $\tau$ and $r$, which change the power spectrum shape only at $\l\ll 10^2$.

By horizontal (a.k.a.~``standard ruler'') information, we mean the acoustic angular scale $\lA\equiv\pi\dA(\zrec)/\rsound(\zrec)$ 
defined in Table~{\ParameterTable}. The $\l$-values of CMB power spectrum peaks and troughs are all equal to $\lA$ times constants depending 
on $(\om,\ob)$, so changing $\lA$ by some factor by altering $(\Ok,\Ol,w)$ 
simply
shifts the CMB peaks horizontally by that factor and alters the late integrated Sachs Wolfe effect at $\l\ll 10^2$.
Although this single number $\lA$ is now measured to great precision ($\sim 0.3\%$), it depends on multiple parameters, and it is popular to break this degeneracy with assumptions rather than measurements.
The sound horizon at recombination $\rsound(\zrec)$ in the denominator depends only weakly on $(\om,\ob)$, 
which are well constrained from the vertical information, and Table~{\ParameterTable} shows that it is now known to about 1\%.
In contrast, the comoving angular diameter distance to recombination $\dA(\zrec)$ depends sensitively on both 
the spatial curvature $\Ok$ and the cosmic expansion history $H(z)$, which in turn depends on the history of 
the dark energy density:
\beq{Heq}
{H(z)\over H_0}=\left[X(z)\Ol + (1+z)^2\Ok + (1+z)^3\Om + (1+z)^4\Or)\right]^{1/2}.
\eeq
Here $X(z)$ is defined as the dark energy density relative to its present value \cite{supernovae}, 
with vanilla dark energy (a cosmological constant) corresponding to $X(z)=1$. 
The most common (although physically unmotivated) parametrization of this function in the literature has been a simple power law
$X(z)=(1+z)^{3(1+w)}$, although it has also been constrained with a variety of other parametric and non-parametric approaches
(see \cite{Wang06} and references therein). The parameter $\Or$ refers to the radiation contribution from 
photons and massless neutrinos, 
which is given by $h^2\Or\approx 0.0000416 (\Tcmb/2.726\K)^4$ and makes a negligible contribution at low redshift.

Using the vertical WMAP information alone gives a tight constraint on $\om\equiv h^2\Om$, corresponding to the white 
band in \fig{Omh_assumptionsFig}, independent of assumptions about curvature or dark energy.\footnote{To obtain 
this $\om$-constraint, we marginalized over $\lA$ by marginalizing over
either $\Ok$ or $w$; Table~{\ComparisonTable} shows that these two approaches give essentially identical answers.}
To this robust measurement, we can now add two independent pieces of information 
if we are willing to make the vanilla assumptions that curvature vanishes and dark energy is a cosmological constant:
If we add the WMAP horizontal information, the allowed region shrinks to 
the thin ellipse hugging the $h\simpropto\Om^{-0.3}$ line of constant $\lA$ (dotted).
If we instead add the LRG information (which constrains $h\Om^{0.93}$ via the $P(k)$ turnover scale and $h\Om^{-0.37}$ via the acoustic oscillation 
scale\footnote{The origin of these scalings can be understood as follows. 
The matter-radiation equality horizon scale $\req\propto\om^{-1}$.
The sound horizon scales as $\rsound(\zeq)\propto\om^{-0.25}$ with a weak dependence on $\ob$ that is negligible in this context \cite{Hu04}.
For the LRG mean redshift $z=0.35$, 
the power law fit $\dA(z,\Om)= 0.3253 (\Om/0.25)^{-0.065}c H_0^{-1}\propto h^{-1}\Om^{-0.065}$ is quite good within our range of interest,
accurate to within about $0.1\%$ for $0.2<\Om<0.3$.
For $z=1100$, the power law fit $\dA(z,\Om)\approx 3.4374(\Om/0.25)^{-0.4}c H_0^{-1}\propto h^{-1}\Om^{-0.4}$ 
retains $0.1\%$ accuracy for for $0.19<\Om<0.35$.
The $P(k)$ turnover angle $\propto\req/\dA(0.35)\propto (h^2\Om)^{-1}/h^{-1}\Om^{-0.065}$ is therefore constant for $h\propto\Om^{-0.93}$,
the $P(k)$ acoustic angle $\propto\rsound/\dA(0.35)\propto (h^2\Om)^{-0.25}/h^{-1}\Om^{-0.065}$ is constant for $h\propto\Om^{0.37}$, and
the $C_\l$ acoustic angle $\propto\rsound/\dA(\zrec)\propto (h^2\Om)^{-0.25}/h^{-1}\Om^{-0.4}$ is constant for $h\propto\Om^{-0.3}$.}), 
the allowed region shrinks to the thick ellipse.

These two independent pieces of horizontal information are seen to be not only beautifully consistent, but also complementary:
the joint constraints
are significantly tighter than those from using either separately. 
When going beyond vanilla models below, the thin CMB-only ellipse is of course no longer relevant, making the LRG constraints even more valuable.

\begin{figure} 
\centerline{\epsfxsize=\figsize\epsffile{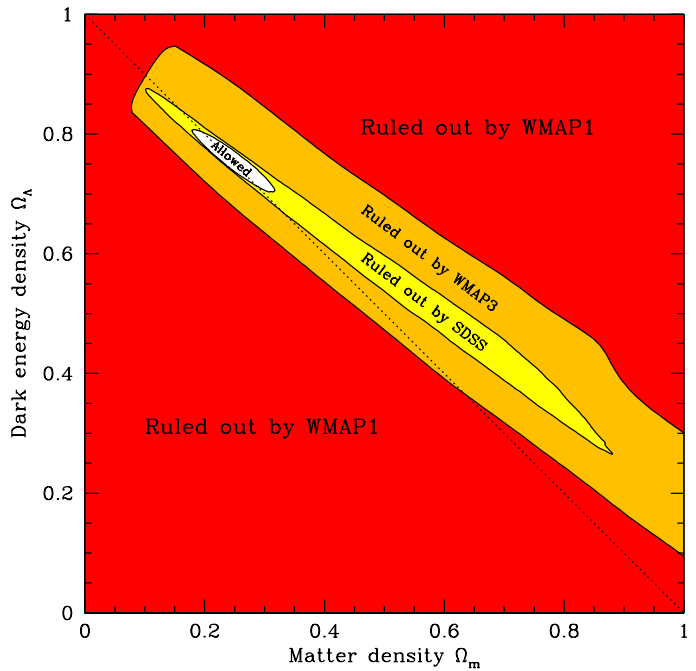}}
\caption[1]{\label{2d_OmOl_7par_fig}\footnotesize%
95\% constraints in the $(\Om,\Ol)$ plane. 
The large shaded regions are ruled out by 
WMAP1 (red/dark grey) and 
WMAP3 (orange/grey)
when spatial curvature is added to the 6 vanilla parameters, 
illustrating the well-known 
geometric degeneracy between models that all give the same 
acoustic peak locations in the CMB power spectrum.
The yellow/light grey region is ruled out when adding SDSS LRG information, breaking the degeneracy mainly by 
measuring the acoustic peak locations in the galaxy power spectrum.
Models on the diagonal dotted line are flat, those below are
open and those above are closed.
Here the yellow banana has been cut off from below by an $h\simgt 0.4$ prior in the CosmoMC software.
}
\end{figure}

\begin{figure} 
\centerline{\epsfxsize=\figsize\epsffile{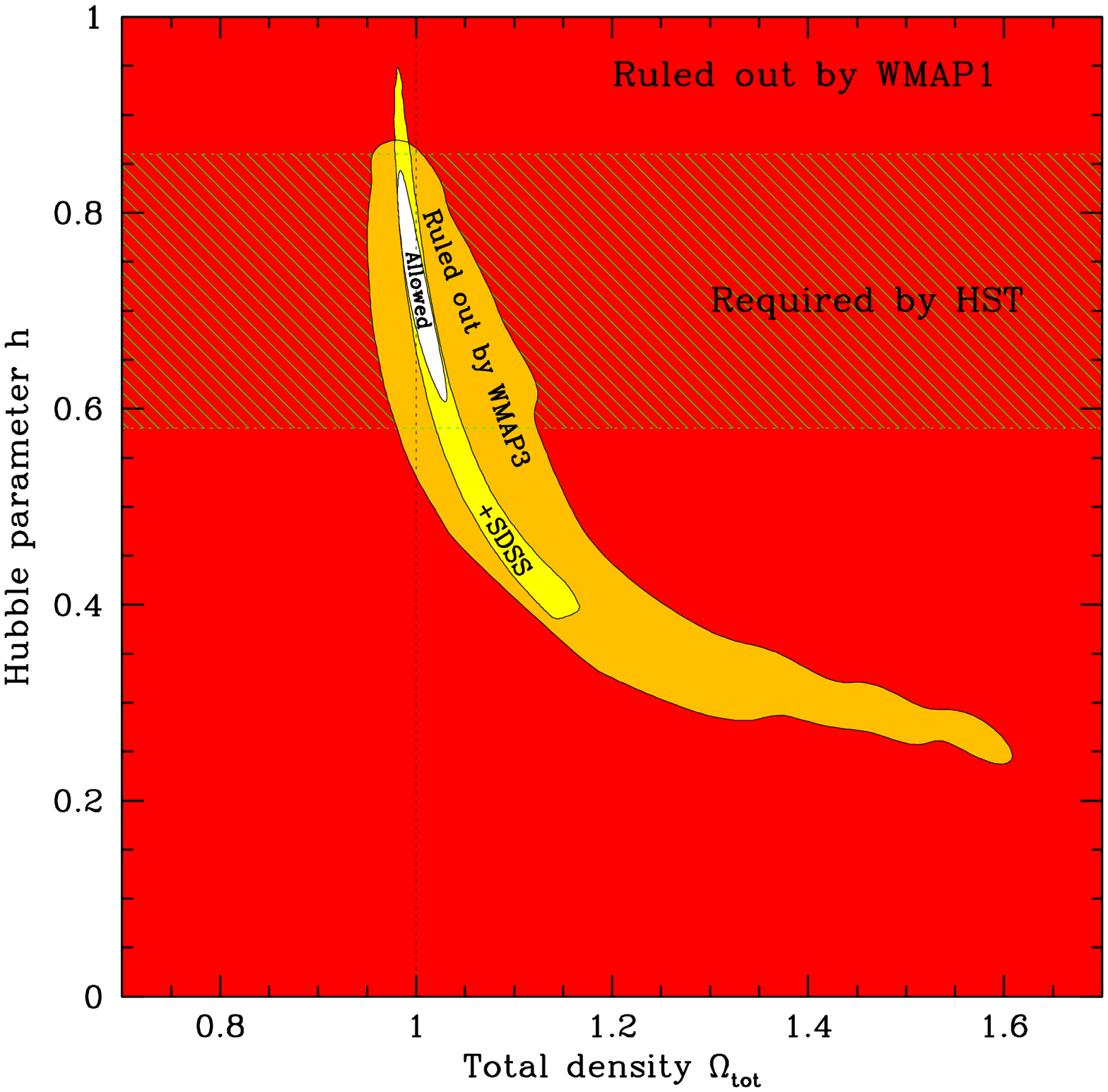}}
\caption[1]{\label{2d_Oth_7par_fig}\footnotesize%
95\% constraints in the $(\Ot,h)$ plane for 7-parameter curved models. 
The shaded red/dark grey region was ruled out by WMAP1 alone, and 
WMAP3 tightened these constraints (orange/grey region), illustrating that 
CMB fluctuations alone do not simultaneously show space to be flat and measure the Hubble parameter.
The yellow/light grey region is ruled out when adding SDSS LRG information.
Here the yellow banana has been artificially cut off for $h\simgt 0.4$ by a hardwired prior in the CosmoMC software.
}
\end{figure}

\begin{figure} 
\centerline{\epsfxsize=\figsize\epsffile{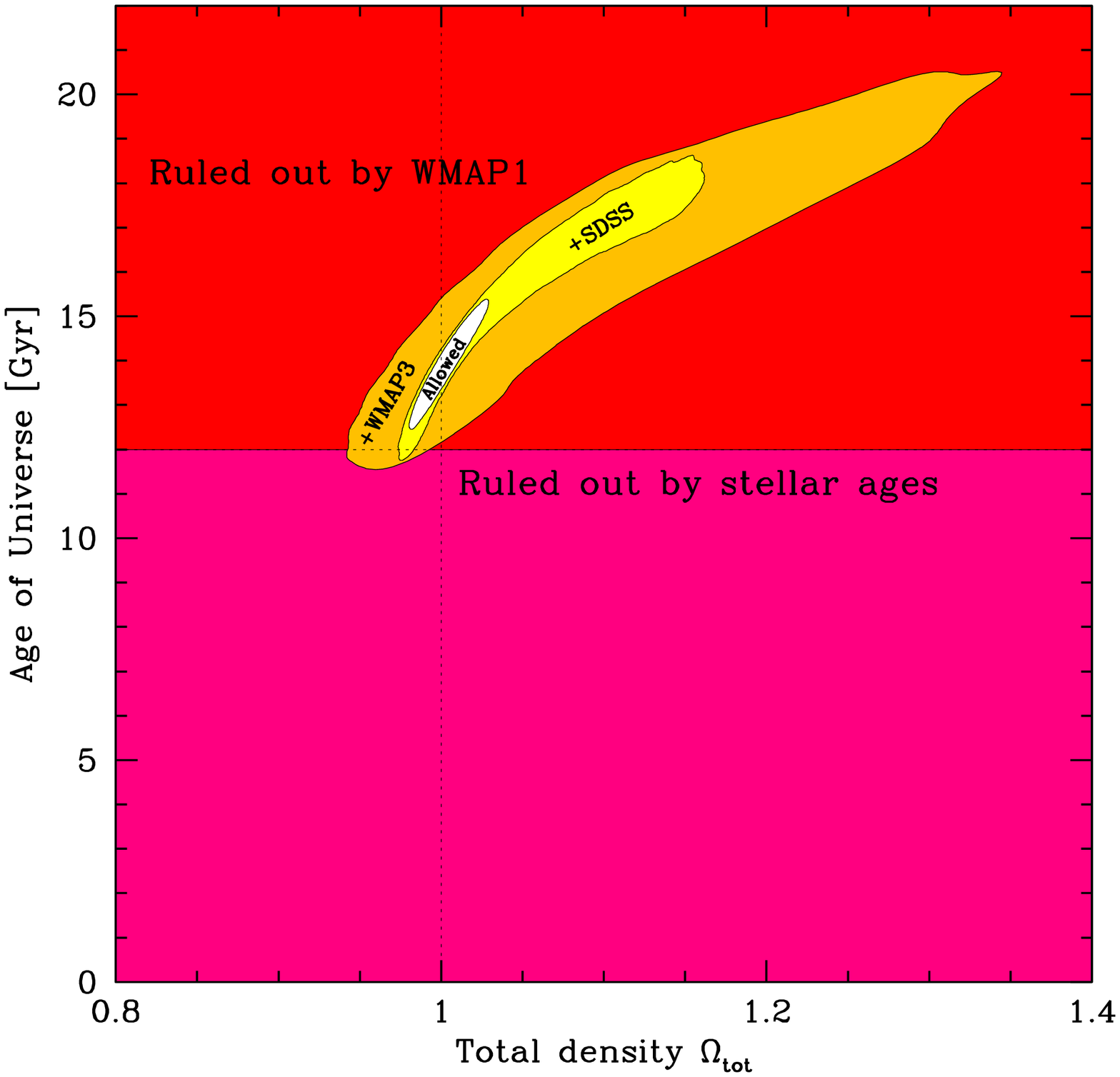}}
\caption[1]{\label{2d_Ott0_7par_fig}\footnotesize%
95\% constraints in the $(\Ot,\tnow)$ plane for 7-parameter curved models.
The shaded red/dark grey region is ruled out by WMAP1 alone, and 
WMAP3 tightened these constraints (orange/grey region), illustrating that 
CMB fluctuations do not simultaneously show space to be flat and measure the age of 
the Universe.
The yellow/light grey region is ruled out when adding SDSS LRG information.
The age limit $\age>12$ Gyr shown is the 95\% 
lower limit from
white dwarf ages by \cite{Hansen02}; for a review of 
recent age determinations, see \cite{Spergel06}.
}
\end{figure}

\subsection{Spacetime geometry}

To zeroth order (ignoring perturbations), the spacetime geometry is simply the Friedmann-Robertson-Walker metric 
determined by the curvature $\Ok$ and the cosmic expansion history $H(z)$.
The vanilla assumptions imply the special case of no curvature $(\Ok=0)$ and constant dark energy 
($H(z)$ given by \eq{Heq} with $X(z)=1$).

Let us now spice up the vanilla model space by including spatial curvature $\Ok$ 
and a constant dark energy equation of state $w$
as free parameters, both to constrain them and to quantify how other
constraints get weakened when dropping these vanilla assumptions. 

\subsubsection{LRGs as a standard ruler at $z=0.35$}
\label{StandardRulerSec}

Before constraining specific spacetime geometry parameters, let us review the relevant physics to 
intuitively understand what CMB and LRGs do and do not teach us about geometry.
As discussed in the previous section, current CMB data accurately 
measure only a single number that is sensitive to the
spacetime geometry information in $\Omega_k$ and $H(z)$.
This number is the peak angular scale $\lA$, and it in turn depends on the four independent 
parameters ($\Om,\Ok,w,h$). ($\Ol$ is of course not independent, fixed by the identity $\Ol=1-\Ok-\Om$.)
Since the sound horizon size $\rsound$ is now accurately known independently of spacetime geometry from CMB peak ratios, the CMB $\lA$-measurement
provides a precise determination of the comoving angular diameter distance to the last scattering surface, $\dA(\zrec)$, 
thus allowing {\it one} function of $(\Om,\Ok,w,h)$ to be accurately measured.

As emphasized in \cite{hubble,BlakeGlazebrook03,sdssbump,supernovae}, measuring the acoustic angular scale at 
low redshift in galaxy clustering similarly 
constrains a second independent combination of $(\Om,\Ok,w,h)$, and measuring $\dA(z)$ at multiple redshifts with
future redshift surveys and current and future SN Ia data can break all degeneracies and allow 
robust recovery of both $\Ok$ and the dark energy history $X(z)$.
For the galaxy approach, the point is that 
leaving the early universe physics ($\ob$, $\om$, $\ns$, \etc) fixed, 
changing the spacetime geometry merely scales the horizontal axis of the angular power spectrum of galaxies at a given redshift 
$z$ as $d_A(z)$.
More generally, as described in detail in \cite{sdssbump}, 
the main effect of changing the spacetime geometry 
is to shift our measured three-dimensional power 
spectrum horizontally by rescaling the 
$k$-axis. The $k$-scale for angular modes dilates as the comoving angular diameter 
distance $\dA(z)$ to the mean survey redshift $z\approx 0.35$, 
whereas that for radial modes dilates as $d(\dA)/dz=c/H(z)$ for the flat case.
For small variations around our best fit model,
the change in $H(0.35)$ is about half that of the angular diameter distance.
To model this, \cite{sdssbump} treats the net dilation as the cube root of 
the product of the radial dilation times the square of the transverse dilation, defining 
the distance parameter 
\beq{dVdefEq}
\dV(z)\equiv\left[d_A(z)^2{cz\over H(z)}\right]^{1/3}.
\eeq
Using only the vertical WMAP peak height information as a prior on $(\ob,\od,\ns)$, our LRG power spectrum gives
the measurement
$\dV(0.35)=1.300\pm 0.088$ Gpc, 
which agrees well with the value measured in \cite{sdssbump} using the LRG correlation function.
It is this geometric LRG information that explains most of the degeneracy breaking seen in the 
Figures~\ref{2d_OmOl_7par_fig}, \ref{2d_Oth_7par_fig}, \ref{2d_Ott0_7par_fig} and~\ref{2d_Omw_7parw_fig} below.

As more LRG data become available and strengthen the baryon bump detection from a few $\sigma$ to $> 5\sigma$, this measurement should become even more robust, 
not requiring any $\om$-prior from WMAP peak heights.

\subsubsection{Spatial curvature}
\label{CurvatureSec}

Although it has been argued that closed inflation models require particularly ugly fine-tuning \cite{Linde0303245}, 
a number of recent papers have considered nearly-flat models either 
to explain the low CMB quadrupole \cite{Efstathiou03},
in string theory landscape-inspired short inflation models,  
or for anthropic reasons \cite{Linde95,Vilenkin97,Q}, 
so it is clearly interesting and worthwhile to continue sharpening observational tests of the flatness assumption.
In the same spirit, measuring the Hubble parameter $h$
independently of theoretical assumptions about curvature and
measurements of galaxy distances at low redshift provides a powerful consistency check 
on our whole framework.  

Figures~\ref{2d_OmOl_7par_fig}, \ref{2d_Oth_7par_fig} and~\ref{2d_Ott0_7par_fig} illustrate the well-known CMB
degeneracies between the curvature $\Ok \equiv 1 - \Ot$ and dark energy $\Ol$, the Hubble parameter $h$, and the age of the universe
$\tnow$; without further information or priors, one
cannot simultaneously demonstrate spatial flatness and accurately measure $\Ol$, $h$ or $\tnow$, since the CMB accurately constrains
only the single combination $\lA$.
Indeed, the WMAP3 degeneracy banana extends towards even larger $\Ot$ than these figures indicate; the plotted banana has been artificially
truncated by a hardwired lower limit on $h$ in the CosmoMC software used to compute this particular MCMC.

Including our LRG information is seen to reduce the curvature uncertainty by
about a factor of five, providing a
striking vindication of the standard inflationary prediction $\Ot=1$. The physical reason for this LRG improvement is obvious from 
the thick ellipse in \fig{Omh_assumptionsFig}: WMAP vertical peak height information combined with LRG standard ruler 
information on $\dV(0.35)$ measures $\Om$ rather independently of curvature.

Yet even with WMAP+LRG information, the figures show that a strong degeneracy  
persists between curvature and $h$, and curvature and $\age$, leaving the measurement uncertainty on $h$ 
comparable with that from the HST key project \cite{Freedman01}.
If we add the additional assumption that
space is {\it exactly} flat, then uncertainties shrink by factors around 
4 and 10 for $h$ and $\age$, respectively, still in beautiful agreement with other measurements.

In conclusion, within the class of almost flat models, 
the WMAP-only constraints on $h$, $\age$, $\Ol$ and $\Ot$ remain weak, and 
including our LRG measurements provides a huge improvement in precision.

\begin{figure} 
\centerline{\epsfxsize=\figsize\epsffile{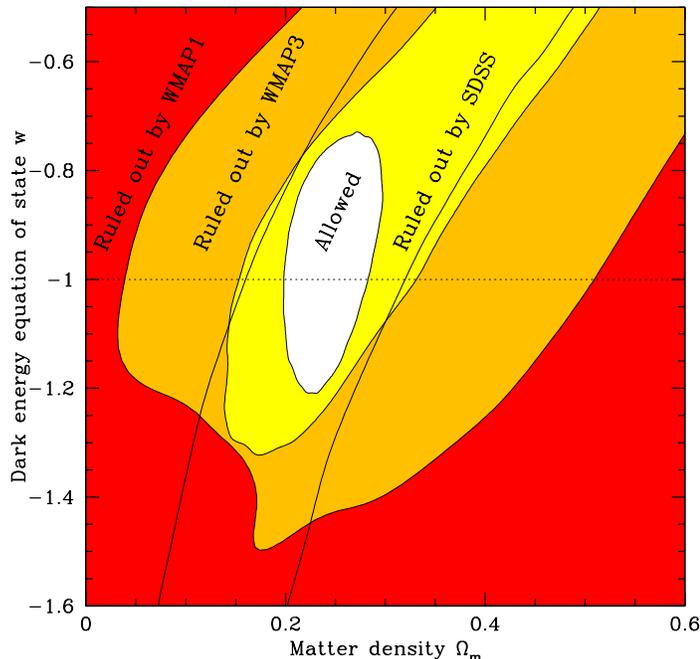}}
\caption[1]{\label{2d_Omw_7parw_fig}\footnotesize%
95\% constraints in the $(\Om,\w)$ plane. 
The shaded red/grey region is ruled out by WMAP1 alone
when the dark energy equation of state $w$ is added to the 6 vanilla parameters.
The shaded orange/grey region is ruled out by WMAP3.
The yellow/light grey region is ruled out when adding SDSS LRG information.
The region not between the two black curves is ruled out by WMAP3 when dark energy is assumed to cluster.
}
\end{figure}

\subsubsection{Dark energy}

Although we now know its present density fairly accurately, we still know precious little 
else about dark energy, and much interest is
focused on understanding its nature.
Assuming flat space1, Table~{\ComparisonTable} and \fig{2d_Omw_7parw_fig} show our constraints on constant $w$ for two cases: 
assuming that dark energy is homogeneous (does not cluster) and that it allows spatial perturbations (does cluster) as modeled in \cite{Spergel06}. 
We see that adding $w$ as a free parameter does
not significantly improve $\chi^2$ for the best fit, and all data are consistent with
the vanilla case $w=-1$, with $1\sigma$ uncertainties in $w$ in the 10\% - 30\% range, depending on dark energy clustering assumptions.
 
As described above, the physical basis of these constraints is similar to those for curvature, since 
(aside from low-$\l$ corrections from the late ISW effect and dark energy clustering), 
the only readily observable effect of the dark energy density history $X(z)$ is 
to alter $\dA(\zrec)$ and $\dA(0.35)$, and hence the CMB and LRG acoustic angular scales.
(The dark energy history also affects fluctuation growth and hence the power spectrum amplitude, but we do not measure this 
because our analysis marginalizes over the galaxy bias parameter $b$.)

It has been argued (see, \eg, \cite{Wright06}) that it is inappropriate to assume $\Ok=0$ when constraining $w$, since there is currently no experimental 
evidence for spatial flatness unless $w=-1$ is assumed.
We agree with this critique, and merely note that no interesting joint constraints can currently be placed on 
as many as four spacetime geometry parameters ($\Om,\Ok,w,h$) from WMAP and our LRG measurements alone, 
since they accurately constrain only
the two combinations $\dA(\zrec)$ and $\dV(0.35)$. Other data such as SN Ia need to be included for this; \cite{Spergel06}
do this and obtain $w=-1.06^{+0.13}_{-0.08}$. 

One can also argue, in the spirit of Occam's razor, that the fact that vanilla works so well can be taken as evidence 
against {\it both} $\Ok\ne 0$ and $w\ne -1$, since it would require a fluke coincidence for them to both have significantly non-vanilla
values that conspire to lie on the same $\dV(0.35)$ and $\dA(\zrec)$ degeneracy tracks as the vanilla model.

\subsection{Inflation}
\label{InflationSec}

\begin{figure} 
\centerline{\epsfxsize=\figsize\epsffile{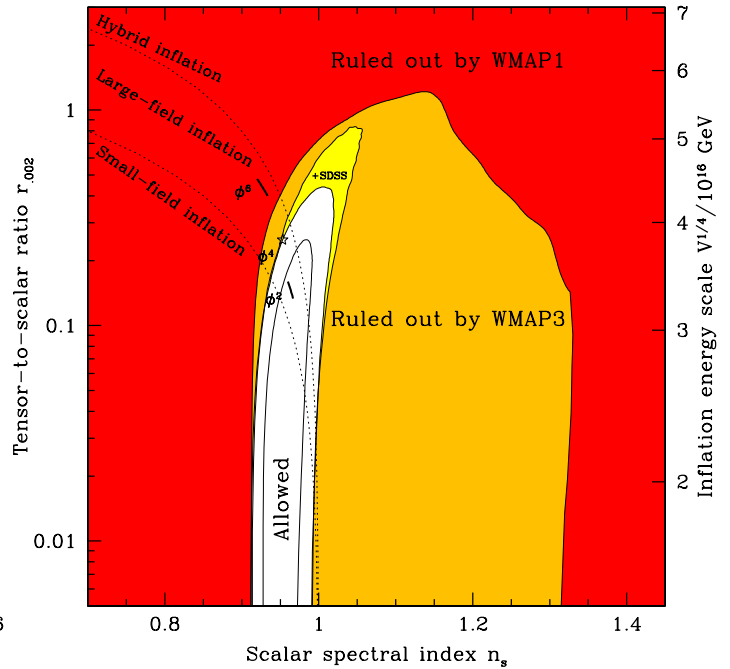}}
\caption[1]{\label{2d_nsr_7parr_fig}\footnotesize%
95\% constraints in the $(\ns,r_{.002})$ plane 
for 7-parameter tensor models (the vanilla parameters plus $r$).
The large shaded regions are ruled out by 
WMAP1 (red/dark grey) and 
WMAP3 (orange/grey).
The yellow/light grey region is ruled out when adding SDSS LRG information, 
pushing the upper limit on $r_{.002}$ down by a factor of two to 
$r_{.002}<0.33$ (95\%). The solid black curve without shading shows the 68\% 
limit.
The two dotted lines delimit the three classes of inflation models known 
as small-field, large-field and hybrid models.
Some single-field inflation models make highly specific predictions in this plane as indicated.
From top to bottom, the figure shows the predictions for
$V(\phi)\propto\phi^6$ (line segment; ruled out by CMB alone), 
$V(\phi)\propto\phi^4$ (star; a textbook inflation model; on verge of exclusion) and
$V(\phi)\propto\phi^2$ (line segment; the eternal stochastic inflation model; still allowed). 
These predictions assume 
that the 
number of e-foldings between horizon exit of the observed fluctuations and 
the end of inflation is 64 for the $\phi^4$ model and between 50 and 60 for the others
as per \cite{LiddleLeach03}.
}
\end{figure}

Inflation \cite{Guth81,Starobinsky1980,Linde82,AlbrechtSteinhardt82,Linde83}
remains the leading paradigm for what happened in the early universe because it can solve
the flatness, horizon and monopole problems (see, \eg, \cite{cmbtaskforce})
and has, modulo minor caveats, 
successfully predicted that $\Ot\approx 1$, $\ns\approx 1$, $|\alpha|\ll 1$ and $r\simlt 1$
as well as the facts that the seed fluctuations are mainly Gaussian and adiabatic. 
For the ekpyrotic universe alternative \cite{Ekpyrotic01}, controversy remains about 
whether it can survive a ``bounce'' and  whether it predicts $\ns\approx 1$ \cite{Ekpyrotic03} or $\ns\approx 3$ \cite{Creminelli05}.

In the quest to measure the five parameters $(Q,\ns-1,\al,r,\nt)$ characterizing inflationary seed fluctuations, 
the first breakthrough was the 1992 COBE discovery that $Q\sim 10^{-5}$ and that the other four quantities were 
consistent with zero \cite{Smoot92}.
The second breakthrough is currently in progress, with WMAP3 suggesting $1-\ns>0$ at 
almost the $3\sigma$ level ($1-\ns=0.049^{+0.019}_{-0.015}$) \cite{Spergel06}. 
This central value is in good agreement with classic (single slow-rolling scalar field) inflation models, 
which generically predict non-scale invariance in the ballpark $1-\ns\sim 2/N\sim 0.04$, assuming that the 
number of e-foldings between the time horizon the observed fluctuations exit the horizon and 
the end of inflation is $50<N<60$ as per \cite{LiddleLeach03}.
This central value of $\ns$ agrees well with numerous measurements in the recent 
literature (\eg, \cite{b03pars}); it is merely the error bars that have changed.

As illustrated in \fig{2d_nsr_7parr_fig} and discussed in \cite{Spergel06}, $\ns=1$ becomes allowed if 
the tensor fluctuation parameter $r$ is included (as it clearly should be when constraining inflation models),
but the ``vanilla lite'' Harrison-Zeldovich model $(\ns=1,r=0)$ remains ruled out. In contrast, the arguably simplest of all 
inflation models, a single slow-rolling scalar field with potential $V(\phi)\propto\phi^2$, 
remains viable: it predicts
$(\ns,r)=(1-2/N,8/N)\approx (0.96,0.15)$. The string-inspired ``N-flation'' model makes a similar prediction \cite{nflation,Easther05}.

Our constraints on the inflation parameters $(Q,\ns,\al,r,\nt)$ in Table~{\ComparisonTable} and \fig{2d_nsr_7parr_fig} are seen to confirm 
those reported in \cite{Spergel06} ---
the main addition of our LRG analysis is simply to provide a clean way of tightening the 
WMAP-only constraints on both $\Ot$ and $r$ (by factors of 5 and 2, respectively).
Lyman $\alpha$ Forest (Ly$\alpha$F) constraints provide valuable complementary information on smaller scales, constraining the 
running of the spectral index 
to vanish at the percent level \cite{Seljak06,Viel06}.

Since the WMAP3 announcement, there has been substantial discussion of how strong the evidence against Harrison-Zeldovch ($\ns=1,r=0$)
really is \cite{Peiris06,Lewis06,Seljak06,Kinney06,Martin06,Eriksen06,Huffenberger06,Liddle06,Feng06}. 
For example, the WMAP team marginalized over the SZ-amplitude on small scales, which lowered
the $\ns$-estimate by about 0.01, but did not model the CMB lensing effect, 
which would raise the $\ns$-estimate by a comparable amount \cite{Lewis06}.
It has also been argued that improved modeling of point source contamination increases the $\ns$-estimate \cite{Huffenberger06}.
Inclusion of smaller-scale CMB data and Ly$\alpha$F information clearly affects the significance as well.
The bottom line is therefore that even modest improvements in measurement accuracy over the next few years can significantly improve our 
confidence in distinguishing between competing early-universe models --- even without detecting $r>0$.

\begin{figure} 
\centerline{\epsfxsize=\figsize\epsffile{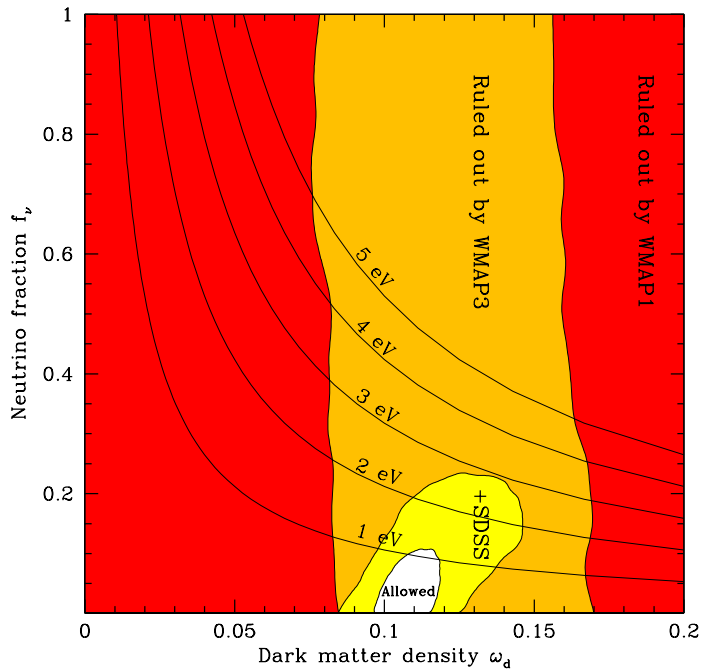}}
\caption[1]{\label{2d_odfn_7parnu_fig}\footnotesize%
95\% constraints in the $(\od,\fn)$ plane. 
The large shaded regions are ruled out by 
WMAP1 (red/dark grey) and 
WMAP3 (orange/grey)
when neutrino mass is added to the 6 vanilla parameters.
The yellow/light grey region is ruled out when adding SDSS LRG information.
The five curves correspond to $M_\nu$, the sum of the neutrino masses,
equaling 1, 2, 3, 4 and 5 eV, respectively --- barring sterile neutrinos,
no neutrino can have a mass exceeding 
$\sim M_\nu/3\approx 0.3$ eV (95\%).
}
\end{figure}

\subsection{Neutrinos}
 
It has long been known \cite{neutrinos} that galaxy surveys are sensitive probes of 
neutrino mass, since they can detect the suppression of small-scale 
power caused by neutrinos streaming out of dark matter overdensities. 
For detailed discussion of post-WMAP3 astrophysical neutrino constraints, see 
\cite{Spergel06,Goobar06,Lesgourges06,Fukugita06,Seljak06,Cirelli06,HannestadRaffelt06,Feng06a}.

Our neutrino mass constraints are shown in \fig{2d_odfn_7parnu_fig} and in the $\Mnu$-panel of
\fig{1d_fig}, where we allow our standard 6 vanilla
parameters and $f_\nu$ to be
free\footnote{It has been claimed that the true limits on neutrino masses from the  
WMAP1 (but not WMAP3) CMB maps are tighter than represented in these figures \cite{Sanchez06,Lesgourges06,Fukugita06}.}.
Assuming three active neutrinos with standard freezeout abundance, 
we obtain a 95\% 
upper limit $\Mnu<0.9\>$eV, so combining this with the atmospheric 
and solar neutrino oscillation results \cite{King03,superk05},
which indicate small mass differences between the neutrino types,
implies that none of the three masses can exceed $\Mnu/3\approx 0.3\>$eV. In other words, 
the heaviest neutrino (presumably in a hierarchical model mostly a linear combination of 
$\nu_\mu$ and $\nu_\tau$) would have a mass in the range $0.04-0.3$ eV.

If one is willing to make stronger assumptions about the ability to model smaller-scale physics, notably involving the Ly$\alpha$F,
one can obtain the substantially sharper upper bound $\Mnu<0.17\>$eV \cite{Seljak06}.
However, it should be noted that \cite{Seljak06} also find that these same assumptions 
rule out the standard model with three active neutrino species at $2.5\sigma$, preferring more than three species.

\subsection{Robustness to data details}
\label{DataRobustnessSec}

Above, we explored in detail how our cosmological parameter constraints depend on assumptions about physics in the form of parameter priors ($\Ok=0$, $w=-1$, \etc).
Let us now discuss how sensitive they are to details related to data modeling.

\subsubsection{CMB modeling issues}
\label{CMBrobustnessSec}

With any data set, it is prudent to be extra cautious regarding the most recent additions and the parts with the lowest signal-to-noise ratio.
In the WMAP case, this suggests focusing on the $T$ power spectrum 
around the third peak and the large-scale $E$-polarization data, which as discussed in 
\Sec{WMAPaddsSec} were responsible for tightening and lowering the constraints on $\om$ and $\tau$, respectively.

The large-scale $E$-polarization data appear to be the most important area for further investigation, because they are single-handedly responsible 
for most of the dramatic WMAP3 error bar reductions, yet constitute only a
$3\sigma$ detection after foregrounds an order of magnitude larger have been 
subtracted from the observed polarized CMB maps \cite{Page06}. As discussed in \cite{Lewis06} and \Sec{WMAPaddsSec}, 
all the WMAP3 polarization information is 
effectively compressed into the probability distribution for $\tau$, since
using the prior $\tau=0.09\pm 0.03$ instead of 
the polarized data leaves the parameter constraints essentially unchanged.
This error bar $\Delta\tau=0.03$ found in \cite{Spergel06} and Table~{\ParameterTable} 
reflects only noise and sample variance and does not include foreground uncertainties.
If future foreground modeling increases this error bar substantially, it will reopen the vanilla banana degeneracy 
described in \cite{sdsspars}:
Increasing $\tau$ and $\As$ in such a way
that $A_{\rm{peak}}\equiv\As e^{-2\tau}$ stays constant, the peak heights remain unchanged and the only effect is 
to increase power on the largest scales. The large-scale power relative to the first peak can  
then be brought back down to the observed value by increasing $\ns$, after which the second peak
can be brought back down by increasing $\ob$. 
Since quasar observations of the Gunn-Peterson effect 
allow $\tau$ to drop by no more than about $1\sigma$ (0.03) \cite{Chiu03,Fan06},
the main change possible from revised foreground modeling is therefore that
$(\tau,\Ol,\od,\ob,\As,\ns,h)$ all increase together \cite{sdsspars}. 
For a more detailed treatment of these issues, see \cite{Zahn06}.

A separate issue is that, as discussed in \Sec{InflationSec}, 
reasonable changes in the CMB data modeling can easily increase $\ns$ by of order $0.01$ 
\cite{Peiris06,Lewis06,Seljak06,Kinney06,Martin06,Huffenberger06}, weakening the significance 
with which the Harrison-Zeldovich model ($\ns=1,r=0$) can be ruled out.

With the above-mentioned exceptions, parameter measurements now appear rather robust to WMAP modeling details.
We computed parameter constraints using the WMAP team chains available on the LAMBDA archive.
We created our own chains using the CosmoMC package \cite{CosmoMC} for the vanilla case 
(of length 310{,}817) as a cross-check and 
for the case with curvature (of length 226{,}456) since this was unavailable on LAMBDA. The parameter constraints were in excellent agreement between 
these two vanilla chains.
For a fair comparison between WMAP team and CosmoMC-based chains, the best-fit $\chi^2$ values listed in Table~{\ComparisonTable} have been offset-calibrated so that 
they all give the same value for our best fit vanilla model.

\begin{figure} 
\centerline{\epsfxsize=\figsize\epsffile{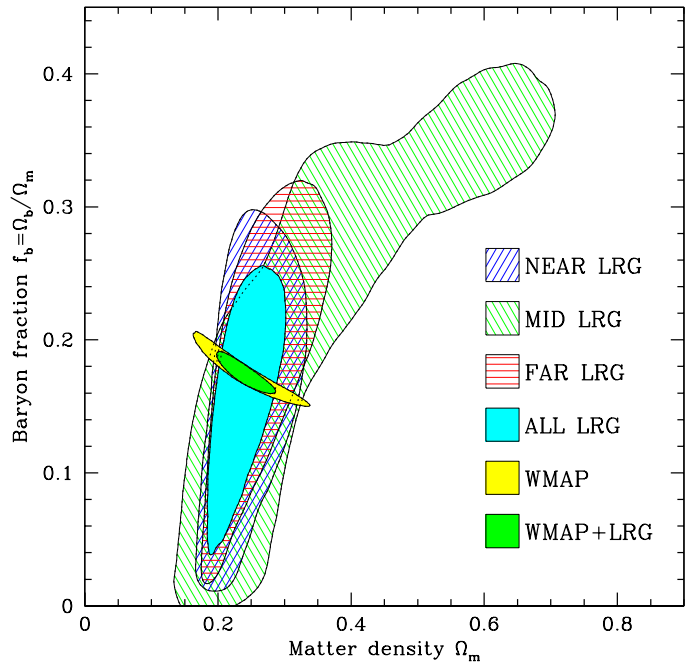}}
\caption[1]{\label{2d_Omfb_fig}\footnotesize%
The key information that our LRG measurements add to WMAP 
comes from the power spectrum shape.
Parametrizing this shape by $\Om$ and the baryon fraction $\Ob/\Om$ for vanilla models 
with $\ns=1$, $h=0.72$, the 95\% 
constraints above are seen to be nicely consistent between the various radial subsamples.
Moreover, the WMAP+LRG joint constraints from our full 6-parameter analysis are seen to be essentially
the intersection of the WMAP and ``ALL LRG'' allowed regions, indicating 
that these two shape parameters carry 
the bulk of the cosmologically useful LRG information.
}
\end{figure}

\begin{figure} 
\centerline{\epsfxsize=\figsize\epsffile{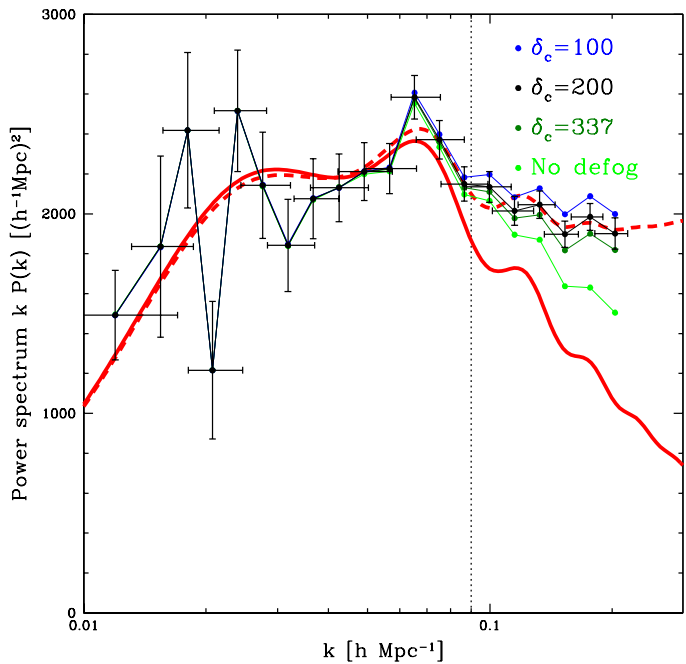}}
\caption[1]{\label{defogFig}\footnotesize%
Effect of finger-of-god (FOG) compression.
Raising the FOG compression threshold $\delta_c$ means that fewer FOGs are identified and compressed, 
which suppresses small-scale power while leaving the large scale power essentially unchanged.
}
\end{figure}

\subsubsection{LRG modeling issues}
\label{LRGrobustnessSec}

Since we marginalize over the overall amplitude of
LRG clustering  
via the bias parameter $b$, the LRG power spectrum adds 
cosmological information only through its shape.
Let us now explore how sensitive this shape is to 
details of the data treatment.
A popular way to parametrize the power spectrum shape in the literature has been in terms of the
two parameters $(\Om,\fb)$ shown in \fig{2d_Omfb_fig}, where $\fb\equiv\Ob/\Om$ is the baryon fraction.
Since we wish to use $(\Om,\fb)$ merely to characterize this shape here, 
not for constraining cosmology, we will ignore all CMB data and 
restrict ourselves to vanilla models with $\ns=1$, $h=0.72$ and $\As=1$, 
varying only the four parameters ($\Om,\fb,b,\Qnl)$.
\Fig{2d_Omfb_fig} suggest that for vanilla models, the two parameters
$(\Om,\fb)$ do in fact capture the bulk of this shape information,
since the WMAP+LRG joint constraints from our full 6-parameter analysis are seen to be essentially
the intersection of the WMAP and ``ALL LRG'' allowed regions in the $(\Om,\fb)$-plane.

\paragraph{Sensitivity to defogging}

\Fig{2d_Omfb_fig} shows good consistency between the power spectrum shapes recovered from the three radial subsamples.
Let us now explore in more detail issues related to our nonlinear modeling.
Our results were based on the measurement using 
FOG compression 
with threshold $\delta_c=200$ defined in \cite{sdsspower}.
Applied to the LRG sample alone, the FOG compression algorithm (described in detail in \cite{sdsspower}) finds
about 20\% of the LRGs in FOGs using this threshold;
77\% of these FOGs contain two LRGs, 16\% contain three, and $7\%$ contain more than three.
Thus not all LRGs are brightest cluster galaxies that each reside in a separate dark matter halo. 
\Fig{defogFig} shows a substantial dependence of $P(k)$ on this $\delta_c$ identification threshold for $k\simgt 0.1h/$Mpc. 
This is because FOGs smear out galaxy clusters along the line of sight, thereby strongly 
reducing the number of very close pairs, suppressing the small-scale power.
\Fig{defogFig} shows that on small scales, the approximate scaling $P(k)\simpropto k^{-1.3}$ seen for our default FOG compression matches the 
well-known correlation function scaling $\xi(r)\simpropto r^{-1.7}$, which also agrees with the binding energy considerations of 
\cite{FukugitaPeebles04}.
Fitting linear power spectra to these $P(k)$ curves would clearly give parameter constraints strongly dependent on $\delta_c$, with
less aggressive FOG-removal (a higher threshold $\delta_c$) masquerading as lower $\Om$.
Using our nonlinear modeling, however, we find that $\delta_c$ has almost no effect on the cosmological parameters, 
with the change seen in \fig{defogFig} being absorbed by a change in the $\Qnl$-parameter.
For the three cases $\delta_c=(100,200,337)$, our above-mentioned 4-parameter fits give
highly stable best-fit values $\Om=(0.244,0.242,0.243)$ and $\fb=(0.168,0.169,0.168)$ together with the strongly 
varying best-fit values $\Qnl=(27.0,30.9,34.2)$. 
If we fix the baryon density at the best fit WMAP3 value and vary only the three parameters ($\Om,b,\Qnl$),
the corresponding results are 
$\Om=(0.246,0.243,0.244)$ and $\Qnl=(27.1,31.0,34.3)$. 
Note that the cosmological parameter values do not show a rising or falling trend with $\delta_c$.
For comparison, the $1\sigma$ uncertainty on $\Om$ from 
Table~{\ParameterTable} is $\Delta\Om\approx 0.02$, an order of magnitude larger than these variations.
In other words, the $\Qnl$-parameter closely emulates the effect of changing $\delta_c$, so that 
marginalizing over $\Qnl$ is tantamount to marginalizing over $\delta_c$, making our treatment rather 
robust to the modeling of nonlinear redshift distortions.
 
\begin{figure} 
\vskip-1.45cm
\centerline{\epsfxsize=\figsize\epsffile{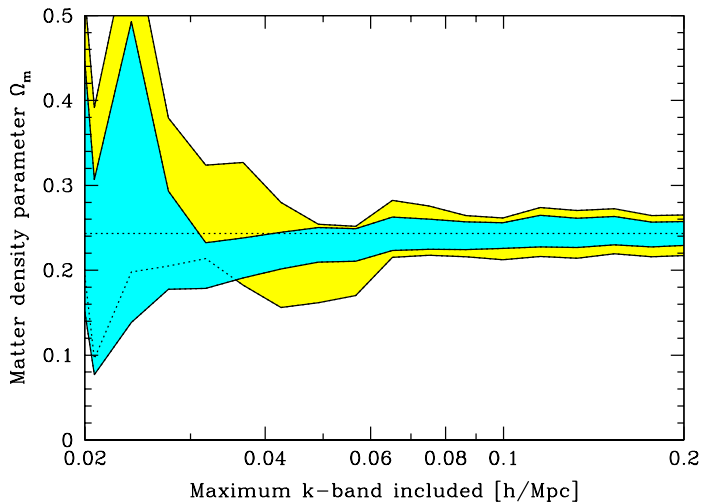}}
\vskip-1.45cm
\caption[1]{\label{OmFig}\footnotesize%
$1\sigma$ constraints on $\Om$ as a function of the largest $k$-band included in the analysis. 
The yellow band shows the result when marginalizing over the baryon density $\ob$, the thinner cyan/grey band shows the
result when fixing $\ob$ at the best-fit WMAP3 value. 
}
\end{figure}

\paragraph{Sensitivity to $k$-cutoff}

This is all very reassuring, showing that our cosmological constraints are almost completely unaffected by major changes in the
$k\simgt 0.1h$/Mpc power spectrum. (The reason that we nonetheless perform the $\Qnl$-marginalization is if course that we wish to 
immunize our results against any small nonlinear corrections that extend to $k\simlt 0.1h$/Mpc.)
To further explore this insensitivity to nonlinearities, 
we repeat the above analysis for the default $\delta_c=200$ case, including measurements for 
$0.01h/\Mpc\le k\le\kmax$, and vary the upper limit $\kmax$. We apply a prior $0\le\Qnl\le 50$ to prevent 
unphysical $\Qnl$-values for small $\kmax$-values (where $\Qnl$ becomes essentially unconstrained).  
If no nonlinear modeling is performed, then as emphasized in \cite{Percival06}, 
the recovered value of $\Om$ should increase with $\kmax$ as nonlinear effects become important.
In contrast, \fig{OmFig} shows that with our nonlinear modeling, the recovered $\Om$-value is strikingly 
insensitive to $\kmax$. For $\kmax\ll 0.07h/\Mpc$, the constraints are weak and fluctuate noticeably as each new
band power is included, but for $\kmax$ beyond the first baryon bump at $k\sim 0.07h/\Mpc$, both the central 
value and the measurement uncertainty 
remain essentially constant all the way out to $\kmax=0.2h/\Mpc$.

The above results tells us that, to a decent approximation, our $k\simgt 0.1/$Mpc data are not contributing information 
about cosmological parameters, merely information about $\Qnl$. Indeed,
the error bar $\Delta\Om$ is larger when using $k<0.2h/\Mpc$ data and marginalizing over $\Qnl$ then 
when using merely $k<0.09h/\Mpc$ data and fixing $\Qnl$.
In other words, our cosmological constraints come almost entirely from the
LRG power spectrum shape at $k\simlt 0.1 h/$Mpc.

\paragraph{Comparison with other galaxy $P(k)$-measurements}

Let us conclude this section by briefly comparing with $\Om$-values obtained from other recent galaxy clustering 
analyses. 

Our WMAP3+LRG measurement $\Om=0.24\pm 0.02$ has the same central as that from WMAP3 alone \cite{Spergel06}, merely with a smaller error bar,
and the most recent 2dFGRS team analysis also prefers $\Om\approx 0.24$ \cite{Sanchez06}. 
This central value is $1.5\sigma$ below the result $\Om=0.30\pm 0.04$ reported 
from WMAP1 + SDSS main sample galaxies in \cite{sdsspars}; part of the shift comes from the lower third peak in WMAP3 as discussed in
\Sec{VanillaSec}.
Post-WMAP3 results are also consistent with ours. Analysis of an independent SDSS LRG sample with photometric redshifts gave best-fit $\Om$-values
between 0.26 and 0.29 depending on binning \cite{sdsslrgcl}, while an independent analysis including acoustic 
oscillations in SDSS LRGs and main sample galaxies preferred $\Om\approx 0.256$ \cite{Percival06b}.

The galaxy power spectra measured from the above-mentioned data sets are likely to be reanalysed as nonlinear modeling methods improve.
This makes it interesting to compare their statistical constraining power. \cite{sdsslrgcl} do so by comparing the error bar 
$\Delta\Om$ from fitting two parameter $(\Om,b)$-models
to all $k\le 0.2 h/\Mpc$ data, with all other parameters, including $\Qnl$ 
or other nonlinear modeling parameters, 
fixed at canonical best fit values. This gives $\Delta\Om\approx 0.020$ for 2dFGRS and
$\Delta\Om\approx 0.012$ for for the SDSS LRG sample with photometric redshifts \cite{sdsslrgcl}.
Applying the same procedure to our LRGs yields $\Delta\Om=0.007$.
This demonstrates both the statistical power of our sample, and that our cosmological analysis has been 
quite conservative in the sense of 
marginalizing away much of the power spectrum information
(marginalizing over $\Qnl$ doubles the error bar to $\Delta\Om=0.014$).

\subsubsection{Other issues}

A fortunate side effect of improved cosmological precision is that priors now matter less.
Monte Carlo Markov Chain generators usually assume a uniform Bayesian prior in the space of its ``work parameters''. For example, 
if two different papers parametrize the fluctuation amplitude with $\As$ and $\ln As$, respectively, they implicitly assign 
$\As$-priors that are constant and $\propto 1/\As$, respectively (the new prior picks up a factor from the Jacobian of the parameter transformation).
Such prior differences could lead to substantial ($\sim 1\sigma$) discrepancies on parameter constraints a few years ago, when some parameters were still only known to a factor 
of order unity. In contrast, Table~{\ParameterTable} shows that most parameters are now measured with relative errors in the range $1\%-10\%$. 
As long as these relative measurement errors are $\ll 1$, such priors become unimportant:
Since the popular re-parametrizations in the literature and in Table~{\ParameterTable} involve smooth functions that do not blow up except perhaps 
where parameters vanish or take unphysical values, the relative variation of their Jacobian across the allowed parameter range will be of the same order
as the relative variation of the parameters ($\ll 1$), \ie, approximately constant.
Chosing a uniform prior across the allowed region in one parameter 
space is thus essentially equivalent to choosing a uniform prior across the allowed region of 
anybody else's favorite parameter space.

\section{Conclusions}
\label{ConclSec}

We have measured the large-scale real-space power spectrum $P(k)$ using luminous red 
galaxies in the Sloan Digital Sky Survey (SDSS) with 
narrow well-behaved window functions and uncorrelated minimum-variance errors. The results are 
publicly available in an easy-to-use form 
at \url{http://space.mit.edu/home/tegmark/sdss.html}.

This is an ideal sample for measuring the large-scale power spectrum, since its 
effective volume exceeds that of the SDSS main galaxy sample by a factor of six and that of the 2dFGRS by an order of magnitude.
Our results are robust to omitting purely angular and purely 
radial density fluctuations and are consistent between different
parts of the sky. 
They provide a striking model-independent confirmation 
of the predicted large-scale $\Lambda$CDM power spectrum.
The baryon signature is clearly detected (at $3\sigma$), and the acoustic 
oscillation scale provides a robust measurement of the distance
to $z=0.35$ independent of curvature and dark energy assumptions. 

Although our measured power spectrum provides independent cross-checks on $\Om$ and the baryon fraction, in good agreement with WMAP,
its main utility for cosmological parameter estimation lies in complementing CMB measurements by breaking their degeneracies;
for example, Table~{\ComparisonTable} shows that it cuts 
error bars on $\Om$, $\om$ and $h$ by about a factor of two for vanilla models 
(ones with a cosmological constant and negligible curvature, tensor modes, neutrinos and running spectral index)
and by up to almost an order of magnitude
when curvature, tensors, neutrinos or $w$ are allowed.
We find that all these constraints are essentially independent of scales $k>0.1h$/Mpc and associated nonlinear complications.

Since the profusion of tables and figures in \Sec{CosmoSec} can be daunting to digest, 
let us briefly summarize them and discuss both where we currently 
stand regarding cosmological parameters and some outstanding issues.

\subsection{The success of vanilla}
 
The first obvious conclusion is that ``vanilla rules OK''. 
We have seen several surprising claims about cosmological parameters
come and go recently, such as a running spectral index, very early
reionization and cosmologically detected neutrino mass ---
yet the last two rows of Table~{\ComparisonTable} show that there is no strong evidence in the data for any non-vanilla behavior:
none of the non-vanilla parameters reduces $\chi^2$ significantly relative to the vanilla case.
The WMAP team made the same comparison for the CMB-only case and came to the same conclusion \cite{Spergel06}.
Adding a generic new parameter would be expected to reduce $\chi^2$ by about unity by fitting random scatter.
Although WMAP alone very slightly favor spatial curvature, this preference disappears when SDSS is included.
The only non-vanilla behavior that is marginally favored 
is running spectral index $\alpha<0$, although only at $1.6\sigma$.
This persistent success of the vanilla model may evoke disturbing parallels with the enduring success of the standard model of particle physics, which has 
frustrated widespread hopes for surprises. However, 
the recent evidence for $\ns<1$ 
represents a departure from the $\ns=1$ ``vanilla lite'' model that had been
an excellent fit ever since COBE \cite{Smoot92}, and as we discuss below, 
there are good reasons 
to expect further qualitative progress soon.

\subsection{Which assumptions matter?}

When quoting parameter constraints, it is important to know how sensitive they are to assumptions about both data sets and priors.
The most important data assumptions discussed in \Sec{DataRobustnessSec} are probably those about 
polarized CMB foreground modeling for constraining $\tau$ and those about nonlinear 
galaxy clustering modeling for constraining the power spectrum shape.
The effect of priors on other parameters is seen by comparing the seven columns of Table~{\ComparisonTable}, and the effect 
of including SDSS is seen by comparing odd and even rows.

WMAP alone has robustly nailed certain parameters so well that that neither adding SDSS information nor changing priors
have  any significant effect. Clearly in this camp are
the baryon density $\ob$ (constrained by WMAP even-odd peak ratios) and 
the reionization optical depth $\tau$ (constrained by WMAP low-$\l$ E-polarization); indeed, Table 1 in \cite{Seljak06} 
shows that adding Ly$\alpha$F and other CMB and LSS data does not help here either.
The spectral index $\ns$ is also in this nailed-by-WMAP category as long as we assume that $\al$ is negligible; generic slow-roll inflation models 
predict $|\al|\simlt 10^{-3}$, well below the limits of detectability with current data sets.

For many other parameters, \eg, $\Om$, $h$ and $\tnow$, 
the WMAP-only constraints are 
extremely sensitive to priors, with the inclusion of SDSS information tightening them by factors 2 - 10.
The prior assumptions of the vanilla model ($\Ok=r=\fn=\al=0$, $w=-1$)
matter a lot with WMAP alone, and when one of them is dropped, the
best fit values of $\Om$ and $h$ are typically very different,
with much larger errors. These assumptions no longer matter much
when SDSS is included,
greatly simplifying the caveat list that the cautious cosmologist needs to keep in mind.
This is quite different from the recent past, when the joint constraints from older WMAP and SDSS data were sensitive to 
prior assumptions such as spatial flatness \cite{sdsspars}; a major reason for this change is clearly 
the SDSS measurement of the baryon acoustic scale.
Indeed, one of the most interesting results of our analysis
is the strengthened evidence for a flat universe, with
the constraint on $\Ot$ tightening from
$1.054^{+0.064}_{-0.046}$ (WMAP3 only) to $1.003^{+0.010}_{-0.009}$ (WMAP3+SDSS).

In other words, large-scale cosmic clustering data now robustly constrain {\it all} the vanilla parameters, 
even when any one of $(\fn,\Ok,r,\fn,w)$ are
included as in Table~{\ComparisonTable}. If $w$ is varied jointly with $\Ok$ (as it arguably should be \cite{Wright06}), one expects 
dramatically weakened constraints on the two (since two standard rulers cannot determine the three parameters $(\w,\Ok,\Om)$), but rather 
unaffected degradation for the rest.

\subsection{Other data}
\label{OtherDataSec}

Our cosmological parameter analysis has been very conservative, using the bare minimum number of data sets (two) needed to break all 
major degeneracies, and using measurements which mainly probe the large-scale linear regime.
It is therefore interesting to compare our results with the complementary approach of \cite{Seljak06} of pushing the envelope by using essentially {\it all} 
available data
(including Ly$\alpha$F, supernovae Ia and smaller-scale CMB experiments), which gives tighter constraints at the cost of more caveats.
Comparing with the error bars in Table 1 of \cite{Seljak06} shows that the additional data give 
merely modest improvements for $(\ob,\od,\ns,r,h)$, a halving of the error bars on $\Ot$
(still consistent with flatness), and great gains for $\alpha$ and $\Mnu$.
These last two parameters are strongly constrained by the small-scale Ly$\alpha$F information, 
with \cite{Seljak06} reporting $\alpha=-0.015\pm 0.012$ and $\Mnu/3<0.06$ eV (95\%), a factor of six below our constraint and bumping right up against 
the atmospheric lower bound $\sim 0.04$ eV.
On the other hand, the same analysis also rules out the standard model with three active neutrino 
species at $2.5\sigma$ \cite{Seljak06};
one can always worry about pushing the envelope too far by underestimating modeling uncertainties and systematics. 
\cite{Seljak06} also highlight interesting tension at the $2\sigma$-level 
between the Ly$\alpha$F and WMAP3 data regarding the fluctuation amplitude $\sigma_8$, and weak gravitational 
lensing may emerge as the decisive arbiter here, by directly pinning down the matter 
fluctuation amplitude independently of bias \cite{Jarvis05,Heavens06}.

\begin{figure} 
\vskip-1.45cm
\centerline{\epsfxsize=\figsize\epsffile{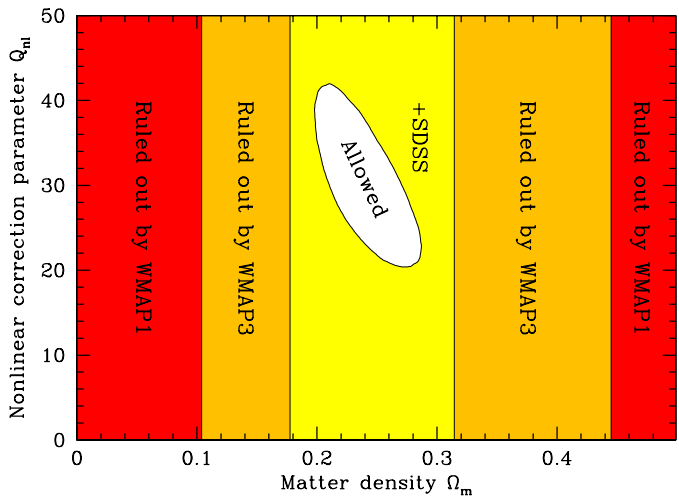}}
\vskip-1.45cm
\caption[1]{\label{2d_OmQnlFig}\footnotesize%
95\% constraints in the $(\Om,\Qnl)$ plane for vanilla models. 
The shaded regions are ruled out by 
WMAP1 (red/dark grey), WMAP3 (orange/grey) and when adding SDSS LRG information.
}
\end{figure}

\subsection{Future challenges}
\label{FutureSec}
 
The impressive improvement of cosmological measurements is likely to continue in coming years.
For example, the SDSS should allow substantially better cosmological constraints from LRGs for several reasons.
When the SDSS-II legacy survey is complete, the sky area covered should be about 50\% larger than the DR4 sample we have 
analyzed here, providing not only smaller error bars, but also narrower window functions as the 
gaps in \fig{AitoffFig} are filled in. 
Global photometric calibration will be improved \cite{PadmanabhanUebercal}.
Various approaches may allow direct measurements of the bias parameter $b$, 
\eg, galaxy lensing \cite{sdssbias}, higher-order correlations \cite{Verde02}, halo luminosity modeling \cite{OstrikerVale05} and reionization physics \cite{ChiuFanOstriker03}.
A bias measurement substantially more accurate than our $11\%$ constraint from redshift space distortions would be a powerful degeneracy breaker.
\Fig{2d_OmQnlFig} shows that our other galaxy nuisance parameter, $\Qnl$, is somewhat degenerate with $\Om$, so improved nonlinear modeling that 
reliably predicts the slight departure from linear theory in the quasilinear regime from smaller scale data would 
substantially tighten our cosmological parameter constraints. More generally, any improved modeling that allows inclusion of higher $k$ will help.
 
As a result of such data progress in many areas, parameter constraints will clearly keep improving. How good is good enough?
The baryon density $\ob$ is a parameter over which it is tempting to declare victory and move on: The constraints on it from 
cosmic clustering are in good agreement, and are now substantially tighter than those from the most accurate competing technique against which it can be cross-checked
(namely Big Bang nucleosynthesis), and further error bar reduction appears unlikely to lead to qualitatively new insights.
In contrast, there are a number of parameters where cosmic clustering constraints are only now beginning to bump up against 
theory and other measurements, so that further sensitivity gains give great discovery potential.
We have $(\ns,r,\al,\Ok)$ to test inflation,
$\Mnu$ to cosmologically detect neutrino mass, $w$ and more generally $X(z)$ to constrain dark energy,
and $\sigma_8$ to resolve tension between different cosmological probes.

Cosmology has now evolved from Alan Sandage's ``search for two numbers" $(h,\Om)$ 
to Alan Alexander Milne's 
``Now we are six''
$(h,\Ob,\Oc,\sigma_8,\ns,\tau)$.
Each time a non-trivial value was measured for a new parameter, nature gave up a valuable clue. For example, $\Oc>0$ revealed the existence of dark matter, 
$\Ol>0$ revealed the existence of dark energy and the recent evidence for $\ns<1$ may sharpen into a powerful constraint on inflation.
Milestones clearly within reach during the next few years include a measurement of 
$\ns<1$ at high significance and $\Mnu>0$ from cosmology to help uncover the neutrino 
mass hierarchy. If we are lucky and $r\sim 0.1$ (as suggested by classic inflation and models such as \cite{nflation}), 
an $r>0$ detection will push the frontier of our ignorance back to $10^{-35}s$ 
and the GUT scale. Then there is always the possibility of a wild surprise
such as $\Ot\ne 1$, large $|\al|$, $X(z)\ne 1$, demonstrable non-Gaussianity, 
isocurvature contributions, or something totally unexpected.
Our results have helped demonstrate that challenges related to survey geometry, bias and potential systematic errors can be overcome, giving galaxy clustering 
a valuable role to play in this ongoing quest for greater precision 
measurements of the properties of our universe.

\bigskip
{\bf Acknowledgments:}

We thank Angelica de Oliveira-Costa, Kirsten Hubbard, Oliver Zahn and Matias Zaldarriaga for helpful comments, 
and Dulce Gon\c{c}alves de Oliveira-Costa for ground support.
We thank the WMAP team for making data and Monte Carlo Markov Chains public via
the Legacy Archive for Microwave Background Data Analysis (LAMBDA) 
at \url{http://lambda.gsfc.nasa.gov}, and 
Anthony Lewis \& Sarah Bridle for making their CosmoMC software \cite{CosmoMC} available 
at \url{http://cosmologist.info/cosmomc}.
Support for LAMBDA is provided by the NASA Office of Space Science.

MT was supported by NASA grants NAG5-11099 and NNG06GC55G,
NSF grants AST-0134999 and 0607597, the Kavli Foundation, and fellowships from the David and Lucile
Packard Foundation and the Research Corporation. DJE was supported by 
NSF grant AST-0407200 and by an Alfred P.\ Sloan Foundation fellowship.

Funding for the SDSS has been provided by the Alfred P.~Sloan Foundation, the Participating Institutions, the National
Science Foundation, the U.S.~Department of Energy, the National Aeronautics and Space Administration, the Japanese Monbukagakusho, the
Max Planck Society, and the Higher Education Funding Council for England. The SDSS Web Site is \url{http://www.sdss.org}.

The SDSS is managed by the Astrophysical Research Consortium for the Participating Institutions. The Participating Institutions are the
American Museum of Natural History, Astrophysical Institute Potsdam, University of Basel, Cambridge University, Case Western Reserve
University, University of Chicago, Drexel University, Fermilab, the Institute for Advanced Study, the Japan Participation Group, Johns
Hopkins University, the Joint Institute for Nuclear Astrophysics, the Kavli Institute for Particle Astrophysics and Cosmology, the
Korean Scientist Group, the Chinese Academy of Sciences (LAMOST), Los Alamos National Laboratory, the Max-Planck-Institute for
Astronomy (MPIA), the Max-Planck-Institute for Astrophysics (MPA), New Mexico State University, Ohio State University, University of
Pittsburgh, University of Portsmouth, Princeton University, the United States Naval Observatory, and the University of Washington.

\appendix

\section{Power spectrum estimation details}

\subsection{Relation between methods for measuring the power spectrum and correlation function}
\label{MethodComparisonSec}

In this section, we clarify the relationship between different popular techniques for quantifying galaxy clustering with pair-based statistics, 
including correlation function estimation with the ``DD-2DR+RR'' method \cite{LandySzalay93,Hamilton93}
and power spectrum estimation with the FKP \cite{FKP}, FFT \cite{Percival01,Cole05,Huetsi06a,Huetsi06b,Percival06} and PKL 
\cite{uzc,pscz,2df,sdsspower,sdss3dkl,Hamilton05} methods.

Suppose we have $N_d$ data points giving the comoving redshift space position vectors $\r_i$ of galaxies numbered $i=1,N_d$, and
$N_r$ random points $\s_i$ from a mock catalog which has the same selection function $\nbar(\r)$ as the real data.
The number densities of data points and random points are then sums of Dirac $\delta$-functions:
\beqa{nDefEq}
n_d(\r)&=&\sum_{i=1}^{N_d}\delta(\r-\r_i),\\
n_r(\r)&=&\sum_{i=1}^{N_r}\delta(\r-\s_i).\label{nDefEq2}
\eeqa
By definition of the selection function $\nbar(\r)$, the quantity 
\beq{deltahatDefEq}
\deltahat(\r)\equiv {n_d(\r)-\alpha n_r(\r)\over\nbar(\r)},
\eeq
where $\alpha\equiv N_d/N_r$, is then an unbiased estimator of the underlying density fluctuation field $\delta(\r)$ in the sense that
$\expec{\deltahat}=\delta$, where the averaging is over Poisson fluctuations as customary.
Except for the PKL method, all techniques we will discuss take the same general form, weighting galaxy pairs
in a form that depends only on the position of each galaxy and on the distance between the two, so we will now describe them all with a unified notation.
(For an even more general pair-weighting formalism that also incorporates the PKL method, see \cite{galpower}.)
As long as one uses $N_r\gg N_d$ random points, they will contribute negligible Poisson noise; their role is in effect to evaluate certain
cumbersome integrals by Monte Carlo integration.

Let us define the quantity 
\beq{xihatDefEq}
\xihat[f]\equiv \int\!\int w(\r)\deltahat(\r)w(\r')\deltahat(\r')f(|\r-\r'|)d^3r d^3r'.
\eeq
Here $w(\r)$ and $f(d)$ are the above-mentioned weight functions that depend on position and distance, respectively.
As we will see, the ``DD-2DR+RR'', FKP and FFT methods simply correspond to different choices of $w$ and $f$.
Substituting equations\eqn{nDefEq}-(\ref{deltahatDefEq}) into \eq{xihatDefEq}, we find that 
\beq{xihatEq}
\xihat[f]=\xihat_{dd}[f] - 2\xihat_{dr}[f] + \xihat_{rr}[f],
\eeq
where we have defined
\beqa{xihatddEq}
\xihat_{dd}[f]&\equiv&\sum_{i=1}^{N_d}\sum_{j=1}^{N_d}{w(\r_i)w(\r_j)\over\nbar(\r_i)\nbar(\r_j)}f(|\r_i-\r_j|),\\
\xihat_{dr}[f]&\equiv&\alpha\sum_{i=1}^{N_d}\sum_{j=1}^{N_r}{w(\r_i)w(\s_j)\over\nbar(\r_i)\nbar(\s_j)}f(|\r_i-\s_j|),\\
\xihat_{rr}[f]&\equiv&\alpha^2\sum_{i=1}^{N_r}\sum_{j=1}^{N_r}{w(\s_i)w(\s_j)\over\nbar(\s_i)\nbar(\s_j)}f(|\s_i-\s_j|),\label{xihatrrEq}
\eeqa

\begin{figure} 
\centerline{\epsfxsize=\figsize\epsffile{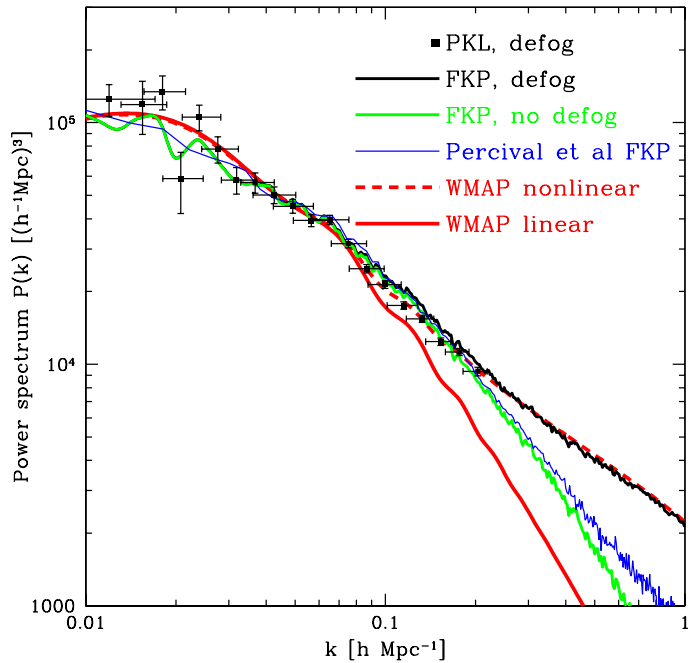}}
\caption[1]{\label{PercivalComparisonFig}\footnotesize%
Comparison of power spectrum estimation techniques.
Our FKP measurement without defogging is seen to agree quite well with the measurement 
of \cite{Percival06} considering that the latter includes also main sample galaxies with different
$\beta$ and small-scale clustering properties.
These curves cannot be directly compared with the PKL measurements or theoretical models, 
because they are not corrected for the effects of redshift distortions,
window functions and the integral constraint; the qualitative agreement that is nonetheless seen 
is as good as one could expect given these caveats.
}
\end{figure}

As a first example, let us consider the FKP method \cite{FKP}.
This corresponds to \cite{galpower}
\beqa{FKPeq1}
f(d)&=&j_0(kd),\\
w(\r)&\propto&{\nbar(\r)\over 1+\nbar(\r)P(k)},\label{FKPeq2}
\eeqa
and turns $\xihat$ into the FKP estimator of the window-convolved power spectrum $P(k)$.
Here $j_0(x)\equiv\sin(x)/x$, $w$ is normalized so that $\int w(\r)^2 d^3r=1$ and $P$ 
is an {\it a priori} guess as to what the galaxy power spectrum is.
For details, see \cite{galpower} around equations (25) and (56).
The main point is that Fourier transforming $\deltahat$ and averaging $|\deltahat(\k)|^2$ over a spherical shell in $k$-space gives
the factor $\int e^{-i\k\cdot|\r-\r'|}d\Omega_k/4\pi=j_0(k|\r-\r'|)=f$.
We apply this method to our LRG data and compare the results with those of \cite{Percival06}
in \Fig{PercivalComparisonFig}, finding good agreement.

The FFT method \cite{Percival01,Cole05,Huetsi06a,Huetsi06b,Percival06} is identical to the FKP 
method except for two simplifications: $P$ in \eq{FKPeq2} is taken to be a $k$-independent constant and the 
density field is binned onto a three-dimensional grid to replace the time-consuming double sums above with a fast Fourier transform.

The ``DD-2DR+RR'' method \cite{LandySzalay93,Hamilton93}
estimates the correlation function $\xi(r)$ by the Landy-Szalay estimator 
\beq{xihatEq2}
\xihat_{LS}={\xihat_{dd} - 2\xihat_{dr} + \xihat_{rr}\over \xihat_{rr}},
\eeq
which is often written informally as $(DD-2DR+RR)/RR$.
Here two common weighting choices in the literature are 
$w(\r)=\nbar(\r)$ \cite{LandySzalay93}
and $w(\r)=\nbar(\r)/[1+\nbar(\r)J]$ \cite{Hamilton93}, where $J\equiv\int_0^r\xi(r')d^3r'$
tends to be of the same order of magnitude as $P(k)$.
To measure the binned correlation function using equations~(\ref{xihatddEq})-(\ref{xihatrrEq}),
one thus sets $f(d)=1$ when $d$ is inside the bin and $f(d)=0$ otherwise. 

These close relationships between the FKP, FFT and ``DD-2DR+RR'' methods lead to interesting conclusions
regarding all three methods.

First, it can be interesting for some applications to replace $J$ by $P$ when measuring the correlation function, 
using $w(\r)=\nbar(\r)/[1+\nbar(\r)P]$, as was done for the analysis of the QDOT survey in \cite{FKP}
and for the LRG analysis in \cite{sdssbump}.
For instance, one could use a constant $P$ evaluated at the baryon wiggle scale if the goal is
to measure the baryon bump in the correlation function.

Second, there is an interesting equivalence between the methods.
For reasons that will become clear below, let us refer to the numerator of \eq{xihatEq2}, 
$\xihat_{dd} - 2\xihat_{dr} + \xihat_{rr}$, 
as the ``convolved'' correlation function estimator and full expression $\xihat_{LS}$ as the
``deconvolved'' estimator.
The information content in the convolved and decovolved estimators is clearly the same, since 
dividing by $\xihat_{rr}$ in \eq{xihatEq2} is a reversible operation. 
Moreover, it is straightforward to show that the FKP estimator of $P(k)$ is simply the 3D Fourier transform 
of the convolved correlation function estimator 
as long as the same weighting function $w(r)$ is used for both. \cite{SzapudiSzalay98} also comment on this.
(Note that this is a quite different statement from the well-known fact that $P(k)$ is the 3D Fourier transform of 
the correlation function $\xi(r)$.)
This implies that the measured FKP power spectrum and the measured correlation function contain exactly the same 
information. In particular, it means that cosmological constraints from one are no better than cosmological constraints 
from the other, since they should be identical as long as window functions, covariance matrices, \etc, are handled
correctly. (An analogous correspondence for purely angular data is discussed in \cite{Szapudi01}.)
In contrast, the information content in the PKL measurement of the power spectrum is {\it not} identical; 
it uses a more general pair weighting than \eq{xihatDefEq} and
by construction contains more cosmological 
information; a more detailed discussion of this point is given in Appendix A.3 in \cite{sdsscl}.

Third, this Fourier equivalence between the convolved correlation function estimator and the FKP power spectrum estimator
sheds light on the fact that the deconvolved correlation function estimator $\xihat_{LS}$ is unbiased
($\expec{\xihat}_{LS}(d)=\xi(d)$, the true correlation function), whereas the expectation value of the FKP estimator is 
merely the true power spectrum convolved with a so-called window function.
This difference stems from the division by $\xihat_{rr}$ in \eq{xihatEq2}: 
Multiplication by $\xihat_{rr}$ in real space corresponds
to convolution with the Fourier transform of $\xihat_{rr}$ (the window function) in Fourier space.
The reason that one cannot deconvolve this windowing in Fourier space is that one cannot Fourier transform $\xihat_{LS}$,
as it is completely unknown for large $d$-values that exceed all pair separations in the survey.

Fourth, this equivalence implies that gridding errors in the 3D FFT method (which become important at large $k$ \cite{Huetsi06a}) 
can be completely eliminated by
simply computing the correlation function with $w(\r)=\nbar(\r)/[1+\nbar(\r)P]$ by summation over pairs and 
then transforming the convolved correlation function with the kernel $j_0(kr)$.

\Fig{PercivalComparisonFig} compares the LRG power spectra measured with the 
different techniques discussed above.
A direct comparison between our PKL $P(k)$-measurement and that of \cite{Percival06} is complicated both 
by window function effects and 
by the fact that the latter was performed in redshift space without FOG compression, with SDSS MAIN galaxies mixed in with the LRG sample. 
To facilitate comparison, we performed our own FKP analysis using the direct summation 
method as described above, with constant $P=30000(h^{-1}\Mpc)^3$ and $\alpha\approx 0.06$.
This is seen to agree with the measurement of \cite{Percival06} to within a few percent 
for $0.04h/\Mpc<k<0.2h/\Mpc$ for the case of no defogging, with the remaining differences presumably 
due mainly to the inclusion of main-sample galaxies, particularly on small scales where nonlinear
behavior becomes important. 
\Fig{PercivalComparisonFig} also shows that 
our defogged FKP measurements agree qualitatively between the PKL and FKP techniques, 
and that the FKP power spectrum continues to track our nonlinear WMAP model beautifully
all the way out to $k=1h/\Mpc$ even though $\Qnl$ was only fit to the $k<0.2h/\Mpc$ PKL data.

An important caveat must be borne in mind when interpreting \fig{PercivalComparisonFig}:
The PKL points are constructed in such a way as to allow direct visual comparison with 
a model power spectrum \cite{sdsspower}, but the FKP and \cite{Percival06} curves are not, and should
not be expected to fall right on top the PKL points or the best-fit cosmological model
because they are not corrected for the effects of redshift distortions, 
window functions and the integral constraint.
Redshift distortions should boost the FKP LRG curves slightly above the true real space power
spectrum (see \Sec{BetaSec}), and should boost the curve from \fig{PercivalComparisonFig} slightly more
because the main sample galaxies have a higher $\beta$ than the LRGs.
The FKP window functions are broader than their PKL counterparts, and the steeper the power spectrum is, 
the more power leaks in from larger scales, causing the plotted measurements to lie above the true
power spectrum. Finally, the integral constraint suppresses the plotted FKP power on the largest scales.
In conclusion, the agreement seen in \fig{PercivalComparisonFig} is as good as one could
expect given these many caveats.

\subsection{Numerical acceleration of the PKL method}

In this section, we describe a numerical improvement over the PKL power spectrum estimation method described in \cite{sdsspower} that enables us to increase the number 
of modes from 4000 to 42{,}000.

The cosmological information content in a galaxy redshift survey, quantified by the Fisher information matrix 
\cite{Karhunen47,karhunen,galfisher}, scales approximately as the effective volume $\Veff$ defined in \eq{VeffDefEq}, with error bars 
on cosmological parameters optimally measured from the survey scaling as $\Veff^{-1/2}$.
However, actually extracting all this information in a numerically feasible way is far from trivial, contributing to the
extensive literature on power spectrum estimation methods.

Our PKL method expands the galaxy density field in $N$ functions (``PKL modes'') that probe successively smaller scales, 
and the number of modes needed to retain all information down to some length scale $\lambda=2\pi/k$ is clearly 
of order $\Veff/\lambda^3$.
In \cite{sdsspower}, $N=4000$ modes were used, and it was empirically determined that this retained essentially all information
for $k\simlt 0.1h/$Mpc with a gradual tapering off towards smaller scales.
This was a convenient coincidence, since using $N\gg 4000$ becomes numerically painful:
because of the many $N\times N$ matrix operations involved in the analysis, the disk usage is 
about 80 GB times $(N/4000)^2$ and the CPU time required on a current workstation is 
about 20 days times $(N/4000)^3$. 

The effective volume of our LRG sample is about ten times larger than that of the above-mentioned main galaxy
analysis because the sky area covered has increased and because the sample is significantly deeper. 
To extract all the $k\simlt 0.1h$/Mpc information, we would therefore like to use about ten times more modes,
but without the analysis taking $10^3$ times longer ($\sim 50$ years).

We therefore combine the method of \cite{sdsspars} with a divide-and-conquer approach, performing a separate
$2000$-mode analysis on each of the 21 sub-volumes described in \Sec{DataSec} (3 radial $\times$ 7 angular subsets)
and combining the results with minimum-variance weighting (which, following the notation of \cite{sdsspower}, 
corresponds to simply summing both 
the $\F$-matrices and the $\q$-vectors). Although this combined analysis with its $21\times 2000=42{,}000$ modes becomes 
lossless in the information theory sense on scales 
substantially smaller than each of the 21 sub-volumes,
it destroys most of the information on scales comparable to these volumes, because the mean density in each volume is projected out 
(effectively marginalized over) \cite{sdsspower}. It also becomes suboptimal on these largest scales because it neglects correlations between
different sub-volumes when optimizing the pair weighting.
We therefore complement the combined analysis with a $4000$-mode global analysis of the entire volume, which is optimal on the largest scales.

Both of these analyses produce uncorrelated band power estimators, and we use the first 8 (with $k<0.04h/$Mpc) 
from the global analysis and the remaining ones from the combined analysis.
This splice point was chosen because the Fisher matrices show that the global analysis contains the most information
(gives the smallest power spectrum error bars) for smaller $k$, and the combined analysis contains the most information
for larger $k$. 
For the radial subsamples, the corresponding splice points are after bands 11 (NEAR), 10 (MID) and 8 (FAR).
We confirm that, as the above scaling arguments suggest, the two analyses give essentially identical
results in the intermediate $k$-range where they both retain virtually all the information.
For example, the two analyses agree for band number 9 to about $0.7\%$ in power, 
a difference which is completely negligible compared to the statistical error bars.

\subsection{Redshift space distortion details}
\label{BetaSec}

As described in detail in \cite{sdsspower}, our PKL method produces three estimators 
$(\Pgghat(k),\Pgvhat(k),\Pvvhat(k))$ of the galaxy-galaxy, galaxy-velocity and velocity-velocity 
power spectra $(\Pgg(k),\Pgv(k),\Pvv(k))$.
These estimators are uncorrelated, both with each other and between different $k$-bands, but not unbiased: 
the expectation value of $\Pgghat(k)$, say, includes contributions from all three power spectra. 
As explained in \cite{sdsspower}, we therefore construct our final power spectrum estimator $\Pghat$
as a linear combination of $\Pgghat(k)$, $\Pgvhat(k)$ and $\Pvvhat(k)$ that makes it an
an unbiased estimator of the real-space galaxy power spectrum $\Pgg(k)$.
This linear combination corresponds to the process of marginalizing over the relative amplitudes of $\Pgv(k)$ and $\Pvv(k)$,
which according to equations\eqn{KaiserLimitEq1} and\eqn{KaiserLimitEq2} are $\beta\rgv$ and $\beta^2$, respectively, 
so it can also be thought of as a marginalization over $\beta$ and $\rgv$.

Two ways of forming this linear combination were explored in \cite{sdsspower}, referred to as the 
modeling method and the disentanglement method, respectively. 
The former corresponds to marginalizing over $\beta$ and $\rgv$ globally, treating them as scale-independent constants,
whereas the latter corresponds to treating them as arbitrary functions of $k$ and marginalizing over them separately for each $k$-band.
We used the former approach for the ``official'' $P(k)$-measurement in \cite{sdsspower} that was used for
cosmological parameter estimation, and we make the same choice in the present paper, using only 
$k<0.09h/\Mpc$
data to find the
best-fitting values $(\beta,\rgv)\approx (0.3,1)$.
The latter approach is more conservative, at the price of producing much larger error bars.

To facilitate the interpretation of our thus-measured power spectrum $\Pghat(k)$, it is helpful to re-express it in terms of 
multipoles of the redshift space power spectrum.
In the small-angle (distant observer) approximation where all galaxy pairs subtend a small angle relative to the line of sight,
$(\Pgg,\Pgv,\Pvv)$ reduce to simple linear combinations of the
monopole, quadrupole and hexadecapole power spectra in redshift space
\cite{ColeFisherWeinberg94,Hamilton98}:
\beq{multipole2flavorEq}
\left(\nskip\begin{tabular}{c}
$\Pgg(k)$\\[4pt]
$\Pgv(k)$\\[4pt]
$\Pvv(k)$
\end{tabular}\nskip\right)
=
\left(\begin{tabular}{ccc}
$1$	&$-{1\over 2}$	&${3\over 8}$\\[4pt]
$0$	&${3\over 4}$	&$-{15\over 8}$\\[4pt]
$0$	&$0$	&${35\over 8}$
\end{tabular}\right)
\left(\nskip\begin{tabular}{c}
$\Pmono(k)$\\[4pt]
$\Pquad(k)$\\[4pt]
$\Phexa(k)$
\end{tabular}\nskip\right).
\eeq 
Inverting \eq{multipole2flavorEq} gives
\beq{flavor2multipoleEq}
\left(\nskip\begin{tabular}{c}
$\Pmono(k)$\\[4pt]
$\Pquad(k)$\\[4pt]
$\Phexa(k)$
\end{tabular}\nskip\right)
=
\left(\begin{tabular}{ccc}
$1$	&${2\over 3}$	&${1\over 5}$\\[4pt]
$0$	&${4\over 3}$	&${4\over 7}$\\[4pt]
$0$	&$0$	&${8\over 35}$
\end{tabular}\right)
\left(\nskip\begin{tabular}{c}
$\Pgg(k)$\\[4pt]
$\Pgv(k)$\\[4pt]
$\Pvv(k)$
\end{tabular}\nskip\right).
\eeq
\Eq{multipole2flavorEq} tells us that, in the small angle approximation, the disentanglement method would correspond to 
measuring $\expec{\Pghat(k)}=\Pmono(k)-{1\over 2}\Pquad(k)+{3\over 8}\Phexa(k)=\Pgg(k)$.
The corresponding weights for the modeling method are found by minimizing the variance among the class of all unbiased estimators, 
and thus depend on the detailed survey geometry, the shot noise level, {\etc}
Empirically, we find $\expec{\Pghat(k)}\approx 0.8\Pmono(k)-0.07\Pquad(k)+0.006\Phexa(k)$,
with the weights roughly independent of $k$.
This can be intuitively understood from the fact that the estimators of $\Pquad$ and $\Phexa$ are much noisier
than that for $\Pmono$, and thus get assigned low statistical weight.
If $\Pquad$ and $\Phexa$ were so noisy that they were discarded altogether, only the estimator of $\Pmono$ would be used.
The relation 
$\Pmono(k)=\left(1+{2\over 3}\rgv\beta+{1\over 5}\beta^2\right)\Pgg(k)$ following from 
\eq{flavor2multipoleEq} would then give the simple estimator
$\Pghat(k)=\Pmonohat(k)/\left(1+{2\over 3}\rgv\beta+{1\over 5}\beta^2\right)\approx 0.8\Pmonohat(k)$
for $\beta=0.3$, $\rgv=1$, \ie, weights close to those we find empirically.
Our measured uncertainty in this normalization factor $\left(1+{2\over 3}\rgv\beta+{1\over 5}\beta^2\right)$ is about 
$3\%$ (see \fig{betarFig}), in good agreement with the exact numerical calculation described in \cite{sdsspower}, 
and this translates into an overall $3\%$ calibration uncertainty of our 
measured power spectrum which is perfectly correlated between all $k$-bands.

The fact that the quantity measured by our power spectrum estimator $\Pghat(k)$ is so similar to 
the rescaled redshift space monopole spectrum is convenient,
since it implies that nonlinear simulations of the redshift space power spectrum (as discussed in \Sec{NonlinearModelingSec})
should apply rather well to our results. 
However, it is important to keep in mind that our measurement $\Pghat(k)$ is a more accurate estimator of $\Pgg(k)$ than
the rescaled redshift space power spectrum would be, for several reasons.
First, it never resorts to the small-angle approximation. 
Second, full account is taken of the fact that anisotropic survey geometry can skew the relative abundance 
of galaxy pairs around a single point that are aligned along or perpendicularly to the line-of-sight.
These two caveats matter because $\Pquad(k)$ and $\Phexa(k)$ are undefined except in the small angle limit, 
which could cause the correction factor 
$\left(1+{2\over 3}\rgv\beta+{1\over 5}\beta^2\right)$ to be inaccurate on large scales.
Finally, our estimator $\Pghat(k)$ by construction has smaller error bars than a standard FKP estimator of the redshift space power
spectrum, and one expects this advantage to be most important on the largest scales, comparable to and exceeding the thickness
of the slices seen in \fig{AitoffFig}.

\subsection{How spacetime geometry affects the power spectrum measurement}
\label{LikelihoodSec}

We performed our power spectrum analysis in comoving three-dimensional space, with the conversion of 
redshifts into comoving distances performed for a fiducial flat $\Lambda$CDM model with $\Om=0.25$.
As described in \Sec{StandardRulerSec}, 
the conversion between redshift and comoving distance (measured in $h^{-1}\Mpc$) depends on the 
cosmological parameters $(\Om,\Ot,w)$, so if 
a different fiducial model had been used for the conversion, then the inferred three-dimensional galaxy 
distribution in comoving coordinates would be radially dilated. As discussed in \cite{sdssbump} and 
\Sec{StandardRulerSec}, this would approximately dilate the dimensionless power spectrum $k^3 P(k)$ by scaling the
$k$-axis by a factor
\beq{aDefEq}
a\equiv{\dV(z)\over\dV^{\rm fiducial}(z)},
\eeq
where $\dV(z)$ is given by \eq{dVdefEq}
and $z=0.35$ is the median survey redshift.
For the parameter range allowed by WMAP3 and our LRG data, 
\beq{dVApproxEq}
a\approx\left({\Om\over 0.25}\right)^{-0.087} (-w \Ot)^{0.19}.
\eeq
This means that the typical correction is very small: the rms scatter in the scaling factor $a$ is 0.7\% for vanilla models, 
1\% for curved models and 3\% for $w$-models.
For example, increasing the fiducial $\Om$-value by 25\%, from 0.24 to 0.30, alters the scaling factor by 2\%
and, since the power spectrum turnover scale $\propto\Om$, ignoring this correction could potentially bias the measured $\Om$-value from 
$0.240$ to $0.245$.

To be conservative, we nonetheless correct for this scaling effect in our likelihood software.
Reanalyzing the galaxy data with the fiducial model replaced by the one 
to be tested would shift the measured $P(k)$ curve up to the left on a log-log plot if $a>1$, 
with $k\mapsto k/a$ and $P\mapsto P a^3$.
We therefore apply the opposite scaling ($k\mapsto k a$ and $P\mapsto P/a^3$) to the theoretically predicted power spectrum
$P(k)$ before computing its $\chi^2$ against our measurement power spectrum from Table~\PowerTable.
We repeated our entire power spectrum analysis for $\Om=0.30$ and confirmed that this scaling is accurate.
Our likelihood software, which is available at \url{http://space.mit.edu/home/tegmark/sdss/}, 
evaluates $a$ exactly instead of using \eq{dVApproxEq}.

In summary, the correction discussed in this section is quite small, especially since marginalizing over bias erases the effect of the 
$a^3$ amplitude shift, but we include it anyway to ensure that there is no bias on cosmological parameter estimates.

\end{document}